\pdfoutput=1
\documentclass[preprint2]{aastex}

\usepackage{graphicx}
\usepackage{amssymb}
\usepackage{amsmath}
\usepackage{epstopdf}
\usepackage{fancyhdr}
\usepackage{hyperref}

\newcommand*{\brokenurl}[2]{\href{#1#2}{\texttt{#1}}\par\nopagebreak\href{#1#2}{\texttt{#2}}}
\begin{document}

\title{The IPAC Image Subtraction and Discovery Pipeline for the
       {\it intermediate} Palomar Transient Factory}

\shorttitle{IPAC iPTF Image Differencing and Discovery}

\author{
  Frank J. Masci\altaffilmark{1},
  Russ R. Laher\altaffilmark{2},
  Umaa D. Rebbapragada\altaffilmark{3},
  Gary B. Doran\altaffilmark{3},
  Adam A. Miller\altaffilmark{3,4,7},
  Eric Bellm\altaffilmark{4},
  Mansi Kasliwal\altaffilmark{4},
  Eran O. Ofek\altaffilmark{5},
  Jason Surace\altaffilmark{2},
  David L. Shupe\altaffilmark{1},
  Carl J. Grillmair\altaffilmark{2},
  Ed Jackson\altaffilmark{2},
  Tom Barlow\altaffilmark{4},
  Lin Yan\altaffilmark{1},
  Yi Cao\altaffilmark{4},
  S. Bradley Cenko\altaffilmark{6},\\
  Lisa J. Storrie-Lombardi\altaffilmark{2},
  George Helou\altaffilmark{1},
  Thomas A. Prince\altaffilmark{4}, and
  Shrinivas R. Kulkarni\altaffilmark{4}}

\email{fmasci@caltech.edu}

\altaffiltext{1}{Infrared Processing and Analysis Center, California Institute
                 of Technology, MS 100-22, Pasadena, CA 91125, USA}

\altaffiltext{2}{Spitzer Science Center, California Institute of Technology,
                 MS 314-6, Pasadena, CA 91125, USA}

\altaffiltext{3}{Jet Propulsion Laboratory, California Institute of Technology,
                 Pasadena, CA 91109, USA}

\altaffiltext{4}{Cahill Center for Astrophysics, California 
                 Institute of Technology, Pasadena, CA 91125, USA}

\altaffiltext{5}{Benoziyo Center for Astrophysics, Weizmann Institute of 
                 Science, 76100 Rehovot, Israel}

\altaffiltext{6}{Astrophysics Science Division, NASA/Goddard Space Flight
                 Center, MC 661, Greenbelt, MD 20771, USA}

\altaffiltext{7}{Hubble Fellow}

\date{\today}

\begin{abstract}

We describe the near real-time transient-source discovery engine for the 
{\it intermediate} Palomar Transient Factory (iPTF), currently in operations
at the Infrared Processing and Analysis Center (IPAC), Caltech. We coin this
system the IPAC/iPTF Discovery Engine (or IDE). We review the algorithms
used for PSF-matching, image subtraction, detection, photometry, and
machine-learned (ML) vetting of extracted transient candidates. We also
review the performance of our ML classifier. For a limiting signal-to-noise
ratio of 4 in relatively unconfused regions, {\it bogus} candidates from
processing artifacts and imperfect image subtractions outnumber {\it real}
transients by $\simeq 10:1$. This can be considerably higher for image data
with inaccurate astrometric and/or PSF-matching solutions. Despite this
occasionally high contamination rate, the ML classifier is able to identify
real transients with an efficiency (or completeness) of $\simeq 97$\% for a
maximum tolerable false-positive rate of 1\% when classifying raw
candidates. All subtraction-image metrics, source features, ML probability-based
{\it real-bogus} scores, contextual metadata from other surveys, and possible
associations with known Solar System objects are stored in a relational
database for retrieval by the various science working groups. We review
our efforts in mitigating false-positives and our experience in optimizing
the overall system in response to the multitude of science projects underway
with iPTF. 

\end{abstract}


\maketitle


\section{Introduction}\label{intro}

The Palomar Transient Factory \citep[PTF;][]{rau09} and its successor
survey currently underway, the {\it intermediate} Palomar Transient Factory
\citep[iPTF;][]{srk13} have been advancing our knowledge of the transient,
variable, and dynamic sky at optical wavelengths since March 2009.
From new classes of supernovae \citep{maguire14,white15},
identifying gamma-ray burst optical afterglows \citep{singer15} and
counterparts to gravitational wave triggers \citep{kasliwal16},
exotic stellar outbursts \citep{miller11,tang14}, Milky Way
tomography \citep{sesar13} to near-Earth asteroids \citep{waszczak16}
and comets \citep{waszczak13}, iPTF continues to
deliver\footnote{For a list of all publications to date, see
\brokenurl{http://www.ptf.caltech.edu/iptf}{}}, serving as a
testbed for the development of future time-domain surveys.
iPTF uses a 92-megapixel camera mosaicked into eleven
functional 2048$\times$4096 CCDs covering $7.26\deg^2$ on the Palomar
48-inch Samuel Oschin Schmidt telescope. The single exposures reach a depth of
(Mould) $R\simeq21$ mag (5$\sigma$) in 60 sec. The pixel scale is
$\approx1\arcsec$ and the image quality is $\approx2.2\arcsec$ (median FWHM),
implying the Point Spread Function (PSF) is better than critically sampled
slightly more than 50\% of the time. Further details of the hardware,
survey design, and on-sky performance are described in \citet{law09,law10} and
\citet{ofek12}. An overview of the image pre-processing and photometry
pipelines, and archival system is described in \citet{laher14}.

The near real-time discovery of transients from iPTF imaging data is currently
performed using an image differencing pipeline at the National Energy Research
Scientific Computing Center \citep[NERSC;][]{cao16}. New incoming images
are astrometrically
and instrumentally calibrated, then aligned, PSF-matched, and differenced
with deeper {\it reference} images supplied by the Infrared Processing and
Analysis Center \citep[IPAC, Caltech;][]{laher14}. Transient candidates
are extracted from the differenced images then vetted using a classification
engine \citep{bloom12,rusu14}. The NERSC infrastructure has contributed 
immensely to the success of PTF and iPTF.

We have implemented an enhanced version of the discovery pipeline to complement
the pipeline at NERSC. In 2017, the iPTF project
will be replaced by the Zwicky Transient Facility (ZTF) using a new camera
on the same telescope \citep{bellm14,smith14}. The ZTF camera will have a
field-of-view of $\sim47$ square degrees, enabling a full scan of the
Northern visible sky every night, at a rate $\sim15$ times faster
than iPTF to similar depths. The massive
high-rate data stream and volume expected from ZTF will require advancements
in algorithms and data-management practices despite the (inevitable) growth
in hardware technology. This will pave the way to the Large Synoptic
Survey Telescope \citep[LSST;][]{ivezic14} that is expected to yield at
least $100\times$ as many astrophysical transients per image exposure than ZTF.
In anticipation of this data deluge, we have embarked on a new efficient
discovery pipeline and infrastructure at IPAC.
Our design philosophy is flexibility, i.e., being able to operate
in a range of complex astrophysical environments (including the galactic
plane), robustness to instrumental glitches, adaptability to a wide range of
atmospheric seeing and transparency, minimal tuning (unless warranted by
instrumental changes), optimality (in the signal-to-noise sense), reliability
in extracted candidates to moderately low S/N levels, and fast delivery of
vetted candidates to enable follow-up in near real-time.

Searches for astrophysical transients (by virtue of changes in flux 
and/or position) have traditionally been conducted using
either of two approaches. The first involves differencing of 
astrometrically-aligned, 
PSF-matched images from two epochs: the {\it science} or {\it target} image
containing the potential transient sources, and a deeper {\it reference} or
{\it template} image serving as a static representation of the sky,
for example, defined from an average of images from multiple historical epochs.
The difference image is then thresholded to find
and measure excess signals, i.e, the transient candidates. This approach was
(and in some cases continues to be) used by numerous synoptic surveys, e.g.,
OGLE \citep{wyrzykowski14}, ROTSE \citep{akerlof03}, La Silla-QUEST
\citep{hadjiyska12}, Pan-STARRS \citep{kaiser10}, and PTF \citep{law09}. 
Although simple in theory, a challenging aspect of discovery via image
differencing is the prior matching of PSFs between the input images.
This has lead to an intensive, ongoing research effort
\citep[e.g.,][]{alard98,alard00,wozniak00,yuan08,bramich08,becker12,
bramich16,zackay16}.
The ultimate goal is the elimination of systematic instrumental residuals,
e.g., induced by non-optimal calibrations and/or PSF-matching upstream.
These would otherwise contaminate lists of extracted transient candidates,
i.e., the false positives that would need to be dealt with later (see below).
In practice, one strives to minimize their occurence 
in difference images such that in a global sense, the resulting pixel 
fluctuations and photometric uncertainties of {\it bona fide} flux transients
approach expectations from Poisson noise and/or detector read-noise.

The second approach involves positionally-matching source catalogs extracted
from images at different epochs and searching for large flux differences
between the epochs, e.g., as used by the Catalina Real-time Transient Survey
\citep[CRTS;][]{drake09}. This method avoids systematics from
color-correlated source-position misalignments due to differential chromatic
refraction, an effect that can be severe for some facilities. However,
this method requires a relatively large flux-difference
threshold to ensure reliability. This is at the expense of a higher
missed detection rate (incompleteness) at low flux levels, particularly in
regions with a complex background and/or high source-density
(e.g., the galactic plane) where positional-matching is a
challenge. On the other hand, assuming optimally calibrated and
instrumentally-matched inputs, image differencing excels in regions where
source confusion is high and/or where complex, fast-varying backgrounds
are present (e.g., near or within galaxies).
Due to its adaptability to a wide range of
astrophysical environments, the PTF project adopted image differencing as
its primary means for discovery.

Following the extraction of transient candidates from differenced images,
a somewhat daunting problem is deciding which are {\it bogus}
(i.e., spurious) or {\it real} and worthy of further study. The existing iPTF
discovery pipeline at NERSC accomplishes this using a {\it supervised}
machine-learned (ML) classifier \citep{bloom12,brink13}. Here, a
pre-labelled training set of previously discovered {\it real} transients
are first fit to a two-class ({\it real or bogus}) non-parametric model
described by a number of selected source features (or metrics). This model
is then used to predict the class ({\it real} or {\it bogus}) of future
candidates according to some probability threshold. The probabilities are
also referred to as {\it RealBogus} (or quality) scores.

The iPTF discovery pipeline at NERSC typically yields a few to ten {\it real}
``interesting'' transients per night (excluding Solar-System objects and
periodic or reoccuring variables in regions with a high stellar density).
For ZTF, we expect at least 100 such transients per night.
Currently however, real iPTF transients can be outnumbered by
spurious candidates (false positives) by more than two orders of
magnitude, despite efforts to minimize their incidence through careful
pre-calibration. The problem gets worse if one
is interested in finding the rare gems down to low S/N levels
\citep[e.g.,][]{masci12}. Depending on the science goals,
the vetted candidates need to be delivered in a timely manner to
the respective science working groups for follow-up.
At NERSC, this currently takes $\sim 30$ minutes since observation.
The goal is to get this below $\sim 15$ minutes.
Large numbers of false-positives can strain any machine-learned vetting
process and affect its reliability \citep{brink13}. It is
crucial that the vetting process be efficient and reliable. 

We have developed an automated image-differencing, transient-extraction
and vetting system at IPAC; hereafter, the IPAC/iPTF Discovery Engine
(or IDE). This infrastructure is currently in use for iPTF and is expected
to be a foundation for ZTF in future. We have 6$+$ years of PTF science
data in hand (ongoing with iPTF) and an experienced team at NERSC that aided 
in developing and refining all aspects of a robust discovery
engine---from instrumental calibration to vetted transient candidates.
Guided by previous implementations of the image subtraction problem,
this paper reviews our algorithms, optimization strategies, 
experiences, and liens. We also describe our probabilistic
({\it real}--{\it bogus}) classification scheme for vetting transient
candidates, Quality Assurance (QA) metrics, and database (DB) schema.

We note that two of the core pipeline steps in IDE: (i) image-differencing
(that includes pre-conditioning of image inputs), and (ii) extraction of
{\it raw} transient candidates therefrom, are both implemented in a stand-alone
software module called PTFIDE\footnote{Source code, instructions for
installing external dependencies and examples with test data are available at
\brokenurl{http://web.ipac.caltech.edu/staff/fmasci/home/ptfide}{}}.
In this paper, we use the acronym PTFIDE when referring to
these specific processing steps, otherwise, we use IDE when
referring to the overall processing system. The latter includes all
pre-calibration steps (prior to PTFIDE), machine-learned vetting and
archival steps (post-PTFIDE). Furthermore, when referring to astrophysical
transients, we use the term ``transient'' in a generic sense, i.e., all
types of flux-excesses that can be detected in difference images (in both
the positive and negative sense, relative to a reference image template): 
moving objects, periodic or aperiodic variable sources,
or short-lived (fast) events. The goal of IDE is to deliver reliable
transient candidates to the various science working groups for further
follow-up. From hereon, these science working groups will be
referred to as ``science marshals'', or simply marshals.

This paper is organized as follows. In Section~\ref{realtpl} we give an
overview of IDE and provide references for more information on each subsystem,
both in this paper and elsewhere.  
Section~\ref{revide} gives a broad overview of the image 
differencing and extraction module PTFIDE and its dependencies:
input parameters, reference-image building, and output products.
Section~\ref{psteps} expands on the specific processing steps in
PTFIDE: gain and background matching, astrometric refinement,
reference-image resampling, and PSF-matching. This includes
a summary of all image-based and transient-candidate source metrics, 
and their use in deriving simple initial quality scores.
Section~\ref{schema} reviews the DB schema for storing all
difference image-based and source-based metrics.
The machine-learned vetting infrastructure, which includes training,
tuning, and its overall performance is described in Section~\ref{ml}.
Lessons learned during the course of development and testing 
are given in Section~\ref{lessons}. Future and potential enhancements
are discussed in Section~\ref{enhance} and conclusions are given
in Section~\ref{conc}.

\section{Overview of the Near Real-time Discovery Engine}\label{realtpl}

The raw camera-image files are first sent from the Palomar 48-inch
Samuel Oschin Schmidt telescope to the San Diego Supercomputing Center
via a $\approx100$ Mbit/s microwave link and then pushed to Caltech and IPAC
over the internet (bandwidth is $\gtrsim 1$ Gbit/s).
At IPAC, the camera-image files are ingested into an archive
and associated metadata are stored in a relational database for
fast retrieval and processing soon thereafter (see below).

Figure~\ref{fig:ideflow} gives an overview of the near real-time discovery
pipeline. An executive pipeline wrapper controls the various steps:
preprocessing which performs basic instrumental and astrometric calibration
per CCD image ({\it light purple} boxes); PTFIDE --- the image-differencing
and transient extraction module ({\it red} boxes); then returning to the
pipeline executive for archiving, DB-loading, and machine-learned vetting
({\it light purple} boxes). The preprocessing steps are from
a stripped down version of the PTF/iPTF {\it frame-processing
pipeline}. This executes asynchronously and independently of IDE following
ingestion of an entire night's worth of image data.
The purpose of this pipeline is to provide accurately calibrated images
and source catalogs for future public distribution.
This pipeline and the IDE preprocessing steps borrowed therefrom
({\it light purple} boxes in Figure~\ref{fig:ideflow}) are described
in detail in \citet{laher14}. Below we summarize the major processing
steps. The steps specific to PTFIDE ({\it red} boxes) are expanded in
Sections~\ref{revide}--\ref{ml}. Operational details and tools used
by the various science marshals ({\it green} boxes) will be discussed in
future papers. In particular, the streak-detection functionality that is
designed to detect moving objects in difference images, i.e., that streak 
in individual exposures is described in \citet{waszczak16}. 

\begin{figure*}
\begin{center}
\includegraphics[scale=0.66]{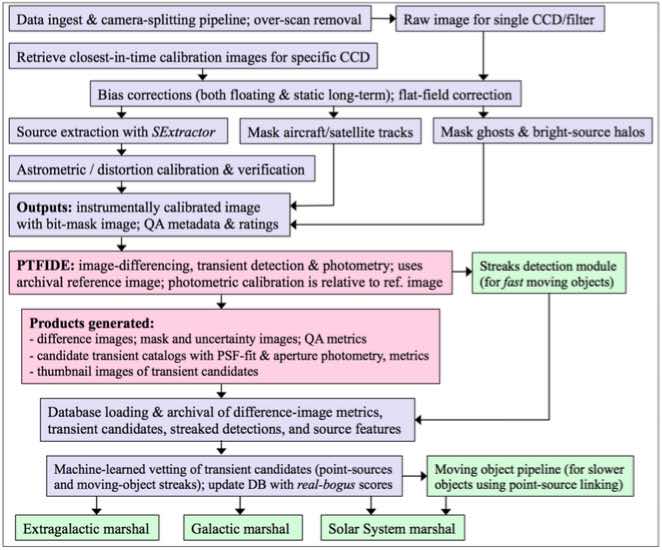}
\caption{Processing flow in the near real-time IPAC/iPTF Discovery Engine
         (IDE). The color-coding separates the various modular steps:
         preprocessing, archival, and machine-learned vetting
         ({\it light purple}); core image-differencing and transient
         extraction module: PTFIDE ({\it red}); external science
         applications and follow-up marshals ({\it green}).
         See Section~\ref{realtpl} for details.}
\label{fig:ideflow}
\end{center}
\end{figure*}

The 92-megapixel raw camera-image files (one per exposure) are processed
by the real-time pipeline soon after they are ingested, check-summed, and
registered in the database at IPAC. The ingest process also loads a
{\it jobs} database table that is automatically queried by the pipeline
executive to initiate the camera-splitting pipeline \citep{laher14}.
This pipeline splits
the camera-images into twelve 17MB CCD image files (with overscan regions
included), and noting that one of the CCDs is defective. An initial
astrometric solution is derived and attached to their
FITS\footnote{FITS stands for Flexible Image Transport System;
see \brokenurl{http://fits.gsfc.nasa.gov}{}} headers. This astrometric
solution is not the final (and best) calibration attached
to the CCD images prior to image-differencing with PTFIDE. It is used to
support source-catalog overlays, quick-look image visualizations, and
quality assurance (QA) from the archive. The individual raw CCD frames in
FITS format are copied to a local sandbox directory and associated metadata
(including quality metrics and image usability indicators) are stored in a
database to facilitate  retrieval for the next processing steps.

A number of preprocessing and instrumental calibration steps are then applied
to the raw CCD image. These include a dynamic (floating) bias correction and
a static bias correction, a flat-field (pixel-to-pixel responsivity)
correction, and cropping to remove overscan regions. The static bias
and flat-field calibration maps are retrieved from an archive. These are
generally the latest (closest-in-time) products available for the night
being processed, i.e., that were made by combining data from a prior 
night. For the flat-field calibration in particular, quality metrics are
used to check that the responsivity pattern falls within the range expected
for a specific CCD and filter. If not, a pristine superflat is used.
This preprocessing also initiates and populates a 16-bit mask image to
record bad hardware pixels for the specific CCD frame, badly calibrated
pixels, and saturated pixels. This mask is further augmented below to
record image artifacts and object detections.

At this stage, we have a bias-corrected, flattened CCD image and
accompanying mask image. Sources are then extracted from the CCD image
using {\it SExtractor} \citep{bertin96, bertin06a} primarily to support
astrometric calibration -- the most important calibration step in the
real-time pipeline since its accuracy is crucial to attaining good quality
difference images (Section~\ref{garef}). The {\it SExtractor} module is
executed twice. The first run is to compute an accurate value of the overall
image seeing (point-source FWHM) from the mode of a filtered distribution of
individual source FWHM values. This estimate is then used to support
source-detection in the second {\it SExtractor} run via a point-source
matched filter. The first {\it SExtractor} run also
folds in the object detections into the image mask, or rather the contiguous
pixels contributing to each object above the specified threshold.
The {\it createtrackimage} module is also executed to detect satellite
and aircraft tracks in the CCD image and record their locations in
the image mask. These occur with a frequency of typically several times
per night and the same track can cross multiple CCDs. Metrics for each
track are also computed (e.g., length and median intensity) and stored in a
database table. For details on track identification and characterization,
see \citet{laher14}. The second {\it SExtractor} run generates a source
catalog for input into the astrometric calibration step.

\begin{figure*}[ht]
\begin{center}
\includegraphics[scale=0.45]{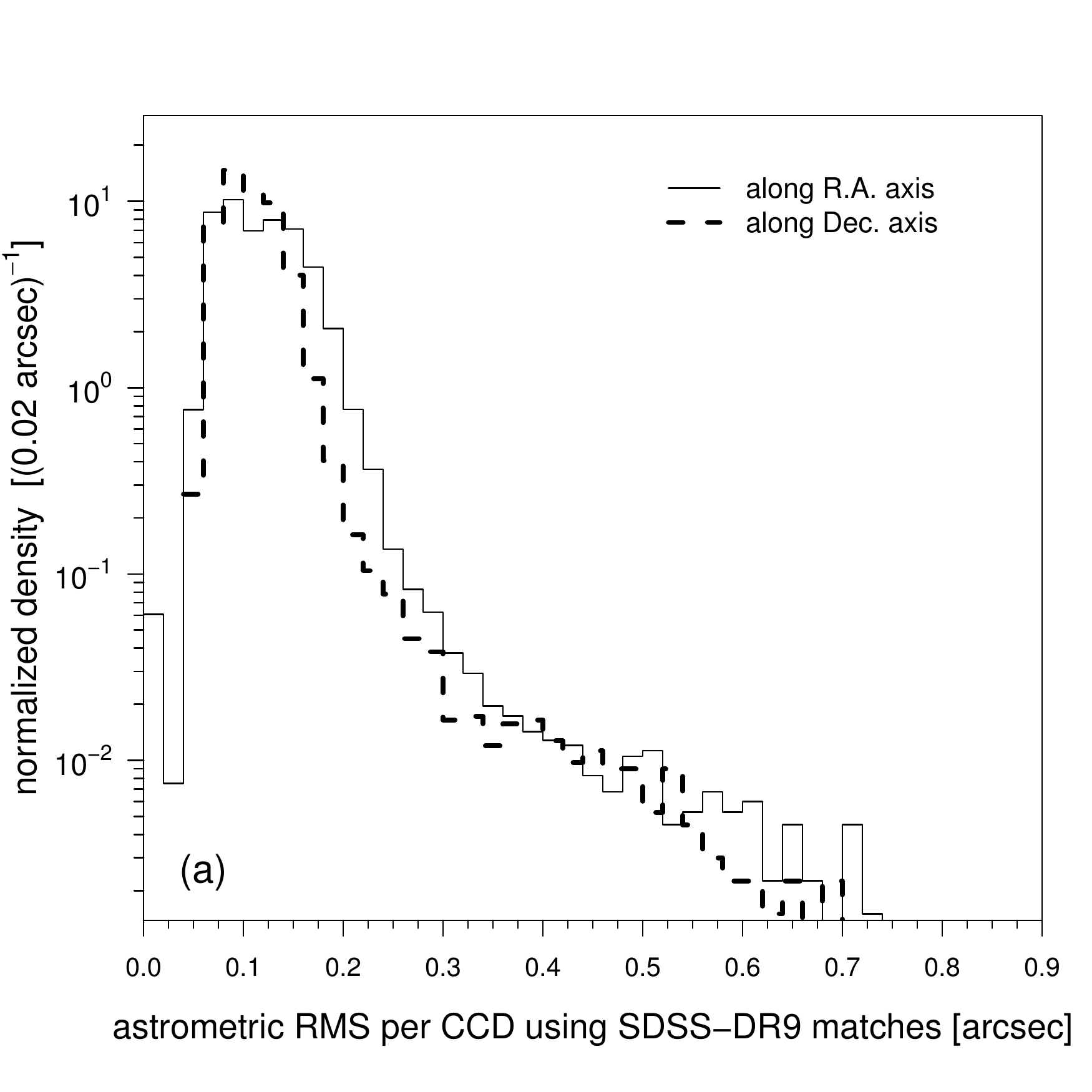}
\includegraphics[scale=0.45]{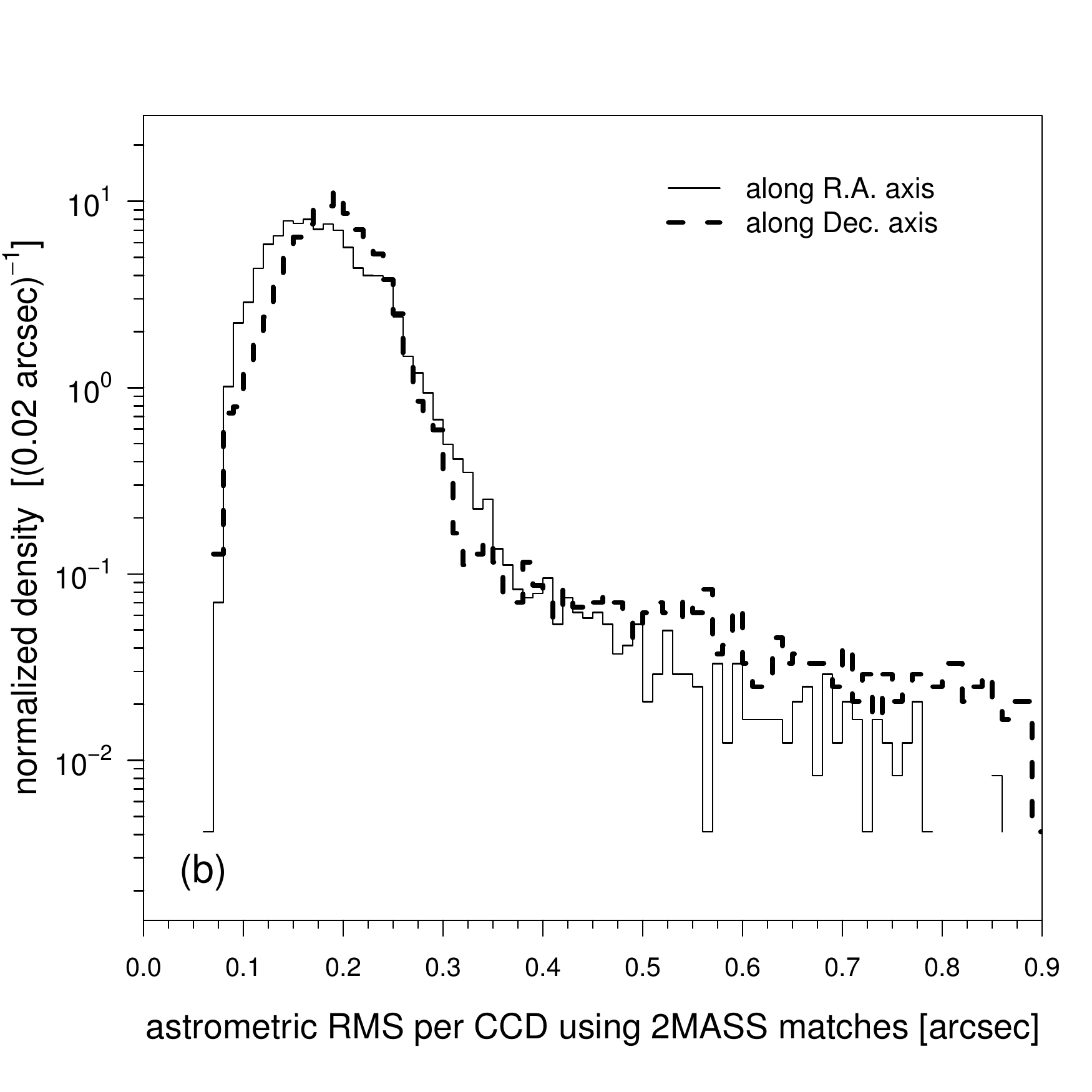}
\caption{Distributions of the astrometric RMS per CCD image along each axis
         with respect to: (a) the SDSS-DR9 Catalog using 68,310 images, and
         (b) the 2MASS PSC using a subset of 24,168 images containing
         sufficient matches. See text for details.}
\label{fig:astrom}
\end{center}
\end{figure*}

Astrometric calibration is initially performed using {\it SCAMP}
\citep{bertin06b, bertin14}. {\it SCAMP} is executed using one of two possible
astrometric reference catalogs as input: if the CCD image overlaps entirely
with a field from the Sloan Digital Sky Survey (SDSS), the {\it SDSS-DR9}
Catalog \citep{ahn12} is used; otherwise, the {\it UCAC4} Catalog
\citep{zacharias13} is used. If {\it SCAMP} fails to find an 
astrometric solution using either of these catalogs, it is rerun
with the {\it USNO-B1} Catalog \citep{monet03}.
In addition to solving for the standard World Coordinate System (WCS)
first-order terms \citep[for a gnomonic sky-projection;][]{calabretta02},
{\it SCAMP} simultaneously solves for field-of-view distortion using
the {\it PV} polynomial convention on a per-image basis. The solution
implicitly captures both the fixed camera-distortion and any variable
atmospheric refraction effects at the time of exposure.
The WCS solution and {\it PV} distortion coefficients are written to 
the CCD image FITS header. To enable other downstream (as well as generic
analysis) software to map from pixel to sky coordinates and vice-versa,
the {\it PV} coefficients are converted to the {\it SIP} representation
\citep{shupe05} using the {\it pv2sip} module \citep{shupe12}. The associated
{\it SIP} coefficients are also written to the FITS header.

The astrometric (and distortion) solution from {\it SCAMP} is then validated.
The first validation step coarsely checks
that the standard first-order WCS terms (pointing, rotation, and scale)
are within their expected ranges according to specific prior values.
The second validation step involves re-extracting sources from the
astrometrically-calibrated CCD image (again using {\it SExtractor}) and
matching them to a filtered subset of sources from the {\it 2MASS} Point
Source Catalog \citep[PSC;][]{skrutskie06}.
A matching radius of $2\arcsec$ is used and a minimum of 20 2MASS matches
must be present. If the number of matches exceeds this minimum, the
axial root-mean-squared (RMS) position differences are root-sum-squared
(RSS'd) and compared against a threshold that is dependent on galactic
latitude. This threshold ($t$) lies in the range
$0.3\arcsec\lesssim t\lesssim 0.7\arcsec$ corresponding to galactic latitudes
$0^{\circ}\lesssim |b|\lesssim 90^{\circ}$. The threshold is interpolated 
from a look-up table of predetermined values according to the galactic
latitude of the input image. The reason for a latitude-dependent threshold
is due to the less reliable RMS estimates in position differences following
source matching when the source density is high. We are less tolerant of
larger RMS estimates in this regime due to the higher probability of false
matches.  

If the above validation checks on the astrometry are not
satisfied, another attempt is made at the astrometric calibration, this time by
executing the {\it Astrometry.net} module \citep{lang10}. This module
uses the 2MASS PSC as the astrometric-reference catalog. {\it Astrometry.net}
also solves for distortion on a per-image basis, however, its
representation is only in the {\it SIP} format. To ensure proper
execution of other downstream pipeline modules that depend exclusively on
the {\it PV} representation, the {\it SIP} coefficients are converted
to {\it PV} equivalents using the {\it sip2pv} module \citep{shupe12}
and written to the FITS header. The solution from {\it Astrometry.net}
is validated in the same manner as above using the 2MASS PSC. If the
acceptability criteria are still not satisfied, a bit-flag is set in
a database table for use downstream. Metrics to assess the astrometric
performance on each image are computed and also stored in the database
to facilitate future analysis and trending \citep[for details, see][]{laher14}.

\begin{figure*}[ht]
\begin{center}
\includegraphics[scale=0.45]{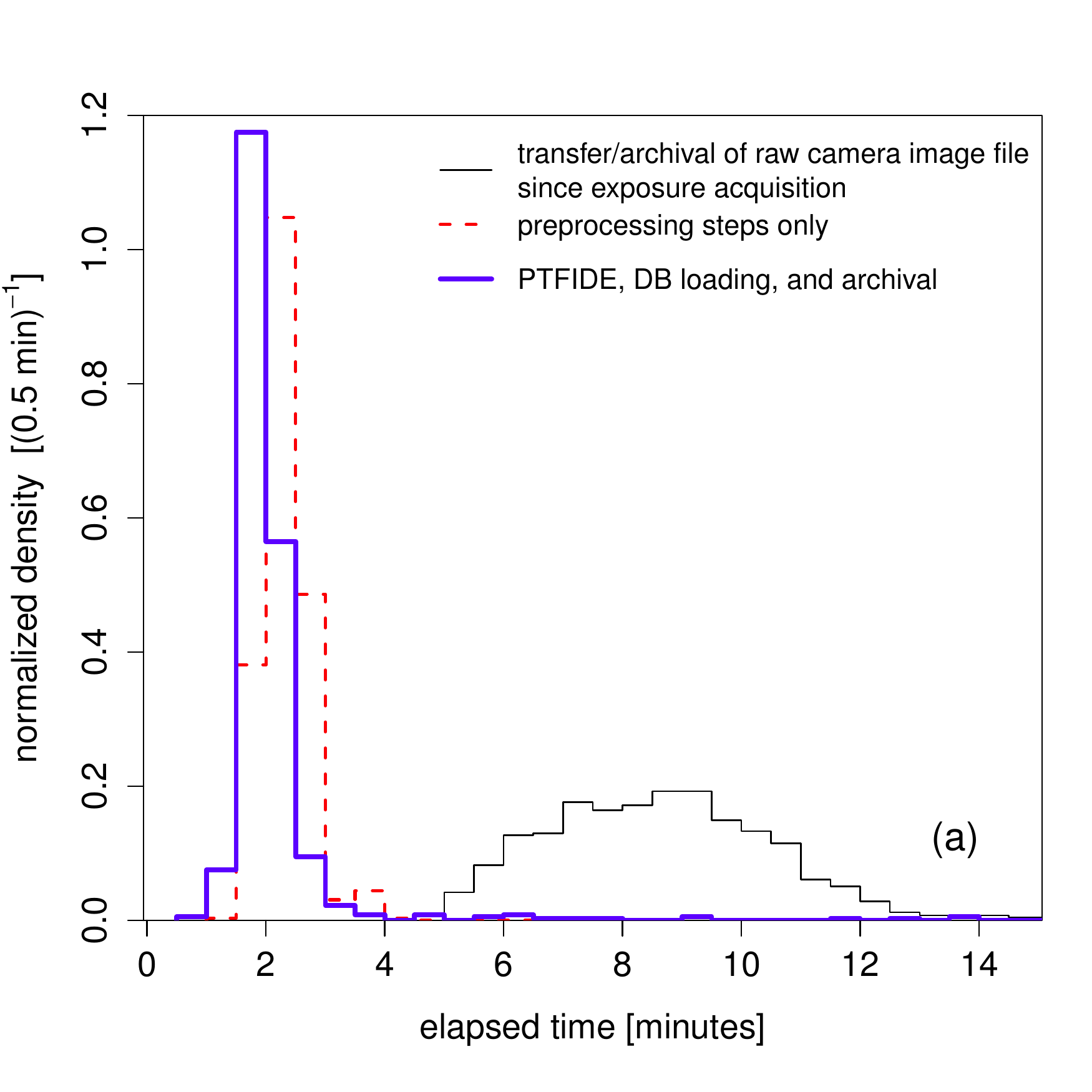}
\includegraphics[scale=0.45]{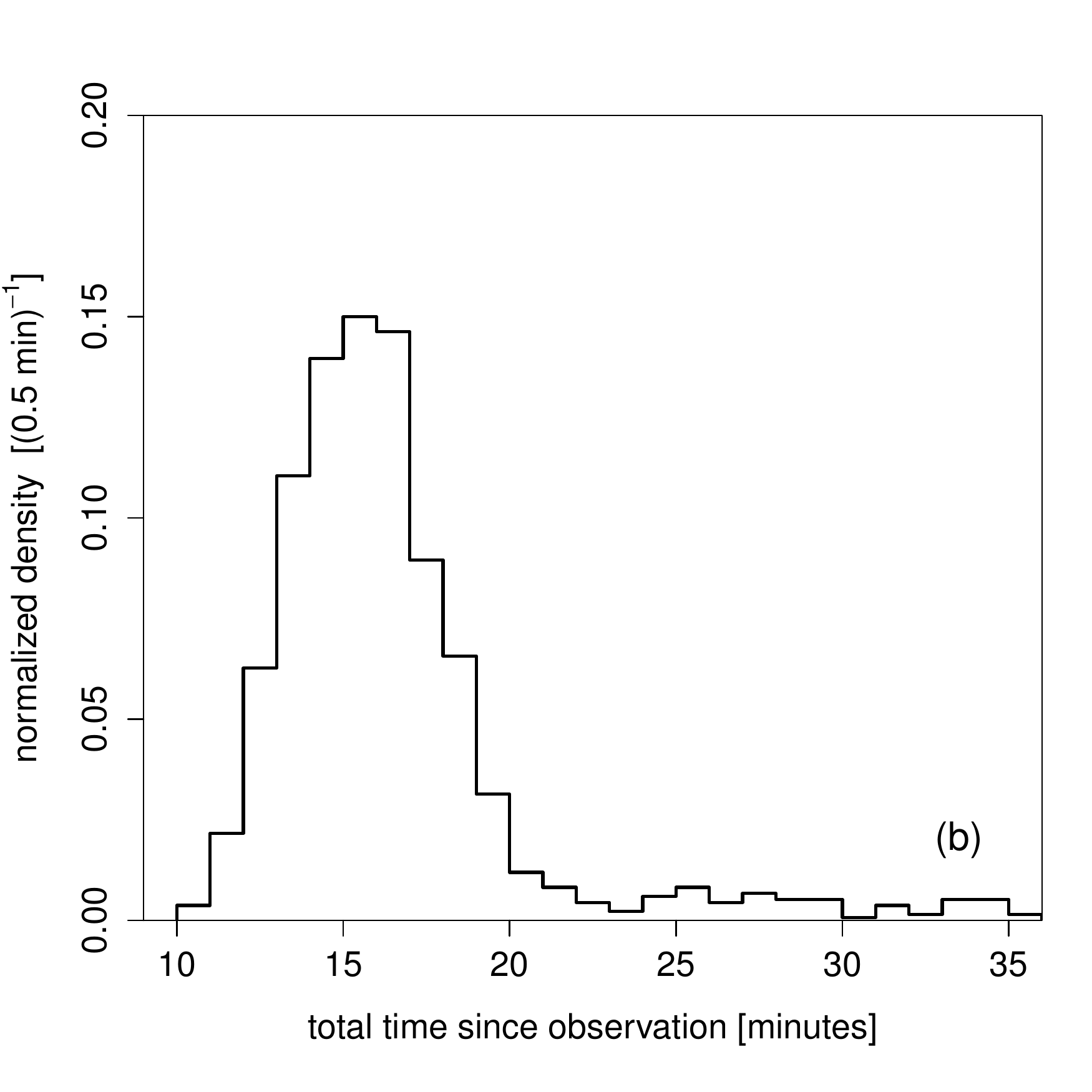}
\caption{(a) Distributions of the elapsed time for various processing steps in
             the (near-)real-time discovery engine {\it per CCD image}.
         (b) Total elapsed time from acquisition of a CCD image exposure
             to vetted candidates extracted therefrom. Metrics were derived
             using 1340 camera-image exposures.}
\label{fig:timing}
\end{center}
\end{figure*}

Figure~\ref{fig:astrom}a quantifies the astrometric performance of the
real-time pipeline for 68,310 iPTF CCD images acquired from 2015 January 1
to 2015 May 1 that used the SDSS-DR9 Catalog in their {\it SCAMP} solution.
This catalog covers $\simeq$ 14,555 $\deg^2$ at galactic latitudes of
typically $|b|\gtrsim 30^{\circ}$ and has an overall astrometric accuracy
(RMS per axis) of $\simeq 50$ milli-arcseconds (mas) with respect to
earlier {\it UCAC} releases \citep{pier03}. The median RMS per iPTF
CCD per axis is typically 115 mas with respect to SDSS-DR9. The astrometric
performance outside the SDSS-DR9 footprint
(calibrated using either UCAC4 or USNO-B1; see above) is similar, except
however for exposures observed in regions with a high source-density (e.g.,
the galactic plane) where systematics are more prevalent. These systematics are
currently being addressed since iPTF includes a number galactic-plane
science programs. Figure~\ref{fig:astrom}b shows
the RMS distributions for a subset of 24,168 CCD images from the same SDSS-DR9
overlap region with respect to the 2MASS PSC. Here, only images
containing $>$ 200 2MASS matches each were used. The median astrometric
accuracy with respect to 2MASS degrades to $\simeq$ 190 mas per axis.
This larger RMS is somewhat expected since the 2MASS PSC has an accuracy
of typically 150-200 mas \citep{skrutskie06}.

Preprocessing in the real-time pipeline also includes a step to detect
and mask artifacts induced by bright-source reflections in the telescope
optics, primarily ghosts and halos. The ghosts are due to bright sources lying
off the telescope's optical axis, while halos are more coincident with
the offending source. These features are located by first searching for
bright (parent) stars from the Tycho-2 Catalog \citep{hog00} with $V<6.2$ mag.
The positions of ghosts are then isolated using a pre-determined
geometric mapping from the parent stars to expected ghost positions.
Circular areas are flagged in the CCD bit-mask image to
indicate probable ghosts and halos. Given the ghost and halo sizes vary with
the brightness of the parent star, a conservative maximally-sized masking
area is used. The positions of ghosts, halos and their parent stars
are stored in the database to facilitate future analysis.
Lastly, the preprocessing phase computes a number of QA metrics for the
image pixels and accompanying mask, a summary of which can be
found in \citet{laher14}. These are also loaded into the database.

There is no absolute photometric calibration in the preprocessing phase of the
real-time pipeline to assign image-specific photometric zeropoints. Instead,
the raw pixel signals are later throughput (gain)-matched to a reference 
image template during PTFIDE processing using sources extracted therefrom
(Section~\ref{garef}).
This reference image has an associated photometric zeropoint and therefore
serves as the generic zeropoint for all real-time products that are matched
to it, including the difference image products downstream. For details, see
Sections~\ref{refcon} and~\ref{psffit}. For an overview on the performance
of the initial photometric calibration of the CCD images (on which the
reference and difference-image products ultimately depend), see \citet{ofek12}.

Figure~\ref{fig:timing} shows the typical durations of the primary steps
in the real-time pipeline: from acquisition of a camera exposure to
vetted candidates, ready to be examined by the science marshals.
The median total time lag since exposure acquisition
(Figure~\ref{fig:timing}b) is $\simeq$ 16 minutes and
the $95^{th}$ percentile is $\lesssim$ 22 minutes. When broken down into
the various steps, the bulk of the lag is in the transfer of image
data from the telescope to IPAC ($\simeq$ 9 minutes). This includes
database ingestion and archiving, which amount to no more than several
seconds per camera exposure. The preprocessing, PTFIDE and final archival
steps amount to no more than $\simeq$ 7 minutes, although there is a
long tail in the PTFIDE runtime which we attribute to the extraction
and processing of transient candidates from ``bad'' difference-images,
i.e., containing an excess of residual artifacts (Section~\ref{train}).
The timing metrics shown in Figure~\ref{fig:timing} are those
inferred at the time of writing using 1340 camera-image files and all
eleven CCD images therein. The overall lag is expected to decrease
in the near future, in particular in the transfer of image-data from the
telescope to IPAC.

At the time of writing (pertaining to iPTF operations), the IDE
pipeline executes on a Linux cluster of 23 machines consisting of 232 64-bit
physical CPU cores in total: 11 machines have 8
Intel\textsuperscript{\textregistered} Xeon\textsuperscript{\textregistered}
cores running at 3.0 GHz each and the remaining 12 machines have 12 similar
cores running at 2.4 GHz each. All the machines, file and
database servers are connected by a 10 Gbit network.
Given that the 12-core machines can admit two threads
per core, this cluster can in principle allow for 376 concurrent processes.
However, since much of the processing involves a considerable amount of
disk I/O, we achieve close to maximum throughput with only one thread per
physical core, and therefore we usually execute at most 232 simultaneous
threads. As raw camera-image files are received during the night,
multiple instances of the camera-splitting pipeline are first run across
all idle processor cores (until filled) to generate the individual raw
CCD-images. These images then enter the processing queue and the level of
core-parallelism now occurs at the CCD-image level through all the
remaining pipeline steps (Figure~\ref{fig:ideflow}).

The IDE pipeline was designed to be flexible enough to also process archival
(preprocessed) image data. This mode facilitates pipeline tuning, iterative
training of machined-learned classifiers in response to changing detector
properties and/or science goals, but it also supports archival research in
general, i.e., ad-hoc discovery projects using different pipeline parameters
and thresholds. This offline execution mode only runs the PTFIDE steps
({\it red} boxes in Figure~\ref{fig:ideflow}) using preprocessed image data
that were previously instrumentally-calibrated and archived by the regular
PTF/iPTF {\it frame-processing pipeline} \citep{laher14}. This is because the
preprocessed intermediate products from the initial phase of the IDE pipeline
({\it light purple} boxes in Figure~\ref{fig:ideflow}) are not stored
in a long-term archive.

To summarize, we have given a general overview of the near real-time IDE
pipeline, with particular emphasis on the preprocessing steps needed
to generate instrumentally and astrometrically calibrated CCD-images
for input into the image-differencing and transient extraction module
(PTFIDE). The primary
outputs from the preprocessing step are a calibrated {\it science} image
exposure, an accompanying {\it bit-mask} image, and metrics that quantify
the astrometric performance and quality of the image-pixel data.
The details on how these metrics and products are used in PTFIDE 
are discussed in Section~\ref{psteps}.

\section{PTFIDE Module Overview and Preliminaries}\label{revide}

This section gives a broad overview of the PTFIDE software\footnote{Source
code, instructions for installing external dependencies and examples with
test data are available at
\brokenurl{http://web.ipac.caltech.edu/staff/fmasci/home/ptfide}{}},
dependencies, design assumptions, input data and formats, tunable parameters,
and outputs -- both primary products for archival and ancillary products
for debug and analysis. PTFIDE is a standalone Unix command-line tool
written in Perl. It calls a number of software executables written
in C, C{}\verb!++! and Fortran. For a summary of the dependencies,
see Section~\ref{swdep}. The software can be built and configured to
run under most Linux or Unix-like operating systems.

\begin{figure*}
\begin{center}
\includegraphics[scale=0.55]{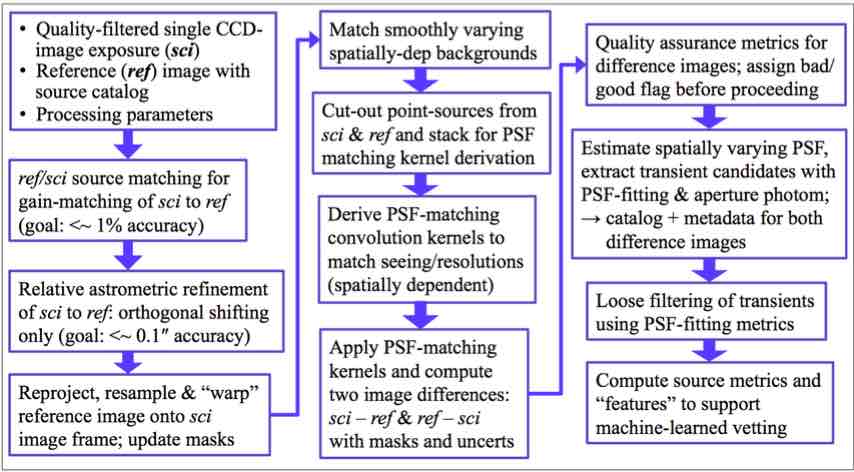}
\caption{Processing flow in the PTFIDE (image differencing and extraction)
         module. These steps are contained in the first red box of the
         real-time pipeline flowchart in Figure~\ref{fig:ideflow}.
         See Section~\ref{psteps} for details.}
\label{fig:ptfideflow}
\end{center}
\end{figure*}

Figure~\ref{fig:ptfideflow} summarizes the main
processing steps in PTFIDE, from preparing the inputs, to extracted
transient candidates and metrics ready for loading into a relational database.
A summary of all input files, parameters, and their default values is given in
Section~\ref{paramsum}. One of the most important inputs is the
{\it reference image} and its accompanying {\it source catalog}. Requirements
regarding its construction are given in Section~\ref{refcon}. Output
products, formats, and their level of importance are summarized in
Section~\ref{prodsum}. The details of each computational step in
Figure~\ref{fig:ptfideflow} are expanded in Section~\ref{psteps}.

\subsection{Software Design Philosophy, Dependencies,
            and Parallelization}\label{swdep}

To expedite the delivery of science quality products following the commencement
of iPTF, some of the processing steps in PTFIDE leverage existing
astronomical software tools. This is mostly heritage software that has been
well tested by the astronomical community and refined over time.
Table~\ref{tab:sw} summarizes the external (third-party) software components
used in PTFIDE and other dependencies.

One of the design goals was robustness against missing or corrupted
input data with appropriate error handling and reporting upon pipeline
termination. Depending on the error, any missing (or out-of-range) data or
associated metadata are replaced with default values in an attempt to salvage
as many products as possible. Warnings are issued and logged if these occur.
Furthermore, different pipeline exit codes are assigned according to
the different anomalies (fatal and benign) encountered in processing.
These status codes are stored as a bit-string in a database table to enable
follow-up or to avoid querying unusable (or non-optimal) science products
in future. Another design consideration was the ability to generate as many
intermediate products and write as much information as possible from each
processing step (Section~\ref{prodsum}). This was to facilitate offline
debugging and tuning since many of the steps have complex interdependencies.
This debug mode is controlled by a command-line switch and is typically
turned off in operations to minimize runtime.

The base language in PTFIDE is Perl. This code executes both the external
software modules and performs its own image-processing computations through use
of the Perl Data Language \citep[PDL;][]{glaze97}. PDL is an object-oriented
extension to Perl5 that is freely available as an add-on module. PDL is
optimized for computations on large multidimensional data sets by making use
of the hyper-threading capabilities of modern processor technologies.
That is, PDL has its own threading engine that uses constructs from linear
algebra to process large arrays as efficiently as possible using parallel 
computations. This is crucial since most of the steps in PTFIDE are CPU-bound.
This low-level parallelism occurs on the individual processor cores where our
basic processing unit is a single CCD-image. A higher level of parallelism is
achieved by using all of the $232$ CPU cores in our Linux cluster (described in
Section~\ref{realtpl}). Here we typically execute 232 simultaneous threads
(one CCD-image per core at any time). This gives us close to maximum throughput.

\begin{deluxetable}{lll}
\tabletypesize{\small}
\tablecaption{External (third-party) software used by PTFIDE\label{tab:sw}}
\tablewidth{0pt}
\tablehead{
\colhead{Software or Library} &
\colhead{Version\tablenotemark{a}} &
\colhead{Purpose} \\
\colhead{} &
\colhead{} &
\colhead{}
}
\startdata
{\it Perl} & $\geq 5.16.2$ & Core language for scripting and performing
                             arithmetic operations. \\
{\it PDL} & $\geq 2.4.10$  & Perl module for vectorized image processing;
                             built with bad-value and GSL support. \\
{\it GSL} & $\geq 1.15$ & GNU Scientific Library (numerical library). \\
{\it Astro-WCS-LibWCS} & $\geq 0.93$ & Perl module to support World
                                       Coordinate System (WCS)
                                       transformations.\\ 
{\it Ptfutils}, {\it Pars} & 1.0 & In-house developed Perl modules specific
                                   to PTF data processing. \\
{\it xy2xytrans} & 2.0 & For fast image-to-image pixel position
                         transformations. \\
{\it libtwoplane} & 1.0 & Library to support {\it xy2xytrans} module. \\
{\it wcstools} & $\geq 3.8.7$ & Contains WCS library to support multiple
                                modules listed here. \\ 
{\it cfitsio} & $\geq 3.35$ & FITS file-manipulation library to support
                              multiple modules listed here. \\
{\it SExtractor} & 2.8.6 & For initial source extraction to support
                           internal source-matching steps. \\ 
{\it SWarp} & 2.19.1 & For image resampling and interpolation using WCS. \\
{\it DAOPhot} & II, 1/15/2004 & Source detection, aperture photometry,
                                and PSF-estimation. \\
{\it Allstar} & II, 2/7/2001 & PSF-fit photometry and support for 
                               PSF-estimation (included in {\it DAOPhot}
                               package). \\ 
\enddata
\tablenotetext{a}{Version number shown is that in use at the time of writing.}
\end{deluxetable}

\subsection{Primary Inputs and Parameter Summary}\label{paramsum}

PTFIDE is driven by the Perl script {\it ptfide.pl}. The inputs can be
broadly separated into the following: an instrumentally-calibrated
CCD-image exposure (the {\it science} image); an accompanying bit-mask
(pixel-status) image; a spatially overlapping {\it reference} image; an
accompanying {\it source catalog} for the reference image; configuration
files for the various external software modules; processing parameters,
thresholds, and control switches.

All image files are in FITS format (defined in Section~\ref{realtpl}). Input
parameters and thresholds may be supplied on either the {\it ptfide.pl}
command-line or in a configuration file, while image FITS-file names, other
configuration files, and switches can only be supplied on the command-line.
Table~\ref{tab:inp} summarizes the inputs to {\it ptfide.pl} with
a brief explanation for each. The default parameter values are those
currently used for iPTF. More details on some of the parameters can be found
in Section~\ref{psteps}.

\clearpage

\begin{deluxetable}{p{1.7cm}p{1.6cm}p{12.2cm}}
\tabletypesize{\footnotesize}
\tablecaption{Inputs to PTFIDE (script {\it ptfide.pl})\label{tab:inp}}
\tablewidth{0pt}
\tablehead{
\colhead{Input\tablenotemark{a}} &
\colhead{Default\tablenotemark{b}} &
\colhead{Purpose\tablenotemark{c}} \\
\colhead{} &
\colhead{} &
\colhead{}
}
\startdata
-cfgide & \nodata &    
Optional input configuraton file listing all numerical parameters and
thresholds defined below; these override those on the command-line, if any. \\

-scilst & \nodata &
Input filename listing science image FITS file(s). \\

-msklst & \nodata &
Input filename listing mask image FITS file(s) accompanying -scilist. \\

-ref & \nodata &
Input FITS filename of reference image (co-add). \\

-catref & \nodata &
Input reference image source catalog file from SExtractor; -cn
specifies required columns. \\

-cn & 2,3,4,42,10,57, 60,63,72,78,27, 48,14,15,45 &
List of integers defining locations of required columns in the 
-catref input file. \\

-catfilt & 0.5,100,19.0, 1.3 &
Thresholds for filtering reference source catalog: min/max tolerable values
for CLASS\_STAR, ISOAREAF\_IMAGE, MAG\_APER, and ratio
AWIN\_WORLD/BWIN\_WORLD. \\

-od & \nodata &        
Directory name for output products (including any debug output). \\

-cfgswp & \nodata &    
Input configuration file for SWarp module. \\

-cfgsex & \nodata &    
Input configuration file for SExtractor to support position/gain matching. \\

-cfgsexpsf & \nodata & 
Input configuration file for SExtractor to support association with
PSF extractions. \\

-cfgcol & \nodata &    
Input SExtractor column name configuration file to support
position/gain matching. \\

-cfgcolpsf & \nodata & 
Input SExtractor column name configuration file to support association with
PSF extractions. \\

-cfgfil & \nodata &    
Input filename for SExtractor convolution kernel filter. \\
  
-cfgnnw & \nodata &    
Input SExtractor neural network configuration file for star/galaxy
classification. \\
  
-cfgdao & \nodata &    
Input generic {\it DAOPhot} parameter file. \\
  
-cfgpht & \nodata &    
Input {\it DAOPhot} photometry parameter file. \\   

-tmaxpsf & 2000.0 &
Threshold [\#bckgnd sigma] above background in {\it reference
image} for maximum usable pixel value when creating PSF. \\

-tdetpsf & 50.0 &
{\it DAOPhot} find-threshold [\#bckgnd sigma] for PSF creation from
reference image. \\

-tmaxdao & 3000.0 &
Threshold [\#bckgnd sigma] above zero-background in {\it difference image} for
maximum usable pixel value for source extraction. \\

-tdetdao & 3.5 &
{\it DAOPhot} find-threshold [\#bckgnd sigma] for source extraction on
difference image. \\

-tchi & 8.0 &
Threshold on {\it chi} metric from {\it Allstar} program below which extractions
on difference image are retained; larger $=>$ more non-PSF-like profiles
are retained. \\

-tshp & 4.0 &
Threshold on {\it sharp} metric from {\it Allstar} program where extractions on
difference image with --tshp $\leq sharp \leq$ $+$tshp are retained; values
of $sharp\simeq 0$ $=>$ sources are more PSF-like. \\

-tsnr & 4.0 &
Threshold on flux signal-to-noise ratio in PSF-fit photometry
above which {\it difference image} extractions are retained. \\

-fatbits & 8,9,10,12 &
Fatal bits to mask as encoded in input mask images (-msklist input);
set to 1 for no masking. \\

-satbit & 8 &
Saturation bit\# in mask images for determining saturation level in
science images. \\

-expnbad & 3 & 
Mask an additional ({\it expnbad} $\times$ {\it expnbad}) - 1 pixels around
each input masked science and reference image pixel; provides more complete
blanketing. \\

-eg & 1.5 &
Native electronic gain of detector [e-/ADU]; used for pixel-uncertainty
estimation. \\

-sxt & 2.0 &
SExtractor detection threshold [\#sigma] to support position/gain matching. \\

-rad & 3.0 &
Match radius [pixels] for associating reference and science frame
extractions for position refinement and gain matching. \\

-nmin & 200 &
Minimum number of reference-to-science image source matches
above which to proceed with position refinement and gain matching. \\

-dgt & 1.5 &
Minimum relative gain factor [\%] above which to proceed with
relative gain correction. \\

-dpt & 0.07 &
Minimum offset [pixels] above which to proceed with
position refinements (dX or dY). \\

-dgsnt & 5.0 &
Minimum S/N ratio in gain factor above which to proceed with relative
gain correction. \\

-dpsnt & 5.0 &
Minimum S/N ratio in deltas above which to proceed with position
corrections (dX or dY). \\

-gridXY & 4,8 &
Number of image partitions per axis to support differential SVB computation. \\

-tpix & 2.0 &
Threshold $t$ [\#sigma] for replacing pixel values $>$ mode $+$ $t*$sigma
in an image partition to support differential SVB computation. \\

-tmode & 500.0 & 
Threshold $t$ [\%] for replacing all pixels of a partition with
global mode if its local mode is $>(1 + t[\%]/100)$ $*$ global mode;
to support differential SVB computation. \\

-tsig & 100 &
Threshold $t$ [\%] for replacing all pixels of a partition with
local mode if its robust sigma is $>(1 + $t$[\%]/100)$ $*$
``median of all partition sigmas''; to support differential SVB
computation. \\

-rfac & 16 &
Image-pixel sampling factor to speed up filtering for
differential SVB computation. \\

-szker & 41 &
Median-filter size for downsampled image to support differential
SVB computation [pixels]. \\
\tablebreak

-ker & LANCZOS3 &
Interpolation kernel type for SWarp module. \\

-zpskey & IMAGEZPT &
Keyword name for photometric zeropoint in {\it science} image FITS headers. \\

-zprkey & IMAGEZPT &
Keyword name for photometric zeropoint in {\it reference} image FITS header.\\

-pmeth & 2 &
Method to derive PSF-matching kernel between {\it sci} and {\it ref} images:
1 $=>$ old Alard-Lupton (1998) method (now deprecated);
2 $=>$ Pixelated Convolution Kernel (PiCK) method. \\

-conv & auto &
For -pmeth 2: image to convolve; can be {\it sci}, {\it ref}, or {\it auto}.
The {\it auto} option uses the {\it sci} and {\it ref} FWHM values to select
the image to convolve. \\
  
-kersz & 9 &
For -pmeth 2: linear size of PSF-matching kernel stamps [pixels]. \\

-kerXY & 3,3 &
For -pmeth 2: number of image partitions along X,Y to represent
spatially-dependent kernel. \\

-psfsz & 25 &
For -pmeth 2 if -rpick was set: linear size of PSF stamps created
from point source cutouts. \\

-apr & 9.0 &
For -pmeth 2 if -rpick was set: source aperture radius [pixels] to
compute flux for normalizing PSFs and background level outside this. \\

-nmins & 20 &
For -pmeth 2 if -rpick was set: minimum number of sources in an
image partition above which PSF-creation is attempted. \\

-nmaxs & 150 &
For -pmeth 2 if -rpick was set: use {\it n} brightest sources per
image partition for PSF-creation. \\

-rpickthres & 4.0,5.0,0.0004, 40,0.045 &
For -pmeth 2 if -rpick was set: list of parameter thresholds for creating PSFs:
{\it N}-sigma threshold for stack-outlier rejection; {\it N}-sigma
threshold for spatial outlier-detection and winsorisation; maximum tolerable
RSS of spatial RMSs of PSF products for (re)assigning partition inputs
for kernel derivation; minimum distance to edge to avoid when
selecting sources from {\it sci} and {\it ref} images; threshold $td$ for
$d = |R-median\{R\}|$ where $R =$ ratio of PSF pixel sums and PSFs with
$d > td$ are rescaled to the $median\{R\}$ of all image partitions. \\
  
-nbreftb & 65 &
For -pmeth 2: number of pixel rows to force as bad at top and
bottom of internal images used for kernel derivation to account for
edge effects in resampled reference image. \\
  
-nbreflr & 35 &
For -pmeth 2: number of pixel columns to force as bad at left and
right of internal images used for kernel derivation to account for
edge effects in resampled reference image. \\    
  
-bckwin & 31 &
For -pmeth 2: linear window size for median filtering of [downsampled]
reference image when computing spatially varying background; note:
downsampling factor is fixed at 16x per axis. \\
  
-tsat & 0.65 &
For -pmeth 2: factor threshold to perform more conservative
tagging of resampled {\it ref} image pixels satisfying $\geq tsat * saturate$
where $tsat \leq 1$ and $saturate$ is from resampled {\it ref} image header.
This assumes the {\it ref} image was made using the {\it mkcoadd.pl} co-addition
software. \\

-goodcuts & 5.3,1.2,5.0, 0.02,22,4,0.8, 35,14.3,7.0,2, 0.07,0.2 &
For -pmeth 2: list of parameter thresholds for performing simple 1-D cuts on
source metrics for assigning {\it goodcand} flag in output extraction tables:
{\it chi, sharp, snrpsf, magfromlim, nneg, nbad, magdiff, mindtoedge, magnear,
dnear, elong, $|1-ksum|$, kpr}. \\

-baddiff & 80,15,15,3,140, 0.2,0.7,0.15, 0.15,1.5,1.5 &
List of parameter thresholds for performing simple 1-D cuts on difference-image
metrics for assigning {\it good} flag in output QA file: {\it diffpctbad,
dmedchi, davgchi, diffsigpixmin, diffsigpixmax, dmedksum, medkpr,
ncandscimrefratio, ncandrefmsciratio, dinpseeing, dconvseeing}. \\

-uglydiff & 80,15,15,140, 0.2,0.7,0.35 &
List of parameter thresholds for performing simple 1-D cuts
on difference-image metrics to decide if should proceed with
source extraction on difference images: {\it diffpctbad, dmedchi,
davgchi, diffsigpixmax, dmedksum, medkpr, maxminksum}. \\

-qas & 41,2008, 41,4056 &
Coordinate range of rectangular region in image for computing
QA metrics in difference images if -qa switch was set; format is:
{\it xmin, xmax, ymin, ymax} where pixel numbering is unit based and
$1 \leq xmin < xmax \leq NAXIS1$; $1 \leq ymin < ymax \leq NAXIS2$. \\

-apnum & 3 &
Internal aperture number for which {\it DAOPhot} aperture photometry information
should be propagated to output PSF-fit photometry table. \\

-forceparams & {\it ra, dec,} 43 &
List of parameters to support ``forced sub-image mode'' if
-forced switch was set; parameters are: ra [deg],
dec [deg], linsize [pixels]. \\

-kerlst & \nodata &    
Input filename listing FITS image cubes storing prior-derived,
spatially-dependent PSF-matching kernels for each science image
to support ``forced sub-image mode''. \\
  
-phtcalsci & \nodata & 
Switch to perform absolute photometric calibration of input science image
after gain-matching to {\it ref}-image by computing a ZP using the calibrated 
MAG\_AUTO values in the {\it ref}-image SExtractor catalog; this ZP will allow 
big-aperture (and PSF-fit) absolute photometry on the input 
science image {\it before} further gain refinements in the
PSF-matching step downstream. \\

-phtcaldif & \nodata & 
Switch to perform absolute photometric calibration on science image {\it after}
possible gain refinement and {\it before} image-differencing with
{\it ref}-image by computing a ZP using the  calibrated MAG\_AUTO values in
the ref-image SExtractor catalog; this ZP will allow big-aperture
(and PSF-fit) absolute photometry on the science and difference images. \\
\tablebreak

-wmode & \nodata &     
Switch to compute image modes (instead of medians) for the differential
SVB correction. \\

-rpick & \nodata &     
Switch to use robust version of the PiCK method (-pmeth 2) when deriving
PSF-matching kernel, i.e., via the construction of image PSFs using
point-source cutouts. \\

-psffit & \nodata &    
Switch to perform PSF-fit photometry on difference images with
prior PSF estimation off [possibly convolved] reference image. \\

-apphot & \nodata &    
Switch to perform fixed-aperture photometry on difference images
using {\it DAOPhot}. \\

-dontextract & \nodata &
Switch to {\it only} estimate spatially-varying PSF; no
extraction or photometry is performed. \\

-forced & \nodata &    
Switch to execute in ``forced sub-image mode'' where only
``{\it sci} minus {\it ref}'' difference image stamps (and ancillary files)
centered on input {\it ra, dec} ({\it -forceparams} inputs) are made. \\

-outstp & \nodata &    
Switch to generate image cutouts of candidates from
``{\it sci} minus {\it ref}'' difference images. \\

-pg & \nodata &        
Switch to compute {\it sci-to-ref} relative astrometric and gain corrections,
and apply if significant. \\

-pcln & \nodata &      
Switch to pre-clean (remove) output products directory specified by -od. \\
  
-qa & \nodata &        
Switch to generate QA metrics on difference images before and
after PSF-matching within image slice defined by -qas string;
results are written to standard output and an ASCII file. \\

-d & \nodata &         
Switch to write debug information to standard output, ASCII files
and FITS images. \\

-v & \nodata & 
Switch to increase verbosity to standard output. \\                                      
\enddata

\tablenotetext{a}{This same name (with prefix ``--'') is used in the
                  {\it ptfide.pl} command-line specification.}
\tablenotetext{b}{Default values, where shown, are optimal for the 
                  iPTF real-time pipeline. Command-line switches are
                  ``off'' by default.}
\tablenotetext{c}{Some of these are further discussed in Section~\ref{psteps}.}
\end{deluxetable}

\clearpage

\subsection{Reference Image Construction and Requirements}\label{refcon}

The purpose of a reference image is to provide a {\it static} representation of
the sky, or more specifically, a historical snapshot as defined by the
state of the sky recorded in previous image exposures. This image provides
a benchmark against which future exposures can be compared (i.e., differenced)
to assist with transient discovery, both temporally (for flux changes) and/or
spatially (for motion changes). The reference images also provide
``absolute anchors'' for assigning a photometric calibration to the incoming
real-time science images and difference images derived therefrom. They are also
used to check and refine astrometric solutions prior to differencing.
Details are given below.

The reference images are co-adds (stack averages; see below) of several to
fifty high-quality CCD-images selected from the image archive. Therefore,
they have a higher S/N than the individual exposures. Besides supporting
transient discovery, they can also benefit other science applications
that require deeper photometry. So far in iPTF, the goal has been to construct
reference images that are optimal for {\it single-exposure} image differencing
and transient discovery. These don't necessarily achieve the highest
possible depths (S/N) by using all available (good quality) images. This may be
performed at a later date on completion of the survey and with different
input image selection criteria.

Reference images are generated by a separate pipeline in iPTF operations that
executes asynchronously and is independent of the real-time pipeline. This
pipeline is only triggered when enough {\it good quality} images are available
for a given field, CCD, and filter in the archive. The generation process is
iterative in that reference images are remade and refined if an existing
product is identified to be of low quality or unusable, provided better quality
image-data are available. The input-image selection criteria for reference
image generation were outlined in \citet{laher14}. Given their importance, we
repeat them below and expand on some of the details. These were derived from
analyses of the distributions of numerous image metrics in June 2013. The
goal was to cover as much of the iPTF-visible sky as possible according to the
available depth-of-coverage across all visited fields at the time.

\begin{enumerate}
\item{The image must have been astrometrically and photometrically
      calibrated in an absolute sense and passed all automated quality
      checks prior to archiving \citep{laher14}.}
\item{The astrometric calibration (including full distortion solution) must
      have passed all validation steps (Section~\ref{realtpl}). This includes
      a separate check on the higher-order terms of the distortion polynomial.}
\item{The spatially-binned photometric zeropoint values (provided
      by the ZP Variations Map or ZPVM from photometric calibration)
      must lie within $\pm 0.15$ mag. Furthermore, the source color-term
      coefficients derived from photometric calibration (as described
      in \citet{ofek12}) must lie between the overall observed
      $1^{st}$ and $99^{th}$ percentiles.}
\item{The seeing (inferred from the mode of the
      point-source FWHM distribution) is $< 3.6\arcsec$.}
\item{The 5-$\sigma$ limiting magnitude, estimated using both theoretical and
      empirically-derived inputs is $R_{lim} > 20$ mag.}
\item{The number of sources extracted from the image (via {\it SExtractor}) is
      $\geq 300$. This reinforces the previous criterion and ensures
      the transparency was not too low or image noise not too excessive.}
\item{The minimum number of input images that must satisfy the above criteria
      before proceeding with reference image generation is $N_{min}=5$.}
\end{enumerate}

If $N_{min}\geq 5$, the image limiting magnitudes $R_{lim}$ are then sorted in
descending order (faintest to brightest). Next, co-add limiting magnitudes
$m^c_{lim}$ are predicted cumulatively and incrementally per-image for this
list of candidate images. The resulting values of $m^c_{lim}$ are then compared
to a predefined set of six target magnitude limits desired for the final co-add;
e.g., for the iPTF $R$ filter, these are defined:
\begin{eqnarray*}
m^t_{lim}(n) & = & R^{med}_{lim} + 2.5\log_{10}\left(\sqrt{N_{min}}\right)
                   + 0.5n \\
             & \simeq & \{21.5, 22.0, 22.5, 23.0, 23.5, 24.0\},
\end{eqnarray*}
where $0\leq n\leq5$, $R^{med}_{lim}$ is the typical (median) 5-$\sigma$
limiting magnitude of a single $R$-band exposure \citep{law09}, and $N_{min}=5$.
The faintest target limit $m^t_{lim}(n_f)$ is then identified as the faintest
$m^t_{lim}$ that just falls below the co-add limit predicted from the
{\it entire} image-list:
$m^c_{lim}\geq m^t_{lim}(n_f)$. The number of images $N$ to co-add
is then the {\it smallest} possible $N$ whose cumulative
$m^c_{lim}$ comes closest to $m^t_{lim}(n_f)$, i.e.,
\begin{equation*}
N = \mbox{min}\{\underset{N}{\operatorname{arg\,min}}
    \left[\left|m^c_{lim} - m^t_{lim}(n_f)\right|\right],\,50\},
\end{equation*}
where 50 is the maximum number of images allowed at this stage.

The requirement of an upper cutoff in the input image FWHM ($3.6\arcsec$;
criterion \#4 above) is an important consideration since it influences
the quality (effective point-source FWHM) of the resulting
reference image and image subtractions derived therefrom.
It is desirable to generate a reference image whose effective FWHM is
smaller than that generally expected in the science (target) images. This
ensures the higher S/N reference image is preferentially convolved (smoothed)
to match the science image PSF prior to subtraction in PTFIDE.
Not only will this minimize the relative fraction of correlated pixel-noise
in the difference images (i.e., since noise will be
dominated by the science image), it ensures robustness and minimizes the
potential for error when using an automatic method to decide on which
image to convolve. This is because the decision metrics themselves are
inherently noisy and one cannot be confident that the correct image will
{\it always} be selected. For the interested reader, \citet{huckvale14} present
an analysis on ways to select the best reference image and convolution
direction for optimal image subtraction in the presence of variable seeing.
The median FWHM of the iPTF science images is $\approx 2.2\arcsec$.
Therefore, it is inevitable that some cases will require the {\it science}
image to be convolved when matching PSFs. This is not detrimental since
PTFIDE can automatically select the image to convolve, with some margin for
error (see below). The desire to have a lower cutoff for the input image
FWHM when constructing reference images is mentioned here as a future
improvement, specifically to optimize image subtraction.
As mentioned, the requirement of FWHM $< 3.6\arcsec$ was driven by data
availability (after accounting for all other selection criteria) and a
need to generate reference images for a large fraction of the iPTF survey
fields in short order.

Before co-addition to create a reference image, the input list of
high-quality overlapping CCD-images (for a given field and filter) are
astrometrically refined as an ensemble. This is performed in a relative
image-to-image sense using {\it SCAMP} with inputs provided by
{\it SExtractor}. Their distortion solutions (in the {\it PV} format)
are also refined self-consistently. This improves the astrometric
solutions of the input images as well as the overall astrometry in
final co-adds.

Following astrometric refinement, the images are fed to an in-house
developed co-addition tool ({\it mkcoadd.pl}) specifically written for iPTF.
This software first determines the WCS geometry of the output co-add
footprint using WCS metadata from all the input images. The co-add pixel
scale is set to the native value determined for the center of the focal plane:
$1.01\arcsec$/pixel, and the footprint X,Y dimensions are fixed at
2500 pixels $\times$ 4600 pixels throughout. These dimensions can accomodate
for slight offsets in the reconstructed image pointing within a field.
Retaining the native pixel scale for co-add images ensures they more-or-less
remain (marginally) critically sampled in median seeing conditions.
A future consideration would be to use half the native pixel scale to
take advantage of the natural dithering offered by random offsets in
telescope pointing across image epochs. This dithering would benefit
input lists that are dominated by undersampled images (i.e., acquired in
better than median seeing) so that the effective PSF can be better
sampled when all images are combined.

Bad and saturated pixels are internally set to NaN in each CCD-image
using their accompanying masks. This facilitates easier omission
and tracking of all bad pixels downstream. Respective image-median
levels are then subtracted. This stabilizes (or homogenizes) the images
against temporally-varying backgrounds before they are combined (see below).
These backgrounds are not always astrophysical, for example, there is
contamination from scattered moonlight and internal scattering from
other bright objects whose line-of-sight may not directly fall on the
focal plane. The individual image background levels are stored for later use.
Each image is then de-warped (distortion-corrected) and interpolated
onto the output co-add grid using its astrometric and distortion solution.
This is accomplished using the {\it SWarp} software \citep{bertin02}.
For an image observed at epoch $t$, the pixel values $p_{ij}^t$ 
at distortion-corrected positions $i,j$ are interpolated and resampled
using a 2D Lanczos kernel of window size three:
\begin{equation}\label{lanc}
\resizebox{0.4\textwidth}{!}
{$
L(x^\prime,y^\prime) = \mbox{sinc}(x^\prime)\mbox{sinc}(x^\prime/3)
                       \mbox{sinc}(y^\prime)\mbox{sinc}(y^\prime/3),
$}
\end{equation}
where $-3 < x^\prime < 3$ and $-3 < y^\prime < 3$, and the signal at pixel
position $x,y$ in the output grid is given by
\begin{equation}
S^t(x,y) = \sum_{i=x-3}^{x+3}\sum_{j=y-3}^{y+3}p_{ij}^t L(x-i,\,y-j).
\end{equation}
This generates a new set of images for epochs $t = 1,2,3...N$ that have 
been corrected for distortion, all sharing the same WCS geometry,
i.e., that of the final co-add footprint.

The choice of a Lanczos kernel (equation~\ref{lanc}), particularly with 
window size three, is motivated by three reasons. First, it is close to optimal 
for PSFs that are sampled close to or above the Nyquist rate, i.e., its 
{\it sinc}-like properties can reconstruct well-sampled signals
to good accuracy. By ``optimal'', we mean in the context of conserving
information content. Second, its {\it sinc}-like nature also ensures that
uncorrelated input noise remains close to uncorrelated on output. 
Third, its relatively compact support minimizes aliasing and
the spreading of bad and saturated pixels on output. 
Given the $\approx1\arcsec$ pixel size, one small downside is that 
localized ringing can occur when the PSF is severely undersampled,
i.e., when the seeing falls below $\approx1.6\arcsec$. 

Since the epochal images will have been observed at different atmospheric
transparencies, their photometric throughput (or effective photon-to-DN gain
factors) will be different. Throughput-matching the images to a common
photometric gain or zeropoint (ZP) value is therefore necessary before
combining them.
This can be done in a relative sense (by computing source-flux ratios
across images and rescaling pixel values therein) or in an absolute sense using 
the image-ZP values derived from photometric calibration upstream.
We have chosen to use the absolute ZP values to compute the gain-factors.
This is accomplished by throughput-matching all images to a common target
zero point of $ZP_c$. This value becomes the final co-add (reference image) ZP,
where currently, all archived PTF reference images have $ZP_c = 27$ magnitudes.
The gain-corrected pixel values in a resampled image at epoch $t$ with
specific zero point $ZP_t$ are given by
\begin{equation}\label{tp}
S_c^t(x,y) = S^t(x,y)\,10^{-0.4(ZP_t - ZP_c)}.
\end{equation}

The resampled and throughput-matched epochal images with pixel signals
$S_c^t(x,y)$ are then combined using a {\it lightly}-trimmed weighted-average.  
Outlier-trimming is performed on the individual pixel stacks (along the $t$
dimension) by first computing robust measures of the location and spread:
respectively the median ($p_{50}$) and $\sigma\simeq 0.5[p_{84} - p_{16}]$,
where the $p_x$ are percentiles. Pixels that satisfy
$|S_c^t - p_{50}| > 9\sigma$ are rejected from their temporal-stack
at position $x,y$ prior to combining the remaining pixels using
a weighted-average (see below). Our choice of a relatively loose trimming 
threshold ($9\sigma$) is driven by our goal to remove the largest outliers
only (e.g., cosmic rays and unmasked satellite trails), therefore preserving
as much information as possible. 

The pixels in a stack are weighted using an inverse power of the 
seeing ($FWHM_t$) in the images they originated from, i.e.,
\begin{equation}\label{wt}
w_t = \left(\frac{FWHM_0}{FWHM_t}\right)^\alpha, 
\end{equation}
where $FWHM_0$ is a constant fiducial value currently set to the modal value
of $2\arcsec$ and is unimportant since it cancels following normalization
in the final weighted average.
$\alpha$ is a parameter that controls the overall importance of the
weighting. This weighting is purely motivated by empirical
and practical considerations as an attempt to handle the time-dependent
seeing in a qualitative sense, i.e., in that relatively more weight is 
given to images acquired in better seeing. There is no theoretical
justification that satisfies some optimality criterion like
maximal S/N, however, it's interesting to note that $\alpha=2$ corresponds to
the case where $w_t\propto1/N_p\propto1/\sigma_{psf}^2$, where $N_p$ is the
effective number of {\it noise pixels}\footnote{see \brokenurl{http://wise2.ipac.caltech.edu/docs/release/allsky/}{expsup/sec4\_6ci.html}}
for a Gaussian-like PSF and $\sigma_{psf}^2$ is the flux-variance that would 
result from PSF-fit photometry on the image \citep[see also][]{masci09}.
Therefore when $\alpha=2$, the weighting is effectively
inverse-variance weighting of the images according to the
expected point-source flux uncertainties from PSF-fitting.
Besides being optimal for PSF-fitting (simultaneously over the entire
image stack), and particularly when the input noise is Gaussian,
we found through simulation and analysis of on-sky data that $\alpha=2$
can lead to significantly distorted PSFs and slight degradations
in the co-add pixel S/N. This is due to the undersampled
nature of the PSF when the seeing is better than average in iPTF exposures.
We found that values of $0.7\leq\alpha\leq1.4$ for the range of seeing
encountered (and a forced cutoff of FWHM $< 3.6\arcsec$;
see above) work best. As a compromise, we assumed $\alpha=1$ throughout.
This choice is similar to that adopted by \citet{jiang14} for combining SDSS 
image data. These authors also included inverse-variance weights in their
weighting scheme, with pixel variances computed from the background RMS in each
input image.

The $N_r$ remaining pixels in a stack following outlier rejection are combined 
using a weighted average to produce the co-added pixel signal:
\begin{equation}\label{wtavg}
S(x,y) = \frac{\sum\limits_{t=1}^{N_r} w_t S_c^t(x,y)}
              {\sum\limits_{t=1}^{N_r} w_t} + median_t\{B_c^t\},
\end{equation}
where $S_c^t(x,y)$ and $w_t$ are given by equations (\ref{tp}) and 
(\ref{wt}) respectively.
The $B_c^t$ are the individual image background levels that were initially
subtracted from each image (see above) then rescaled using the same
throughput-match factors in equation (\ref{tp}). A median of all these levels
is computed and used as a fiducial background for the final co-add.
We also generate an image of the uncertainties in the weighted averages
$S(x,y)$. For co-add pixel $x,y$, this can be written:
$\sigma=\sqrt{\sum_t W_t^2\sigma_t^2}$ where $W_t=w_t/\sum_t w_t$
and $\sigma_t$ is the uncertainty (e.g., a prior) for the input pixel signal
at $x,y,t$. We assume that the noise is spatially and temporally uncorrelated
across images. Instead of using explicit priors for $\sigma_t$ (e.g., from
a pixel-noise model), we approximate $\sigma_t$ using an {\it unbiased} and
unweighted estimate of the population standard-deviation in the stack of
$S_c^t(x,y)$ values. The uncertainty in $S(x,y)$ (equation~\ref{wtavg})
then becomes:
\begin{equation}\label{uncavg}
\begin{aligned}
\sigma_S(x,y) = & \left[\frac{\sum_t w_t^2}{\left(\sum_t w_t\right)^2}
                        \frac{1}{N_r - 1}\right. \\
                & \;\;\times \left.\sum_{t=1}^{N_r}\left[S_c^t(x,y) - S(x,y)
                                                   \right]^2\right]^{1/2}.
\end{aligned}
\end{equation}
The effective $\simeq 1/\sqrt{N_r}$ scaling is implicitly represented
by the fractional term involving $w_t$. An image of the pixel
depth-of-coverage, $N_r(x,y)$, is also generated.

The astrometric solution in the reference image is validated against the
{\it 2MASS} PSC using a procedure similar to that described in
Section~\ref{realtpl}. Sources are extracted and measured from the reference
image using both aperture ({\it SExtractor}) and PSF-fit photometry
({\it DAOPhot}). Ancillary products for the PSF-fit catalog include a
DS9-region file and estimates of the spatially-variable PSF represented in
both {\it DAOPhot}'s look-up-table format and as a grid of FITS-image stamps.
QA metrics for the image and catalog products are also generated.
The product files are archived and their paths$/$filenames and
associated metrics stored in a relational database.

Each reference image product is uniquely identified according to survey field,
CCD, filter, pipeline number, version, and archive status flag. The pipeline
number supports variants of the reference image pipeline tailored for different
science applications, for example, a specific time range, number of input
images, and/or different filtering criteria than the default used to support
real-time processing. As mentioned, the reference image library is
periodically updated as low-quality or unusable products are identified 
from analyses of outputs from the real-time pipeline, provided enough 
{\it good quality} images are available (see above).

The reference image and its {\it SExtractor} catalog for a given survey field,
CCD, and filter are two of the primary products used in PTFIDE
(Section~\ref{psteps}). As mentioned, these provide an absolute anchor
for assigning a photometric ZP to all the {\it new} incoming, spatially
coincident science images and subtractions derived therefrom.
The ZP value in the FITS header of a reference image is the 
target fiducial value $ZP_c$ onto which selected
input images were gain-matched prior to co-addition (equation~\ref{tp}).
The absolute accuracy of $ZP_c$ is therefore determined by the
accuracy of the input image $ZP_t$ values. These were initially
derived from photometric calibration in the {\it frame-processing} 
pipeline using the SDSS-DR9 catalog \citep{ofek12, laher14}. The input
instrumental magnitudes used to perform this calibration are {\it Kron}-like
aperture measurements from {\it SExtractor}, also referred to as $mag\_auto$.
At the time of writing, these are the only instrumental magnitudes
in iPTF products that can be tied to an {\it absolute} photometric system 
via the image $ZP_t$ values. The individual (spatially-averaged) image $ZP_t$ 
values are accurate to 2-4\% \citep[absolute RMS;][]{ofek12}. These could be
less accurate on sub-image scales due to possible residual spatial variations
in the instrumental response. The $ZP_t$-inherent gain-match errors will
propagate into the reference image pixel values following image rescaling
(equation~\ref{tp}). These errors will only be captured by the empirical 
uncertainty estimates in equation (\ref{uncavg}) (with its implicit 
$\simeq1/\sqrt{N_r}$ scaling) assuming no systematics in the $ZP_t$ 
derivations upstream. A future goal is to calibrate the $ZP_t$ 
values to better than 1\%, preferably using PSF-fit photometry.

\subsection{Summary of Output Products}\label{prodsum}

PTFIDE output products are files that are generically named:
{\it InputImgFilename\textunderscore type.ext} where InputImgFilename is
the root filename assigned to the CCD image following pre-calibration
upstream (Section~\ref{realtpl}) and {\it type.ext} is a mnemonic for
the type of PTFIDE product generated. The extension ({\it ext}) can be
either {\it fits} (for FITS-formatted image), {\it tbl} for ASCII table in
the standard IPAC format, {\it psf} for PSF file in {\it DAOPhot}'s
look-up-table format, {\it reg} for DS9 region-overlay file,
{\it log} for logfile, or {\it txt} for other ASCII files.

Table~\ref{tab:pout} lists the primary PTFIDE products generated per CCD image.
By ``primary'', these represent the products that are later used for
real-time transient discovery and/or general archival science applications,
for example, light-curve generation using forced-photometry on the difference
images. The image, PSF, QA, and log files are copied to long-term storage and
their paths/filenames registered in relational database tables.
The table ({\it tbl}) files contain the extracted transient candidates
and associated metrics (Section~\ref{srcqa}), one for the {\it positive} and
another for the {\it negative} difference image. The metadata for each
transient are later stored in database tables (see Section~\ref{schema}).
The metrics in the {\it \_diffqa.txt} QA files (Section~\ref{diffqa}) are
stored in a separate database table. The generation of {\it positive}
({\it sci -- ref}) and {\it negative} ({\it ref -- sci}) difference images may
seem somewhat redundant since one is simply the negative of the other.
The purpose of having a negative difference is to enable
detection of transients that dissapear below the reference image baseline
level, for example, variable stars that are observed in their ``low'' state
relative to their time-averaged (reference image) flux. Our source
detection software is designed to detect positive signals only and
therefore it is necessary to negate the positive difference image and
extract any new transients (or excursions in variable flux) that happened
to be below the reference level at that epoch. 

Table~\ref{tab:sout} lists the secondary or ancillary PTFIDE products
that can be generated per input CCD image. These are diagnostic files 
to support offline analysis, debugging and tuning, and are not
generated by the (real-time) production pipeline. They are generated in
addition to the products in Table~\ref{tab:pout} if the debug ({\it -d})
switch was specified for {\it ptfide.pl}. Furthermore, some products
are only generated when {\it ptfide.pl} is executed in sub-image mode
(with the {\it -forced} switch; Section~\ref{force}).

\clearpage

\begin{deluxetable}{p{3.5cm}p{2.5cm}p{9cm}}
\tabletypesize{\footnotesize}
\tablecaption{Primary Outputs from PTFIDE\label{tab:pout}}
\tablewidth{0pt}
\tablehead{
\colhead{Output file suffix\tablenotemark{a}} &
\colhead{Format} &
\colhead{Description\tablenotemark{b}} \\
\colhead{} &
\colhead{} &
\colhead{}
}
\startdata
\_pmtchscimref.fits & FITS image\tablenotemark{c} &
Final PSF-matched ``science {\it minus} reference'' difference image. \\

\_pmtchscimrefpsffit.tbl & IPAC table &
Table of extracted transient candidates with PSF-fit and aperture photometry,
and source metrics corresponding to the \_pmtchscimref.fits difference image.\\

\_pmtchscimrefpsffit.reg & ASCII &
DS9 region/source-overlay file for all transient candidates in
\_pmtchscimrefpsffit.tbl. \\ 

\_pmtchrefmsci.fits & FITS image\tablenotemark{c} &
Final PSF-matched ``reference {\it minus} science'' difference image. \\

\_pmtchrefmscipsffit.tbl & IPAC table &
Table of extracted transient candidates with PSF-fit and aperture photometry,
and source metrics corresponding to the \_pmtchrefmsci.fits difference image.\\

\_pmtchrefmscipsffit.reg & ASCII &
DS9 region/source-overlay file for all transient candidates in
\_pmtchrefmscipsffit.tbl. \\

\_pmtchdiffunc.fits & FITS image\tablenotemark{c} &
Image storing 1-$\sigma$ pixel uncertainties corresponding to the
\_pmtchscimref.fits and \_pmtchrefmsci.fits difference images. \\

\_pmtchkerncube.fits & FITS cube &
Image stamps of spatially-dependent PSF-matching convolution kernels with
metadata in header. Each plane of cube stores kernel image 
for a specific partition in input science image. \\

\_pmtchconvrefdao.psf & ASCII &
File storing PSF template generated by {\it DAOPhot} from the kernel-convolved
reference image. Only generated if the reference image was convolved to 
match the science image seeing (FWHM). \\

\_resamprefdao.psf & ASCII &
File storing PSF template generated by {\it DAOPhot} directly from the reference
image, with no convolution. Only generated if the science image was convolved
to match the reference image FWHM. \\

\_diffqa.txt & ASCII &
File storing QA metrics on image-differencing process and statistics on number 
of transients extracted. \\

\_ptfide.log & ASCII &
Log file storing processing diagnostics and verbose output. \\
\enddata

\tablenotetext{a}{This is also a mnemonic for the product type; see
                  Section~\ref{prodsum}.}
\tablenotetext{b}{More details are given in Section~\ref{psteps}.}
\tablenotetext{c}{The sizes of these image files are $\simeq$ 34 MB each
                  (2048 $\times$ 4096 pixels with 32 bits per pixel).}
\end{deluxetable}

\clearpage

\begin{deluxetable}{p{4.5cm}p{12cm}}
\tabletypesize{\footnotesize}
\tablecaption{Ancillary (debug-mode) outputs from PTFIDE\label{tab:sout}}
\tablewidth{0pt}
\tablehead{
\colhead{Output file suffix\tablenotemark{a}} &
\colhead{Description\tablenotemark{b}} \\
\colhead{} &
\colhead{}
}
\startdata
\_badmsksci.fits &
Bad pixel mask for science image that includes spatially-expanded bad pixels.\\

\_badmskref.fits &
Bad pixel mask for reference image (mostly showing saturated regions).\\

\_scisatpixels.fits &
Image showing locations of {\it only} saturated pixels in science image.\\

sx\_ref\_filt.tbl &
Table of filtered sources from input reference-image {\it SExtractor} catalog
to support gain-matching and position refinement.\\

sx\_ref\_filt.reg &
DS9 region/source-overlay file corresponding to sx\_ref\_filt.tbl.\\

\_sxrefremap.tbl &
Table of positions and fluxes of filtered reference image sources from
sx\_ref\_filt.tbl with positions mapped onto science image frame to
support source-association in {\it SExtractor} run.\\

\_sx.tbl &
{\it SExtractor} catalog of science image extractions {\it matched} to
filtered and remapped reference image sources from \_sxrefremap.tbl;
to support gain-matching and photometric calibration.\\

\_sx.reg &
DS9 region/source-overlay file corresponding to \_sx.tbl.\\

\_sxbck.fits &
Diagnostic background image computed by {\it SExtractor} when generating 
\_sx.tbl catalog.\\

\_sxbckrms.fits &
Diagnostic background RMS image computed by {\it SExtractor} when generating 
\_sx.tbl catalog.\\

\_sxobjects.fits &
Diagnostic image showing objects extracted by {\it SExtractor} 
when generating \_sx.tbl catalog.\\

\_resampref.fits &
Reference image resampled onto science image frame.\\

\_resamprefunc.fits &
Pixel-uncertainty image corresponding to \_resampref.fits.\\

\_resamprefwt.fits &
Weight image from resampling of reference image using {\it SWarp}.\\

\_newscitmp.fits &
Science image gain-matched and positionally refined relative to reference
image.\\

\_inpsvb.fits &
Regularized image used to compute smoothly-varying differential
background (SVB) image.\\

\_svb.fits &
Image of smoothly-varying differential background; used to correct 
science image.\\

\_newscibmtch.fits &
Science image with differential background, photometric gain, and astrometry
matched to resampled reference image, before PSF-matching.\\

\_newsciuncbmtch.fits &
Pixel-uncertainty image corresponding to \_newscibmtch.fits.\\

\_diffbmtch.fits &
Internal ``science {\it minus} reference'' difference image before any
PSF-matching.\\

\_noconv\_p\textbf{\emph{m}}\_stp\textbf{\emph{n}}.fits &
Point-source image stamp indexed by \textbf{\emph{n}} in partition
\textbf{\emph{m}} of image that {\it is not} convolved.\\

\_toconv\_p\textbf{\emph{m}}\_stp\textbf{\emph{n}}.fits &
Point-source image stamp indexed by \textbf{\emph{n}} in partition 
\textbf{\emph{m}} of image that {\it will be} convolved.\\

\_noconv\_p\textbf{\emph{m}}\_psfcoad.fits &
Final co-added PSF from all point-source stamps in partition
\textbf{\emph{m}} of image that {\it is not} convolved; used to derive
PSF-matching kernel for partition \textbf{\emph{m}}.\\

\_toconv\_p\textbf{\emph{m}}\_psfcoad.fits &
Final co-added PSF from all point-source stamps in partition
\textbf{\emph{m}} of image that {\it will be} convolved; used to derive
PSF-matching kernel for partition \textbf{\emph{m}}.\\

\_noconv\_p\textbf{\emph{m}}\_psfcoaddepth.fits &
Pixel depth-of-coverage map corresponding to
\_noconv\_p\textbf{\emph{m}}\_psfcoad.fits.\\

\_toconv\_p\textbf{\emph{m}}\_psfcoaddepth.fits &
Pixel depth-of-coverage map corresponding to
\_toconv\_p\textbf{\emph{m}}\_psfcoad.fits.\\

\_sxrefremapcorr.tbl &
Equivalent to \_sxrefremap.tbl but performed on regularized science image
(gain-matched, position-refined, and PSF-matched with additional
gain-corrections) prior to differencing.\\

\_scibefdiff.fits &
Regularized science image (gain-matched, position-refined, and PSF-matched with
additional gain-corrections); input for {\it SExtractor} to generate
\_sx\_scibefdiff.tbl catalog.\\

\_sx\_scibefdiff.tbl &
{\it SExtractor} catalog of science image extractions {\it matched} to
filtered and remapped {\it ref} image sources from \_sxrefremapcorr.tbl;
to support photometric calibration of difference image.\\

\_pmtchconvref.fits &
Convolved reference image prior to differencing; only produced if {\it ref}
image was convolved.\\

\_pmtchconvsci.fits &
Convolved science image prior to differencing; only produced  if {\it sci}
image was convolved.\\

\_pmtchdiffmsk.fits &
Bad-pixel mask for final difference images; includes effects of convolution
from PSF-matching.\\

\_pmtchdiffchisq.fits &
Image of binned pseudo-$\chi^2$ values for difference image after
PSF-matching.\\
\tablebreak

\_pmtchconvref.coo\tablenotemark{c} &
{\it DAOPhot} output file listing initial detections from \_pmtchconvref.fits
for PSF generation.\\

\_pmtchconvref.lst\tablenotemark{c} &
{\it DAOPhot} output file listing stars picked from \_pmtchconvref.fits
for PSF generation.\\

\_pmtchconvref.lst.reg\tablenotemark{c} &
DS9 region/source-overlay file corresponding to \_pmtchconvref.lst.\\

\_pmtchconvref.nei\tablenotemark{c} &
{\it Allstar/DAOPhot} output file listing neighbors of the stars listed in 
\_pmtchconvref.lst.\\

\_pmtchconvrefdaosub.fits\tablenotemark{c} &
{\it Allstar/DAOPhot} output image showing PSF-subtracted sources 
from \_pmtchconvref.fits.\\

\_pmtchconvrefdaopsf.fits\tablenotemark{c} &
Image of spatially varying PSF represented as a grid of $16\times32$
postage stamps.\\

\_pmtchscimref.coo &
{\it DAOPhot} output file listing initial detections from \_pmtchscimref.fits
difference image.\\

\_pmtchrefmsci.coo &
{\it DAOPhot} output file listing initial detections from \_pmtchrefmsci.fits
difference image.\\

\_pmtchscimrefdaosub.fits &
{\it Allstar/DAOPhot} output image showing PSF-subtracted sources
from \_pmtchscimref.fits.\\

\_pmtchrefmscidaosub.fits &
{\it Allstar/DAOPhot} output image showing PSF-subtracted sources
from \_pmtchrefmsci.fits.\\

\_pmtchscimrefapphot.tbl &
Table containing concentric aperture photometry for extracted 
transient candidates from the \_pmtchscimref.fits difference image;
only generated if the {\it \textendash{apphot}} switch was set.\\

\_pmtchscimrefapphot.reg &
DS9 region/source-overlay file for all sources in \_pmtchscimrefapphot.tbl.\\

\_pmtchrefmsciapphot.tbl &
Table containing concentric aperture photometry for extracted
transient candidates from the \_pmtchrefmsci.fits difference image;
only generated if the {\it \textendash{apphot}} switch was set.\\

\_pmtchrefmsciapphot.reg &
DS9 region/source-overlay file for all sources in \_pmtchrefmsciapphot.tbl.\\

\_pmtchscimrefsex.tbl &
{\it SExtractor} catalog for \_pmtchscimref.fits difference image to
associate with PSF-fit extractions; used to assign source-shape metrics.\\

\_pmtchrefmscisex.tbl &
{\it SExtractor} catalog for \_pmtchrefmsci.fits difference image to
associate with PSF-fit extraction; used to assign source-shape metrics.\\

\_pmtchconvscistamp.fits\tablenotemark{d} &
Convolved {\it sci} image stamp prior to differencing; only produced
if {\it sci} image was convolved.\\

\_pmtchconvrefstamp.fits\tablenotemark{d} &
Convolved {\it ref} image stamp prior to differencing; only produced
if {\it ref} image was convolved.\\

\_imgtoconvstamp.fits\tablenotemark{d} &
Stamp image that {\it is not} convolved with PSF-matching kernel.
Can be either {\it sci} or {\it ref} image.\\

\_imgnoconvstamp.fits\tablenotemark{d} &
Stamp image that {\it will be} convolved with PSF-matching kernel.
Can be either {\it sci} or {\it ref}.\\

\_msktoconvstamp.fits\tablenotemark{d} &
Mask image stamp corresponding to \_imgtoconvstamp.fits.\\

\_msknoconvstamp.fits\tablenotemark{d} &
Mask image stamp corresponding to \_imgnoconvstamp.fits.\\

\_uncscistamp.fits\tablenotemark{d} &
Pixel-uncertainty image stamp corresponding to regularized science image.\\

\_uncrefstamp.fits\tablenotemark{d} &
Pixel-uncertainty image stamp corresponding to regularized reference image.\\

\_pmtchkernstamp.fits\tablenotemark{d} &
Image of PSF-matching kernel used to convolve image stamp; extracted from
archival \_pmtchkerncube.fits file.\\
\enddata

\tablenotetext{a}{This is also a mnemonic for the product type; listed in
                  approximately the same order as generated by {\it ptfide.pl},
                  along with the primary products in Table~\ref{tab:pout}.}
\tablenotetext{b}{More details are given in Section~\ref{psteps}.}
\tablenotetext{c}{Only generated if the resampled reference image was
                  convolved with the PSF-matching kernel, otherwise, the
                  {\it pmtchconvref} filename string is replaced with
                  {\it resampref} if the science image was convolved.}
\tablenotetext{d}{Only generated in ``forced'' sub-image mode if the
                  {\it -forced} switch was specified in processing;
                  see Section~\ref{force}.}
\end{deluxetable}

\clearpage

\section{PTFIDE Processing Steps}\label{psteps}

The input pre-calibrated CCD images need to satisfy a number of criteria prior
to processing through PTFIDE. These criteria use a number of quality metrics
computed upstream during preprocessing (Section~\ref{realtpl}).
Inputs that do not satisfy these criteria are expected to be of low quality
and are not used for transient discovery. Instead, they are assigned a status
flag (that is encoded into an overall processing bit-string at the end of
processing) and stored in a database table for future reference. The criteria
currently used to declare a CCD image as ``good'' and worthy for image
differencing are as follows:

\begin{enumerate}
\item{The seeing FWHM on the corresponding raw exposure image satisfies 
      $0 < $ FWHM $\leq 4.75\arcsec$. Values of FWHM $> 4.75\arcsec$ are also 
      a good proxy for low atmospheric transparency. Over the course
      of iPTF, $\simeq 0.04\%$ of exposures are above this limit.}
\item{The calculation of the seeing FWHM used in (1) was based on a sufficient
      number of point sources and is not a NaN. The latter may occur due
      to bad inputs.}
\item{At least 500 sources were found by {\it SExtractor} for use in the
      astrometric calibration using {\it SCAMP}.}
\item{The WCS solution from {\it SCAMP} used $> 300$ source matches with the
      astrometric catalog.}
\item{The WCS solution could be derived using {\it SCAMP} with no errors or
      warnings. I.e., the astrometric calibration did not fallback to 
      {\it Astrometry.net}.}
\item{At least 20 sources were matched with the 2MASS PSC for validating
      the WCS.}
\item{The axial RMS position differences using the 2MASS matches are within
      the maximum tolerable value (which depends on galactic latitude;
      see Section~\ref{realtpl}).}
\item{The first-order WCS terms (pointing, rotation, and scale) are within 
      range.}
\item{The higher-order terms of the distortion polynomial are within range.}
\end{enumerate}

Below we expand on the processing steps outlined in the PTFIDE processing flow
of Figure~\ref{fig:ptfideflow}. This includes additional details not shown in
this figure. The descriptions make extensive use of the input parameters and
output products summarized in Tables~\ref{tab:inp}, \ref{tab:pout},
and \ref{tab:sout}. 

\subsection{Mask-creation and Bad-pixel Expansion}\label{bp}

The input bit-mask image for the CCD science image is first AND'd with the
fatal-pixel bit-string template specified by {\it \textendash{fatbits}}. This
identifies those pixels to omit from processing. To enable tracking downstream,
these pixels are forced to NaN and all good (usable) pixels are reset to 1
in an internal image mask. This mask is then further processed and
regularized by forcing an additional $(N\times N) - 1$ pixels around each
masked (NaN'd) pixel to also be bad, where $N$ = input from
{\it \textendash{expnbad}} parameter. This provides more complete blanketing of
bad pixel regions, e.g., for saturated sources in particular whose unmasked
edges and associated bleed artifacts will lead to residuals in the difference
images and hence unreliable extractions. This expansion operation is also
performed on saturated pixels in the {\it resampled} reference image, i.e.,
following reprojection onto the science image frame (Section~\ref{reproj}).
These internal regularized science and reference image masks are propagated
downstream. In debug mode, they can be written to FITS format with filename
suffixes {\it \_badmsksci.fits} and {\it \_badmskref.fits} respectively.

Another reason for spatially expanding all bad input pixels is that both
the reference image resampling and the later PSF-matching step that involves
convolving one of the images (Section~\ref{pkern}) will cause bad-pixel regions
to implicitly grow. A forced expansion provides a more conservative
blanketing that's matched in both images prior to subtraction. Both the
science and reference image masks are later combined (following PSF-matching)
to produce a final effective bad-pixel mask ({\it \_pmtchdiffmsk.fits}) for
both difference images: {\it \_pmtchscimref.fits} and {\it \_pmtchrefmsci.fits}.
Furthermore, all bad pixels in the difference images are tagged with
value -999999.

\subsection{Relative Gain-matching and Astrometric Refinement}\label{garef}

Two important preprocessing steps are photometric throughput (or gain)-matching
and a (possible) astrometric alignment of the science image with the
reference image. As discussed in Section~\ref{refcon}, the reference
image provides an absolute anchor for assigning a photometric zeropoint
(ZP) and a WCS to the final difference image products. The relative
photometric and astrometric corrections are first derived and validated, and
then only applied to the science image if found to be statistically
significant. We describe each in turn below.

First, the input reference image catalog from {\it SExtractor}
({\it \textendash{catref}}) is filtered to retain primarily isolated point
sources using the following {\it SExtractor}-derived metrics: {\it CLASS\_STAR}
(minimum stellarity index); {\it ISOAREAF\_IMAGE} (maximum effective isophotal
area); {\it MAG\_APER} (faintest magnitude based on a fixed 14-pixel diameter
aperture); and the ratio {\it AWIN\_WORLD / BWIN\_WORLD} (maximum effective
source elongation). The thresholds for these metrics are specified by the
{\it \textendash{catfilt}} input. Another requirement is that all sources
be clean and uncontaminated with no bad {\it SExtractor}
flags, i.e., $FLAGS = 0$.
A 14-pixel diameter aperture is used so that integrated source fluxes are
relatively immune to seeing variations for the range of seeing encountered.
This choice however is not optimal for crowded fields (see below).
An intermediate filtered reference image catalog is then generated with
filename {\it sx\_ref\_filt.tbl}.

The source $x, y$ positions in the filtered reference image catalog are then
mapped into the coordinate frame of the science image using the {\it xy2xytrans}
utility. The reason for this is to support efficient source-matching
within {\it SExtractor} when run in source-association mode (below) since it
only supports source-matching in $x, y$ coordinates. A new intermediate catalog
is made with filename suffix {\it \_sxrefremap.tbl} that stores photometric
information for the filtered reference image sources and with $x, y$ positions
in the WCS of the science image. It is not guaranteed that this WCS is correct;
hence any possible astrometric errors will be reflected in the remapped $x, y$
positions. These positions will be used below to refine the overall astrometry.
{\it SExtractor} is then executed in source-association mode using the filtered
and position-remapped reference catalog sources. This entails finding the
{\it nearest} science image sources with S/N above input threshold
{\it \textendash{sxt}} and within a radial tolerance of {\it \textendash{rad}}.
A source-matched catalog table is generated ({\it \_sx.tbl}) with an
accompanying DS9 region file ({\it \_sx.reg}). This table is used to derive
the gain and astrometric corrections.

\subsubsection{Gain and Astrometric corrections}\label{gm}

The relative photometric gain factor $D_g$ is estimated using a median of the
flux ratios of all $N_m$ science-to-reference source matches, where all fluxes
are based on a 14-pixel diameter aperture\footnote{These will
be affected by source confusion in crowded fields. In future, we will use
PSF-fit photometry from both images, that includes source-deblending.}:
\begin{equation}\label{Dg}
D_g = median_i\left\{\left(\frac{f_{sci}}{f_{ref}}\right)_i\right\}.
\end{equation}
The uncertainty in $D_g$ is estimated from the Median Absolute Deviation (MAD),
appropriately rescaled for consistency with Gaussian statistics in the limit
of large $N_m$, and further inflated by $\sqrt{\pi/2}$ to account for the
fact that the median in equation (\ref{Dg}) is noisier than a mean:
\begin{equation}\label{sigDg}
\sigma(D_g) = \sqrt{\frac{\pi}{2N_m}} 1.483\,
              median_i\left\{\left|\left(\frac{f_{sci}}{f_{ref}}\right)_i -
              D_g\right|\right\}. 
\end{equation}

Global position offsets along the $x$ and $y$ axes are also computed
using medians of the source-position differences:
\begin{equation}\label{Dxy}
\begin{aligned}
D_x & = median_i\{\left(x_{ref} - x_{sci}\right)_i\}, \\
D_y & = median_i\{\left(y_{ref} - y_{sci}\right)_i\},
\end{aligned}
\end{equation}
with uncertainties that are also based on the MAD estimator, similar to
equation (\ref{sigDg}).
Note that these represent overall orthogonal offsets between the science
and reference images, and do not account for possible spatially-dependent
offsets, for example, that would result from an erroneous distortion solution
for the science image (as calibrated upstream; see Section~\ref{realtpl}).
Recall that the reference-image pixels have already been corrected for
distortion during the co-addition process (Section~\ref{refcon}). Therefore,
the assumption here is that the distortion solution is reasonably accurate
over the CCD image, and that any systematics in the relative
astrometry are purely global shifts along either $x$ or $y$ or both.

Furthermore, to gauge the spread in the $N_m$ input flux ratios
(equation~\ref{Dg}) and position offsets (equation~\ref{Dxy}),
$5^{th}$ -- $95^{th}$ percentile ranges are also computed for these quantities.
A large spread in the flux ratios for example (relative to some expected
nominal value) may indicate that the flat-fielding was inaccurate upstream.
A large spread in the position offsets may indicate that the
distortion calibration was inaccurate. These metrics are stored
in a database table for trending.

The gain correction factor $D_g$ (equation~\ref{Dg}) is only used to rescale
the science image pixel values to match those in the reference image if the
following criteria are satisfied: the number of matches $N_m$ from
which it was derived exceeds {\it \textendash{nmin}}; the quantity
$100|1 - D_g|$ exceeds {\it \textendash{dgt}}; and its significance or
S/N ratio, $|1 - D_g|/\sigma(D_g)$, exceeds {\it \textendash{dgsnt}}.
Similarly, the orthogonal position corrections $D_x, D_y$ are only applied
to the {\it science} image WCS parameters if the following criteria are
satisfied: the number of matches $N_m$ also exceeds {\it \textendash{nmin}};
either $|D_x|$ or $|D_y|$ exceed {\it \textendash{dpt}}; and their S/N ratios,
$|D_x|/\sigma(D_x)$ or $|D_y|/\sigma(D_y)$, exceed {\it \textendash{dpsnt}}.
Note, since the $D_x, D_y$ are constant corrections (independent of position),
it suffices to simply correct the coordinate origin defining the science
image WCS. These are the FITS keyword values $CRPIX1$ and $CRPIX2$, and are
corrected to the new values $CRPIX1 - D_x$ and $CRPIX2 - D_y$ respectively.
This adjustment then ensures that the reference image is reprojected
(and registered) onto the correct science image WCS later on
(see Section~\ref{reproj}).

As a detail, there are occasions when the input science image was already
absolutely photometrically calibrated and associated with a ZP value, for
example, when PTFIDE is executed in offline mode on processed
archival data. In this case, an initial global gain correction factor $G$ is
computed using the science and reference image ZP values according to
equation (\ref{tp}). The $D_g$ factor (equation~\ref{Dg}) is still computed,
but it becomes a delta-correction on top of $G$. The final effective gain
correction factor for rescaling the science image pixels is then
$D^\prime_g = G/D_g$, where $D_g$ is only applied if the above criteria
are met, otherwise it is reset to 1. Therefore, regardless of whether the
science image had a valid ZP calibration, PTFIDE always computes a relative
gain correction factor in order to place the science image pixels on the same
scale as those in the reference image as best as possible.

\subsubsection{Photometric Zeropoint Refinement}\label{rzp}

After rescaling the science image pixels, the above method then implies
that the reference image ZP will enable absolute photometry on the
science image, and eventually the difference images derived therefrom.
However, it is important to note that the reference image ZP will
only allow an absolute calibration of the same {\it type} of
instrumental photometry that was initially used to calibrate that ZP.
At the time of writing, the instrumental photometry used for the absolute
photometric calibration of PTF/iPTF data are the {\it Kron}-like $MAG\_AUTO$
aperture measurements from {\it SExtractor} \citep{ofek12}.
As discussed in Section~\ref{refcon}, these are used together with sources
from the SDSS-DR9 catalog to derive the absolute ZPs in all archived
image products, including reference images. Unfortunately, the $MAG\_AUTO$
measurements tend to systematically underestimate the total instrumental
flux, with a bias that depends non-trivially on the image seeing. This bias is
of order 4---8\%. Therefore, the current reference image based ZPs are only
applicable to $MAG\_AUTO$-like measurements performed on the gain-corrected
science and difference images. On the other hand, the primary instrumental
photometry extracted from PTFIDE image products is PSF-fitting
(Section~\ref{psffit}). A refinement to the ZP is therefore necessary.

To enable an absolute calibration of other flavors of photometry, for
example PSF-fitting or big-aperture photometry on the {\it real-time}
science and difference images, we compute a new ZP value for
insertion into their FITS headers, denoted {\it ZPSCI}.
This is only performed by PTFIDE if the {\it \textendash{phtcalsci}} switch
was specified. This new ZP is computed using the (absolutely calibrated)
$MAG\_AUTO$ magnitudes of the same filtered reference image point sources
as used for the relative gain-correction above, with matching 14-pixel
diameter aperture measurements from the science image, corrected for any
residual $D_g$. If the number of source-matches exceeds 100, the new ZP is
estimated as
\begin{equation}\label{phtcalsci}
\begin{aligned}
ZPSCI & = median_i\{(MAG\_AUTO_{ref}^{abs}\;\; - \\
      & \qquad\qquad\;\;\;\;\;\; MAG\_APER_{sci}^{inst})_i\}.
\end{aligned}
\end{equation}
A robust RMS based on percentiles, {\it ZPSCIRMS}, is also computed to quote
as a possible systematic uncertainty on {\it ZPSCI}. It's important to note
that the $MAG\_APER_{sci}^{inst}$ instrumental photometry used
here is from a relatively large fixed (14-pixel diameter) aperture.
The {\it ZPSCI} value will also be applicable to PSF-fit instrumental
photometry because analyses have consistently shown that this agrees,
within measurement error, with the instrumental fluxes from large
aperture photometry. I.e., both flavors of photometry catch the {\it same}
total instrumental flux for the range of seeing encountered. Therefore, to
enable an {\it absolute} calibration (on the SDSS system) of the
photometry from PTFIDE image products, involving either PSF-fitting
and/or large apertures, it is advised that the {\it ZPSCI} values be
used. We note that this reverse engineering to recover the correct ZP for
PSF-fit photometry will disappear in the future when our photometric
calibration system is upgraded to use PSF-fit photometry throughout.

To summarize, we have at this stage an internally regularized science image
whose pixels are gain-matched to those in the input reference image,
and with a possibly refined WCS that matches the reference image
astrometric solution. This intermediate science image can be written to a
FITS-formatted file with suffix {\it \_newscitmp.fits}. Other metadata that
depend on the gain-matching operation are recomputed and also propagated
along. These are the science image saturation level and the pixel electronic
gain (used for uncertainty estimation later). Along with the
{\it MAG\_AUTO}-based ZP inherited from the reference image, a new ZP is
also available, {\it ZPSCI}, to enable a more accurate absolute calibration of
PSF-fit photometry downstream. It's important to note that these
corrections represent initial adjustments at the global image level.
Other refinements to the relative photometric gain and/or astrometry are
possible at the {\it local} (sub-image) level later when we apply the
spatially-dependent convolution kernels to match the image PSFs
(Section~\ref{appkern}).

\subsection{Reprojection of Reference Image}\label{reproj}

The reference image is warped onto the native pixel grid of the science
image (accounting for distortion) using its refined WCS as described in
Section~\ref{gm}. The {\it SWarp} utility \citep{bertin02} is used to perform
the reprojection and resampling. This software conveniently uses the science
image's distortion polynomial with coefficients encoded in the PV-format,
derived upstream during astrometric calibration (Section~\ref{realtpl}).
Pixels are interpolated using a 2D Lanczos kernel of window size three
(equation~\ref{lanc}), i.e., the same as that used when constructing the 
reference images. See Section~\ref{refcon} for a discussion of its advantages.

It is imperative that the distortion solution for the science image be as 
accurate as possible over the entire frame to avoid mapping the reference 
image pixels into the wrong locations. Even slight inaccuracies (down to
a tenth of pixel) will lead to systematic residuals in the difference images. 
One could mitigate these spatially-dependent astrometric residuals by fitting
for a differential (relative) distortion between the science and reference 
images and if significant, correcting the science image distortion prior to
reprojection. We found this to be unnecessary for now since for the bulk of 
iPTF fields, the spatially-binned source-position residuals between the science
and reprojected reference image sources have {\it maximal} RMS values of
$\lesssim 0.12\arcsec$ per axis. This maximum RMS per image is computed over
$6\times12$ spatial bins, where the binning is intended to capture
local systematics in the distortion solution of the science image.
$0.1\arcsec$ is typically the maximum tolerable residual to obtain good quality
difference images for iPTF under median seeing or worse, at the location of
$R\gtrsim16$ mag sources. Presently however, the residuals are generally larger
in the densest regions of the galactic plane since that's where the astrometric
calibration is most challenging. This is a work in progress. 

The resampled reference image can be written to a FITS-formatted file
with suffix {\it \_resampref.fits}. An accompanying weight-image file
is also generated ({\it \_resamprefwt.fits}). This weight image is
used to generate a mask image for the resampled reference image
({\it \_basmskref.fits}) that primarily tags saturated pixels. These pixels
are then expanded to account for their growth during the interpolation
and PSF-matching process (see Section~\ref{bp}).

\subsection{Differential Spatially-Dependent Background matching}\label{bckm}

The slowly-varying background (SVB) component in the rescaled
and astrometrically-refined science image is matched to that in the resampled
reference image. This background matching step helps minimize
systematics in the difference-image photometry later on (through estimation 
of the local background), particularly when the differential background
between the science and reference image varies non-linearly with position.
The background correction map is estimated using a robust image-partitioning
method performed on a preliminary {\it science - reference} difference image,
where the inputs are already gain matched (Section~\ref{gm}) and 
astrometrically aligned (Section~\ref{reproj}) but not yet PSF-matched.
Bad and saturated pixels from both images are masked in the difference
image prior to processing. The reason for computing the background correction
map from a {\it raw} difference image is that this minimizes any biases 
from bright extended emission (e.g., galaxies). Furthermore, the presence of
residuals at the point-source level due to the lack of PSF-matching (at high
spatial frequencies) does not impact the estimation process.

The difference image is first partitioned into $M\times N$ rectangles where
$M, N$ are specified by {\it \textendash{gridXY}} (see Table~\ref{tab:inp}).
Pixel modes (or optionally medians)
are computed both globally (for the entire image) and within each
partition using only unmasked pixels. Modes are only computed if the
{\it \textendash{wmode}} switch was specified, otherwise medians are
computed. In the steps described below, {\it mode} can be interchanged with
{\it median} if the latter was used.

First, for each partition, we replace all pixel values therein with the
global mode if its local {\it mode} exceeds or is below the global mode
by a relative percentage specified by {\it \textendash{tmode}}, i.e., if
$100|mode - globalmode|/globalmode > tmode$ is satisfied.  Furthermore, we
replace any outlying pixel values $p_i$ in all partitions with their respective
local {\it mode} if $|p_i - mode| > t\sigma$ is satisfied, where $t$ is a 
threshold specified by {\it \textendash{tpix}} and $\sigma$ is a robust local
RMS estimated from a trimmed standard-deviation of the low-tail pixel
distribution in the partition, i.e., below its local mode.

The resulting outlier-trimmed modal map is further regularized by replacing
all pixels in those partitions with the global image mode whose local
$\sigma$ (robust RMS) is\\$> tsig\times median\{\sigma\text{'s from all
partitions}\}$, where $tsig$ is a relative threshold specified by
{\it \textendash{tsig}}. This avoids noisy partitions (e.g., due to
excessive Poisson noise from bight emission) from affecting the
differential SVB estimate. At this stage, the regularized difference image
can be written to a FITS-formatted file with suffix {\it \_inpsvb.fits}.

Next, we down-sample the regularized difference image using the binning factor
specified by {\it \textendash{rfac}}. This binning uses a local averaging
of pixels and is performed to speed up the filtering in the next step.
The down-sampled map is median filtered using a window size specified by
{\it \textendash{szker}} to smooth out the partition boundaries. The resulting
image is up-sampled back to the original image pixel dimensions
for use downstream. This is the final SVB correction map and can be written
to a FITS-formatted file with suffix {\it \_svb.fits}.

\begin{figure*}
\begin{center}
\includegraphics[scale=0.5]{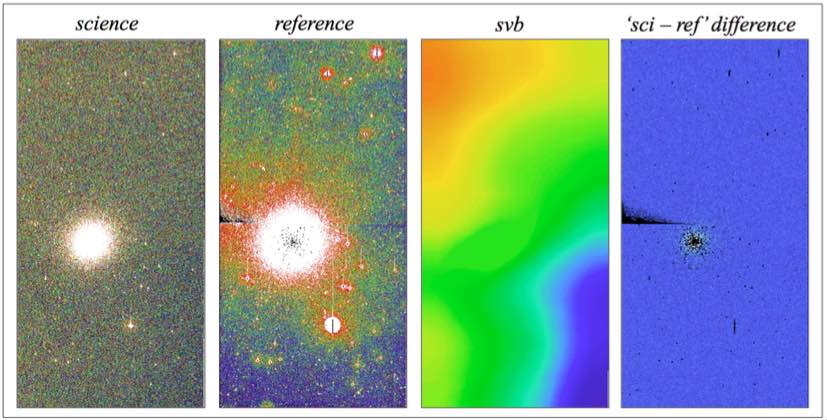}
\caption{Left to right: an example science and reference image containing the
         M13 Globular Cluster. The image stretch is intended to amplify
         background variations. The difference of these images is used to
         generate the differential slowly-varying background (SVB) map where
         the minimum-to-maximum range is $\simeq 18$ DN (or a few percent).
         The SVB map is then subtracted from the science image. When the
         reference image is subtracted from the new (background-matched)
         science image, this results in the spatially uniform difference image
         on the far right. This difference has the same image stretch as the
         SVB map. Black regions are bad and saturated pixels.} 
\label{fig:bmatch}
\end{center}
\end{figure*}

The SVB correction map is subtracted from the input (gain-matched)
science image to produce a new regularized science image (file suffix
{\it \_newscibmtch.fits}). This now has a SVB whose pattern matches that 
in the resampled reference image.
Figure~\ref{fig:bmatch} shows an example of an input science
and resampled reference image, and the differential (low-pass filtered)
SVB image generated therefrom. At this stage, a preliminary difference
image (still with no PSF-matching) can be generated with suffix
{\it \_diffbmtch.fits} to check the quality of the background matching.
An example is shown on the far right of Figure~\ref{fig:bmatch}.

\subsection{Pixel-level Uncertainty Estimation}\label{punc}

At this stage, images of the 1-$\sigma$ uncertainties corresponding to the
gain and background-matched science and resampled reference images
are initialized for propagation downstream. These pixel uncertainties
are later updated to account for additional processing on the images. 

We use a robust semi-empirical method to compute the pixel uncertainties.
First, we find the {\it minimum} background pixel variance $\sigma_{bcksci}^2$
and mode $m_{bcksci}$ over all partitions of the science image. These are
the same partitions from the background matching step above
(Section~\ref{bckm}). The variances are computed from a robust pixel RMS
based on a trimmed standard-deviation of the low-tail pixel distribution
in each partition. The minimum value is used for conservatism in the sense
that any biases from bright emission and/or source-confusion are minimal.
The 1-$\sigma$ uncertainty for a pixel signal $f_{DN}$ in the science image is
approximated by:
\begin{equation}\label{sigsci} 
\sigma_{sci}\approx\left[S\left(\frac{f_{DN}-m_{bcksci}}{g}\right) +
                         \sigma_{bcksci}^2\right]^{1/2},
\end{equation}
where $S$ is a scale factor to account for the gain-matching operation
(Section~\ref{gm}) and is needed since the actual counting of photoelectrons
(for the Poisson term) is always with respect to the native detector ADU 
counts. $g$ is the detector's electronic gain in $e^{-}/DN$
(input parameter {\it \textendash{eg}}). We have subtracted an estimate
of the background from the pixel signals since any Poisson-noise from the
background is already implicitly included in the $\sigma_{bcksci}^2$ term
and we remove any unknown (hidden) bias level that is not induced by
photoelectrons. This avoids overestimating the Poisson contribution 
from the background. Furthermore, $\sigma_{bcksci}^2$ implicitly 
includes the read-noise component. Pixels with $f_{DN} - m_{bcksci} < 0$
are reset to zero.

For the reference image pixel uncertainties, we first compute a robust
background pixel variance $\sigma_{bckref}^2$ from the same 
(minimally-contaminated) partition as used for the science image.
Given that the reference image was created from a co-add of science
images with variable photometric ZPs (that were later used to
throughput-match the images; Section~\ref{refcon}), the Poisson-noise
contribution will be difficult to estimate precisely from first principles.
Instead, we approximate the 1-$\sigma$ uncertainty for a pixel signal
in the reference image by scaling from the science image uncertainties
(equation~\ref{sigsci}) and the relative background RMS estimates:
\begin{equation}\label{sigref}
\sigma_{ref}\approx\sigma_{sci}\left(\frac{\sigma_{bckref}}
                                          {\sigma_{bcksci}}\right).
\end{equation}
This is expected to be a reasonable approximation since the pixel signals
(in DN) are guaranteed to be conserved between the rescaled science image and
resampled reference image, to within measurement error. In other words,
the signals contributing to the Poisson component are not expected to 
change much between these images and any $1/\sqrt{N}$ diminution
in the overall noise in the reference image from co-addition is
effectively handled by the ratio $\sigma_{bckref}/\sigma_{bcksci}$.

An important effect that is not accounted for in the pixel uncertainties
of the resampled reference image at this stage is the possibility of
correlated pixel noise. This could arise from both co-addition (during 
construction of the initial reference image) and the reinterpolation
step (onto the science image pixel grid; Section~\ref{reproj}).
Accounting for spatially correlated noise is more important when estimating
the photometric uncertainties of extracted sources from {\it difference}
images (Section~\ref{psffit}). The contribution of correlated noise
from reference images however is expected to be small. This is because
first, as discussed in Section~\ref{refcon}, correlated noise will be 
negligible due to the choice of a {\it sinc}-like interpolation kernel,
and second, because of the $1/\sqrt{N}$ diminution from co-addition.
The difference image noise will be dominated by that in the science image.
The difference image pixel uncertainty estimates are described in
Section~\ref{appkern}. The science and reference image pixel-uncertainty
images can be written to FITS format with suffix names
{\it \_newsciuncbmtch.fits} and {\it \_resamprefunc.fits} respectively.

\subsection{Preparation of Inputs for PSF-matching}\label{preppsfs}

Our overall goal is to derive a kernel image $K(u,v)$ where $u,v$
are relative pixel coordinates, which when {\it convolved} with
one of the input images (science or reference), will match their PSFs
in some optimal manner. The details of how this kernel is represented and
derived are outlined in Section~\ref{pkern}. Following the {\it global}-matching
steps above (i.e., registration, gain and background-matching), the
PSF-matching problem can be generalized by attempting to model one of the
input images in terms of the other through $K(u,v)$ and some
local differential background $dB$. For instance, let us assume the science
image pixel values $I_{ij}$ can be modeled from those in an overlapping
reference image $R_{ij}$ where the point-source profiles therein are
significantly more narrow (in terms of overall FWHM):
\begin{equation}\label{imgmodel}
I_{ij} = \left[K(u,v) \otimes R_{ij}\right] + dB + \epsilon_{ij}.
\end{equation}
$\epsilon_{ij}$ is a noise term, usually a correction to the random
noise component inherent in the $R$ image since the latter is not
strictly noiseless. Note, if $I$ was determined to be the better seeing
image (according to some $\Delta$FWHM threshold based on prior metrics;
see Section~\ref{pkern}), then $R$ and $I$ would be interchanged
in equation (\ref{imgmodel}) without loss of generality. Aside from
modeling differences in PSF shape, $K(u,v)$ will also (implicitly) model
local residuals in the relative photometric gain and/or astrometry,
for later removal when $K(u,v)$ is applied.

Regardless of how $K(u,v)$ is parameterized (see below), the crucial
inputs for an optimal solution (in the least-squares sense) are accurate
representations of the PSFs as a function of position in both the science
and reference images. These PSFs then effectively take-on the role of $I$ and
$R$ in equation (\ref{imgmodel}). To account for spatial dependencies,
we estimate the PSFs over a grid of $N_m = N_x\times N_y$ image partitions
where $N_x, N_y$ are specified by {\it \textendash{kerXY}} (currently
$= 3\times3$ or $\simeq 11.5^\prime\times 23^\prime$ for iPTF). The boundaries
of the partitions are made to overlap by a length equal to half the kernel
image width (input {\it \textendash{kersz}}; currently $= 9$ pixels).
The partition size is determined from an analysis of the coherency in
PSF-shape versus position, balanced against the typical number of
point-sources expected therein in order to obtain PSFs of reasonable S/N
when all sources are combined. Our derivation of $K(u,v)$ is highly sensitive
to input noise and therefore every attempt is made to maximize the pixel S/N in
the final PSF images, for all image partitions.

\begin{figure*}
\begin{center}
\includegraphics[scale=0.65]{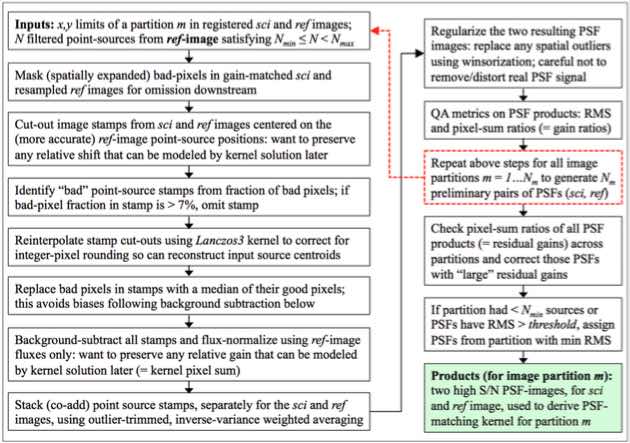}
\caption{Processing flow for the creation and allocation of PSF image products
         for each common science and reference image partition. These are used
         to derive the spatially-dependent PSF-matching kernel between the
         science and reference image. See Section~\ref{preppsfs} for details.}
\label{fig:makepsfs}
\end{center}
\end{figure*}

Figure~\ref{fig:makepsfs} gives an overview of the steps used to construct
the $2\times N_m$ high quality PSF images for all partitions in the
preprocessed science and reference images. The primary input is a list
of clean point-sources from the input reference {\it SExtractor} catalog,
with $x,y$ centroid positions in the {\it resampled} reference image frame.
These are the same filtered point-sources used for the relative
gain-matching and astrometric refinement steps described in
Section~\ref{garef}. The sources are assigned to their specific image
partitions and we require a minimum of $N_{min}=20$ sources (parameter
{\it \textendash{nmins}}) per partition. The maximum is capped at
$N_{max}=150$ ({\it \textendash{nmaxs}}) where if exceeded, the brightest
$N_{max}$ sources in the partition are selected. This maximum is imposed
for runtime reasons, but still provides a sufficient overall S/N when all
sources are combined.

For a given image partition $m$, square stamps of linear size
{\it \textendash{psfsz}} ($= 25$) pixels centered on the reference-image-based
source centroids are cut from the (already registered) science and reference
images. The reason for using the same (reference) position on both images is
that we want to preserve any possible local astrometric shift between
stamps of the same source from each image. This shift (if significant)
will persist into the final respective PSF images and be subsequently
captured by the kernel solution $K(u,v)$. This will allow any local
systematic shifts to be corrected following the application of $K(u,v)$
to its respective image partition (Section~\ref{appkern}). The point-source
cutouts are then filtered to remove cases with large numbers of masked
pixels. Each stamp is then interpolated into a new pixel grid so that
the input (fractional-pixel) source centroids are made to fall close to
their geometric centers, i.e., to correct for the truncation error when
creating the initial cutouts using integer pixel coordinates. This
registration step is important prior to stacking the point-source stamps.
The stamps are then background-subtracted to ensure a zero-median background
level outside an aperture with size specified by {\it \textendash{apr}}.
Pixel signals inside this aperture from {\it only the reference image} cutouts
are then integrated and used to flux-normalize each matching science and
reference image cutout of the same source. The reason is similar to that
mentioned above for source positions -- it preserves any possible residual
in the relative gain that can later be modeled and removed by $K(u,v)$. 

If the debug switch was set, the point-source image cutouts can be
individually written to FITS format with file suffixes
{\it \_noconv\_p\textbf{\emph{m}}\_stp\textbf{\emph{n}}.fits} and
{\it \_toconv\_p\textbf{\emph{m}}\_stp\textbf{\emph{n}}.fits} for the 
science and reference image respectively, where \textbf{m} is the
partition identification index and \textbf{n} is the source index therein. 

The homogenized point-source cutouts for partition $m$ are then separately
stacked for the science and reference images. The first step involves using
robust statistics to identify and mask outlying pixels in the pixel stacks.
The stamps are combined using a weighted average where weights are the
inverse of the pixel variances computed from the background in each stamp
(outside an aperture with radius {\it \textendash{apr}}). This results in
two PSF images for partition $m$, one for the science and another for
the reference image. The PSF images are further cleaned for possible pixel
outliers using Winsorization \citep[e.g.,][]{kafadar01}. Here, pixels that
exceed some threshold (a multiple of the robust spatial RMS above or below
the background) are replaced with the threshold value. This replacement
(when used with a high threshold) does not inadvertently distort
the PSF shape. The input parameters and thresholds for the above
steps are specified by the {\it \textendash{rpickthres}} input string.

The source-cutout and stamp-stacking process is performed on all
$N_m$ partitions. This results in two {\it preliminary}
(science and reference) PSF images per partition, assuming there were 
enough point sources therein (i.e., exceeding $N_{min}$).
If the debug switch was set, these can be written to FITS format with
filename suffixes {\it \_noconv\_p\textbf{\emph{m}}\_psfcoad.fits} and
{\it \_toconv\_p\textbf{\emph{m}}\_psfcoad.fits} for the science and
reference image respectively, where \textbf{m} is the partition
index. Accompanying pixel-depth maps showing the final number of stacked
pixels are also generated with {\it \_psfcoad.fits} replaced by
{\it \_psfcoaddepth.fits} in these filenames.

These PSFs are preliminary in the sense that there is no guarantee that all are
of sufficient quality with good overall S/N across all partitions. For example,
a partition could fall on a highly confused region, exhibit a complex
background, or not have enough good quality point-sources. To account
for this, we further regularize the PSFs for approximate consistency across 
all partitions. This includes replacing a bad PSF with a better quality one
from a neighboring partition. A consequence of this replacement is that the
selected PSFs (for a common science and reference partition) will no
longer represent the true PSF shapes for that partition. This is
a small loss since this replacement does not occur often, and when it does
occur, the overall PSF variation is small enough that the impact to the
PSF-matching is negligible when another partition's PSFs are used. 
These regularization steps are described in more detail below, with control
parameters also specified by the {\it \textendash{rpickthres}} input.

\begin{figure}[ht]
\includegraphics[scale=0.33]{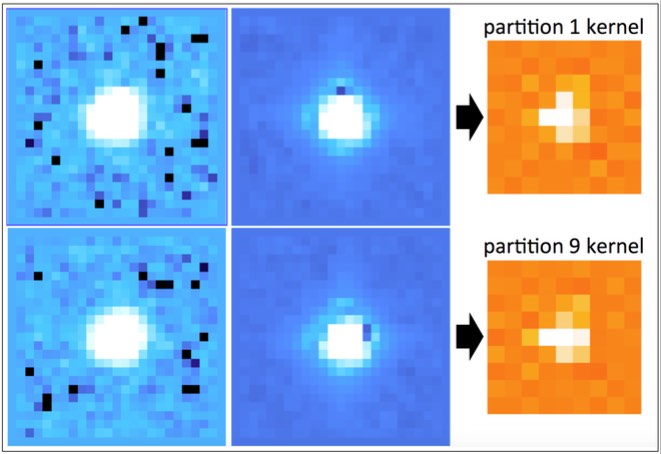}
\caption{Example PSF image products for two separate image partitions:
         {\it top row} corresponds to a partition at the bottom left of
         a CCD image and {\it bottom row} is for a partition at top right of
         the same image. From {\it left to right}: PSF for science image,
         PSF for reference image, and the resulting PSF matching kernel
         derived using the method in Section~\ref{pkern}. Pixel sizes are
         the same throughout. The kernel images are enlarged for clarity.}
\label{fig:egpsfs}
\end{figure}

We first correct any PSF pairs with a large residual gain relative to all other
PSF pairs from other partitions. The residual gain is estimated using
the ratio of the sum of {\it normalized} PSF pixel values for the pair.
Ratios that deviate by more than some threshold from the median pixel-sum
ratio over all partitions have their respective PSF pixels rescaled to match
the median ratio. This median ratio is typically unity due to the global
gain-matching performed earlier (Section~\ref{garef}). Note that this
``gain-homogenization'' is only intended to correct PSF pairs with large
outlying residuals in their relative gain. Next, we allocate the final PSF
pairs to each partition by enuring that first, there were enough sources to
make PSFs in the first place (i.e., $\geq N_{min}$) and second, that the RSS
of their robust spatial RMSs were below some threshold. If either of these
conditions are not satisfied for a partition, we assign PSFs from that 
partition with the lowest RSS'd RMS value.
Each overlapping science and reference image partition is now associated 
with a high quality pair of PSFs, ready for the PSF-matching step. 
Figure~\ref{fig:egpsfs} shows an example of the PSFs for two image
partitions and the resulting matching kernels $K(u,v)$ using the formalism in
Section~\ref{pkern}. As expected, the reference image PSFs will have
a higher S/N and appear smoother since the references were initially
created from a stack of science images.

\subsection{Derivation of PSF-matching Kernel}\label{pkern}

Given high S/N representations of the science and reference-image PSFs for a
spatial partition, with PSF pixel values $I_{ij}$ and $R_{ij}$ respectively at 
common coordinates $i,j$, we outline below the method used to obtain an optimal
solution for the convolution kernel $K(u,v)$ and differential background
$dB$ in equation (\ref{imgmodel}). Following previous approaches, these can be
derived by minimizing a weighted sum of the squared residuals between a model
and some new image (here the science image PSF); for example, using the
objective function:
\begin{equation}\label{chi2}
\chi^2_o = \sum_{i,j}\left[\frac{I_{ij}-\left[K(u,v)\otimes R_{ij}\right] - dB}
                                {\sigma_{ij}}\right]^2,
\end{equation}
where $\sigma_{ij}$ are prior pixel uncertainties in the $I_{ij}$ that may
include some scaled contribution from $R_{ij}$ (see below).
Equivalently, equation (\ref{chi2}) can be recast in vector-matrix notation:
\begin{equation}
\chi^2_o = \left(I - M\right)^T\Omega_{cov}^{-1}\left(I - M\right),
\end{equation}
where $\Omega_{cov}$ is the full error-covariance matrix to account for possible
correlated errors between the input pixels, and $M$ is the model-image with 
elements:
\begin{equation}\label{mod}
M_{ij}=\left[K(u,v)\otimes R_{ij}\right] + dB.
\end{equation}
Strictly speaking, the objective function $\chi^2_o$ can only be compared to
a true $\chi^2$ distribution with degrees-of-freedom $N_{dof}$ for the purpose
of validating model solutions using probabilistic inference if the input
data errors are normally distributed; i.e.,
$\epsilon_{ij}\sim N(0,\sigma^2_{ij})$. These errors can be interdependent
(and if so, need to be captured by $\Omega_{cov}$), but they do need to
be identically distributed for $\chi^2$-validation purposes. The null
hypothesis is that the model (equation~\ref{mod}) ``generated'' the $I_{ij}$.
Furthermore, for linear parameterizations of $K(u,v)$ (see below), the
minimization of $\chi^2_o$ reduces to a generalized linear least-squares
problem with a solution that is {\it unique} and {\it optimal} in the
maximum-likelihood sense for normally distributed errors.

Another important consideration is that the input errors $\epsilon_{ij}$
are {\it heteroskedastic}, i.e., they are not identically distributed with
constant variance over the input pixels $i,j$. This is because we are
exclusively fitting point-source data where Poisson noise dominates. The
noise-variance will have a spatial dependence following the shape of the 
PSF profile ($I_{ij}$).
Even though this dependence can be accounted for by using prior weights derived
from the pixel uncertainties ($1/\sigma_{ij}^2$ or $\Omega_{cov}^{-1}$) that
implicitly include Poisson-noise, their direct use in $\chi^2_o$ will lead
to biased estimates for $K(u,v)$ and $dB$. This was explored in a different
context by \citet{mighell99}.
A number of complex variance-stabilizing methods exist to ensure unbiased
estimates. For simplicity, we omit the use of prior weights when estimating
$K(u,v)$ and $dB$. Our solution will still be optimal in the least-squares
sense and will also be close to the maximum-likelihood solution for
normally-distributed errors in general. Even though the exclusion of weighting
in equation (\ref{chi2}) prohibits the use of goodness-of-fit tests in an
absolute (probabilistic) sense, relative changes in the global $\chi^2_o$ can
be used to validate the performance of different kernel solutions when 
applied to full images (see Section~\ref{diffqa}). For completeness,
we continue to carry the $\sigma_{ij}$ term (as represented in 
equation~\ref{chi2}) in our derivations below. In the end, our estimates
are really solutions to an ordinary linear least-squares (OLS) problem.

Our construction in equations (\ref{imgmodel}) and (\ref{chi2}) assumes that
the image containing the narrower PSF is the one that should be convolved.
However, there is no guarantee that this image is always the reference
PSF ($R_{ij}$) and hence without loss of generality, $I_{ij}$ and $R_{ij}$
can be interchanged. It is therefore important to predefine the convolution
direction. This can be specified by the {\it \textendash{conv}} input
string. The choices are {\it sci}, {\it ref}, or {\it auto}. The {\it sci}
and {\it ref} options always force the science ($I_{ij}$) or reference
($R_{ij}$) image pixels to be convolved respectively. The {\it auto} option
allows an automatic selection of the image to convolve and is our current
operating mode. This selection is based on comparing global-image measures of
the FWHM values of filtered stars from the science and reference images.
The FWHM values are from 2D Gaussian fits performed by {\it SExtractor}.
These values are medianed to yield two measures: $FWHM_{sci}$ and $FWHM_{ref}$.
The automatic selection is based on the value of the relative difference 
\begin{equation}\label{dfwhm}
\delta = 1 - \frac{FWHM_{sci}}{FWHM_{ref}}.
\end{equation}
Currently if $\delta\geq 0.03$, the science image is selected for convolution,
otherwise the reference image is selected, i.e., as depicted in the estimation
equations above. The threshold for $\delta$ was tuned by examining
distributions in $FWHM_{sci}$ and $FWHM_{ref}$ from many images and setting a
conservative value to allow for uncertainty in the median FWHM estimates.
In other words, the fuzzy interval $-0.03<\delta<0.03$ implies the overall
{\it sci} and {\it ref} PSF FWHM measures are consistent within measurement
error. Note that the FWHM measure is assumed to be a good proxy for
PSF shape in general, at least to first order for the purpose of defining
a convolution direction. When the PSF FWHMs are consistent, no
useful information is expected in the kernel solution, i.e., aside from noise
and perhaps variations incurred by higher-order PSF-shape differences.
The kernel then effectively becomes a single spike (i.e., a $\delta$-function)
where in principle, no PSF-matching would be required.

Usually (for $\gtrsim 85$\% of science image exposures encountered), the
reference images are those selected for convolution since by design, these
were constructed from archived science images with moderately better seeing
than average, in addition to being weighted by their inverse-seeing
(Section~\ref{refcon}). However, for cases where the science image ($I_{ij}$) 
is selected for convolution, the roles of $I_{ij}$ and $R_{ij}$ are
interchanged in equations (\ref{chi2}) and (\ref{mod}) so that $I_{ij}$
convolved with $K(u,v)$ becomes the model-image fitted to the data, $R_{ij}$.
This complicates the error and weighting structure if a true $\chi^2$
objective function (based on equation~\ref{chi2}) were to be
used for estimation since the model would be noisier than
the data being fitted. This is another reason for omitting the use of
prior weights in the objective function and treating it as a simple
OLS problem. The application of $K(u,v)$ to the noisier $I_{ij}$ image
also causes a larger fraction of the noise to be correlated in the
final difference image. This is further discussed in Section~\ref{appkern}.

As mentioned above, linear parameterizations for $K(u,v)$ are the simplest
to solve from a computational standpoint, for example, by expanding $K(u,v)$
as a linear combination of $n$ basis functions:
\begin{equation}\label{lin}
K(u,v) = \sum_i^n a_i K_i(u,v),
\end{equation}
and solving for the $n$ coefficients $a_i$. A traditional choice for the
$K_i(u,v)$ are Gaussians of different width, each modified by a 2D
shape-morphing polynomial. The $a_i$ are further expanded into another 
polynomial in $x, y$ to model possible dependencies over the focal
plane. This is the classic PSF-matching algorithm of \citet{alard98} and
\citet{alard00}, and extended by \citet{yuan08}. This algorithm has been
successfully used by several time-domain surveys, e.g.,
OGLE \citep{wyrzykowski14}, La Silla-QUEST \citep{hadjiyska12},
Pan-STARRS \citep{kaiser10} and the SDSS-II Supernova Survey \citep{sako08}.
Initially, we extensively validated this method for PTF (at IPAC, Caltech)
as implemented in the HOTPANTS\footnote{see \brokenurl{http://www.astro.washington.edu/users/becker/v2.0/}{hotpants.html}}
and DIAPL utilities \citep{wozniak00}.
A major limitation was the specification of a number of fixed configuration
parameters. These parameters are not fitted, i.e., the Gaussian widths
(at least four were needed) and the polynomial orders (six more parameters).
The overall performance was sensitive to the precise choice of these
parameters. Furthermore, the basis functions were not complex or flexible
enough to model the bulk of the data encountered in the survey, under a
continuum of atmospheric conditions and unforeseen instrumental behaviors.  
Despite attempts to constrain the basis function constants using
{\it a-priori} information in a dynamic manner (e.g., as prescribed by
\citet{israel07}), a generic-enough representation for $K(u,v)$ under this
framework that kept the false-positive rate amongst transient candidates
appreciably low and at a manageable level eluded us.  

A more flexible ``shape free'' basis representation for $K(u,v)$ was proposed
by \citet{bramich08}. Here the kernel is discretized into $L\times M$ pixels
and each pixel value therein, $K_{lm}$, is treated as a free parameter in the
OLS fitting problem. This free form basis is also referred to as the delta
function representation where the kernel can be expressed as a 2D array of
delta functions:
\begin{equation}
K(u,v) = K_{lm}\delta(u-l)\delta(v-m).
\end{equation}
A kernel size of $9\times 9$ pixels (input parameter {\it \textendash{kersz}})
then has 81 orthonormal basis functions when expanded as a linear combination
according to equation (\ref{lin}). For comparison, the best Gaussian-basis
model mentioned above has 252 free parameters for effectively the same amount
of input data in the estimation process. Apart from the kernel image size and
threshold parameters used to creating the regularized PSF-inputs
(Section~\ref{preppsfs}), there are no shape-based tuning parameters for
the delta-function representation. For this reason, it can accomodate
more generic shapes, as well as capture offsets in the astrometry on
scales used to construct the input PSF data $I_{ij}$ and $R_{ij}$, coming
from effectively a $N_x\times N_y$ image partition. As we shall discuss,
this unconstrained specification is both good and bad.

With this representation, the model image (equation~\ref{mod}) can be written
in terms of the unknown coefficients $K_{lm}$ as follows:
\begin{equation}\label{moddelt}
M_{ij} = dB + \sum_l\sum_m K_{lm}R_{(i+l)(j+m)}
\end{equation}
The objective function to minimize (the equivalent of equation~\ref{chi2})
then becomes:
\begin{equation}\label{chi2new}
\chi^2_o = \sum_{i,j}\frac{1}{\sigma_{ij}}\left[I_{ij} - dB -
                     \sum_l\sum_m K_{lm}R_{(i+l)(j+m)}\right]^2.
\end{equation}

\noindent
The optimal values of $K_{lm}$ and $dB$ are those that minimize $\chi^2_o$,
i.e., where the partial derivaties of $\chi^2_o$ with respect to each
parameter are all zero:
\begin{equation}
\begin{aligned}
& \left(\frac{\partial\chi^2_o}{\partial K_{lm}}\right)_{l=l_o,\,m=m_o,\,dB_o} =
      \,\, 0,\\
& \left(\frac{\partial\chi^2_o}{\partial dB}\right)_{l=l_o,\,m=m_o,\,dB_o} =
      \,\, 0.
\end{aligned}
\end{equation}
These are evaluated at the specific parameter values indexed by $l_o, m_o$
for $K_{lm}$ and $dB_o$ to yield two general relations:
\begin{equation}\label{norm1} 
\begin{aligned}
& K_p\sum_{i,j}\frac{R_{(i+l_o)(j+m_o)}R_{(i+l)(j+m)}}{\sigma_{ij}}\,\, + \\ 
& dB_o\sum_{i,j}\frac{R_{(i+l_o)(j+m_o)}}{\sigma_{ij}}\,\, - \\
& \sum_{i,j}\frac{I_{ij}R_{(i+l_o)(j+m_o)}}{\sigma_{ij}} =\,\, 0 
\end{aligned}
\end{equation}
and
\begin{equation}\label{norm2}
\begin{aligned} 
& \left(\sum_p K_p\right)\sum_{i,j}\frac{R_{(i+l_o)(j+m_o)}}{\sigma_{ij}} +
        dB_o\,\, - \\ 
& \,\,\,\,\sum_{i,j}\frac{I_{ij}}{\sigma_{ij}} =\,\, 0
\end{aligned}
\end{equation}
respectively. The kernel pixel indices in equations (\ref{norm1}) and
(\ref{norm2}) have the ranges:
\begin{equation}\label{ranges} 
\begin{aligned}
& -(L-1)/2 \leq l_o \leq (L-1)/2 \\
& -(M-1)/2 \leq m_o \leq (M-1)/2 
\end{aligned}
\end{equation}  
where $(l_o, m_o) = (0, 0)$ corresponds to the center pixel of the kernel 
with dimensions $L\times M$ pixels and $p$ is a one-dimensional index: 
$p = 1, 2, 3,..., LM$. For a given $l_o, m_o$, 
\begin{equation}\label{pindex}
p = l_o + Lm_o + (LM + 1)/2.
\end{equation}
Equations (\ref{norm1}) and (\ref{norm2}) lead to a simultaneous system
of $LM + 1$ equations in $LM + 1$ unknowns that can be written in the
vector-matrix form: 
\begin{equation}\label{ax} 
A\,X\,=\,B,
\end{equation}
where $X$ is a vector containing the $LM$ kernel-pixel unknowns $K_p$
($= K_{lm}$) and differential background estimate $dB_o$. 

Equation (\ref{ax}) can be inverted using standard techniques, and at first,
$X$ was estimated directly using a $LU$-decomposition of $A$.
However, in accord with previous analyses \citep[e.g.,][]{becker12}, the
unconstrained nature of a pure delta-function representation can make the
$K_{lm}$ solutions very sensitive to noise and contamination from non-PSF 
related signal in the inputs $I_{ij}$ and $R_{ij}$. The model-fit is
therefore  subject to {\it over-fitting}. This is the so-called bias versus
variance tradeoff where one is after a solution that is expressive
enough to avoid biases, but not so complex as to introduce excessive
variance when the solutions are later applied to match the PSFs in 
an entire input image or partition. In the end, one is after the
{\it true} PSF-matching kernel for the two images. Despite our
attempts to mitigate noise in the input PSF co-add stamps
($I_{ij}$ and $R_{ij}$; Section~\ref{preppsfs}), noise is inevitable.
One way to avoid overfitting and still maintain optimality is to invoke
regularization in the estimation process.

\subsubsection{The Regularized PiCK Method}\label{pick}

\citet{becker12} implemented a regularized version of the
delta-function kernel model by introducing a tunable smoothing constraint
in the $\chi^2$ objective function. This function is proportional
to the second spatial derivative in the input pixels and is intended to
penalize fits that are too irregular as a result of high-frequency noise.
This regularization
method was explored and validated in detail by \citet{bramich16} from the
aspect of maximizing photometric accuracy in final difference images.
This extension is attractive, but it did not become known to us until after
we had implemented an alternative regularized version of the delta-function
model (see below). This worked very well on PTF image data. We will refer
to this method as the Pixelated Convolution Kernel method (or PiCK for short).

Instead of imposing a regularizing constraint on the objective function
as in \citet{becker12}, our approach involves regularizing the coefficient
matrix $A$ in equation (\ref{ax}). We perform a spectral decomposition,
also known as an eigendecomposition of $A$:
\begin{equation}\label{decomp}
A\,=\,V\,W\,V^T,
\end{equation}
where $V$ is an orthogonal matrix ($V^T = V^{-1}$) and $W$ is a diagonal
matrix:
\begin{equation}\label{wdiag}
W = \text{diag}\left(w_1,w_2,\ldots,w_i,\ldots,w_{LM+1}\right)
\end{equation}
with eigenvalues $w_1\geq w_2\geq w_3\geq\ldots w_{LM+1}\geq 0$.
The corresponding linearly independent eigenvectors of $A$
reside in the columns of $V$. Given $A$ is a real symmetric
matrix, such a decomposition can always be found.
This decomposition allows us to examine the basis vectors that will contribute
to the kernel solution. The {\it least}-important eigenvectors of $A$ are
those associated with input noise and can be identified by their
relatively small eigenvalues, below some threshold. These can then be
``zeroed-out''. For example, the inverse of $A$ can then be written:
\begin{equation}\label{invA}
A^{-1}\,=\,V\,W^{-1}\,V^T,
\end{equation}
where
\begin{equation}\label{wdiagi}
W^{-1} = \text{diag}\left(w_1^{-1},w_2^{-1},\ldots,w_i^{-1},\ldots,
                          w_{LM+1}^{-1}\right)
\end{equation}
for all $w_i > 0$. For values $w_i = 0$ (within machine precision), the
$w_i^{-1}$ are replaced by zero in $W^{-1}$. This special replacement
corresponds to the classic Singular Value Decomposition (SVD) method for 
handling singular (ill-conditioned) matrices. The inverse defined by
equation (\ref{invA}) then becomes the {\it pseudo-inverse} of $A$.
This approximation is better conditioned for obtaining a solution to the
matrix equation in (\ref{ax}).  

\begin{figure*}[ht]
\begin{center}
\includegraphics[scale=0.37]{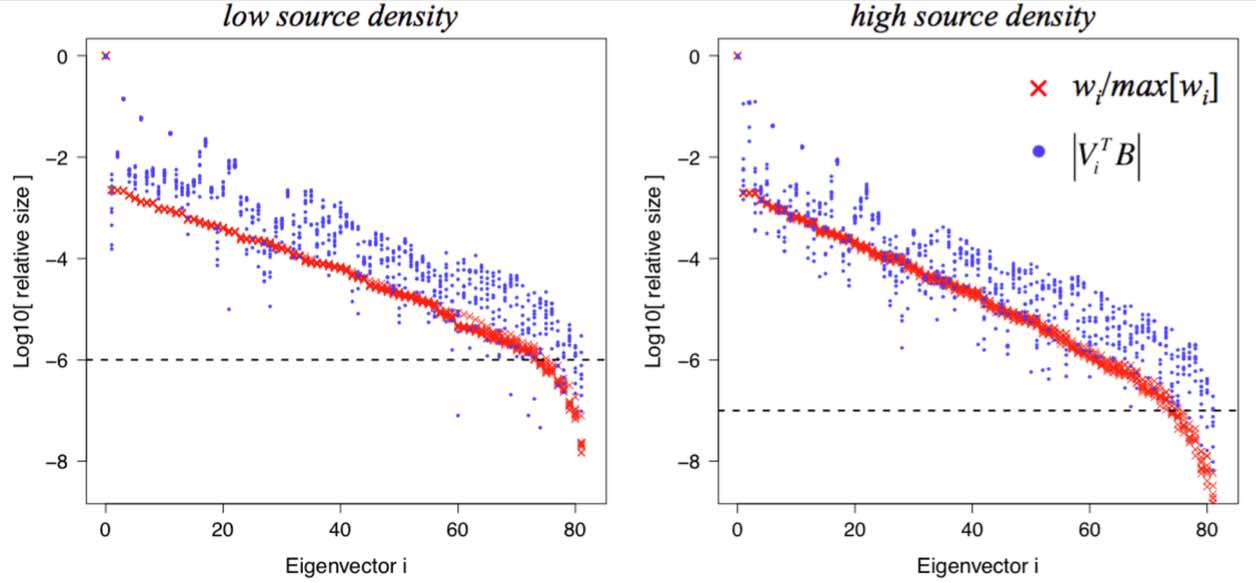}
\caption{Relative eigenvalue strength (crosses) and eigenvector magnitude
         (circles) versus eigenvector index $i$. Increasing $i$ corresponds
         to progressively higher spatial frequencies, eventually approaching
         the pixel noise. Horizontal dashed lines correspond to dynamically
         derived truncation thresholds; see Section~\ref{pick}.
         Results are shown for two kernel solutions derived for image 
         partitions with two different source densities,
         {\it left:} moderate to low-density typical of high galactic
         latitudes; {\it right:} a density that is $\sim 80\times$ higher.}
\label{fig:eigend}
\end{center}
\end{figure*}

The essence of the PiCK method is to make this SVD-like replacement more
generic and less restrictive on the specific $w_i$ to replace.
The goal is to also make $A$ more regularized against noisy input data
(including singular cases) by finding the largest eigenvalue $w_k$ such that
$w_k/\text{max}\{w_i\} < T$ for some threshold $T$ (see below).
For all $i \geq k$, we reset $w_i^{-1}=0$ in $W^{-1}$ and then proceed to
obtain a solution. Following the decomposition in (\ref{invA}), the solution
vector $X$ in equation (\ref{ax}) can be written
\begin{equation}
X\,=\,\sum_i^{LM+1}\left(\frac{1}{w_i}\,V_i^T\,B\right)V_i.
\end{equation}
In this form, it can be seen that the noisiest ({\it least}-relevant)
eigenvectors $V_i$ according to $w_i^{-1}\approx 0$ will not significantly
contribute to $X$. Therefore, they can be eliminated by forcing $w_i^{-1}=0$.
These eigenvectors can be also identified by their relatively small
dot-products with the $B$ vector, $|V_i^TB|$, as shown in
Figure~\ref{fig:eigend}.

The threshold $T$ is dynamically derived on a {\it per} image-partition basis,
i.e., where an inversion of equation (\ref{ax}) via (\ref{invA}) is performed.
This uses the following semi-empirical criterion:
\begin{equation}\label{crit}
T\,=\,\text{min}\left\{10^{-6},\, 10^{th}\text{ percentile in }
       \frac{w_i}{\text{max}\left[w_i\right]}\right\},
\end{equation}
where the second argument corresponds to the low-tail percentile of the
{\it max}-normalized $w_i$ distribution.
The values in equation (\ref{crit}) were tuned by examining the
eigendecompositions of $A$ using iPTF image data acquired
across different environments, from low source-density to densities
approaching those in the galactic plane. A range of $T$ values were
then tested by exploring the impact of the corresponding
regularized solutions on the overall fit $\chi^2$ (equation~\ref{chi2new}).
Examples of these eigendecompositions are shown in Figure~\ref{fig:eigend}.
The criterion in equation (\ref{crit}) corresponds to approximately an
inflexion point in the relative eigenvalue size. This choice is also conservative
in the sense that the amount of legitimate high-frequency information thrown
away (not associated with noise) is expected to be insignificant as determined
by the change in $\chi^2$ with and without regularization ($T=0$). I.e., we
ensure that $\Delta\chi^2\lesssim2\sigma_{\chi^2}=2\sqrt{2\nu}$, where $\nu$
is the effective number of degrees of freedom. In the current setup for iPTF,
\begin{equation}\label{dof}
\begin{aligned}
\nu & = \left(psfsz\,\times\,psfsz\right) - \left(LM + 1\right) \\ 
    & = \left(25\,\times\,25\right) - \left(9\,\times\,9 + 1\right) = 543. 
\end{aligned}
\end{equation}

The regularization threshold $T$ is also dependent on the size of the
kernel assumed ($= L\times M$ free parameters to solve) since this determines
the relative fraction of noise contributing to the $K_{lm}$ estimates.
The kernel image size was tuned beforehand to be small enough to avoid
introducing too many free parameters that would result in overfitting
on noisy backgrounds in the PSF stamps, but large enough to accomodate
the range of seeing (point-source profile widths) encountered. This
ensures both unbiased kernel solutions and minimal variance following
their application, i.e., a compromise in the bias versus variance tradeoff
mentioned above.

\begin{figure*}
\begin{center}
\includegraphics[scale=0.45]{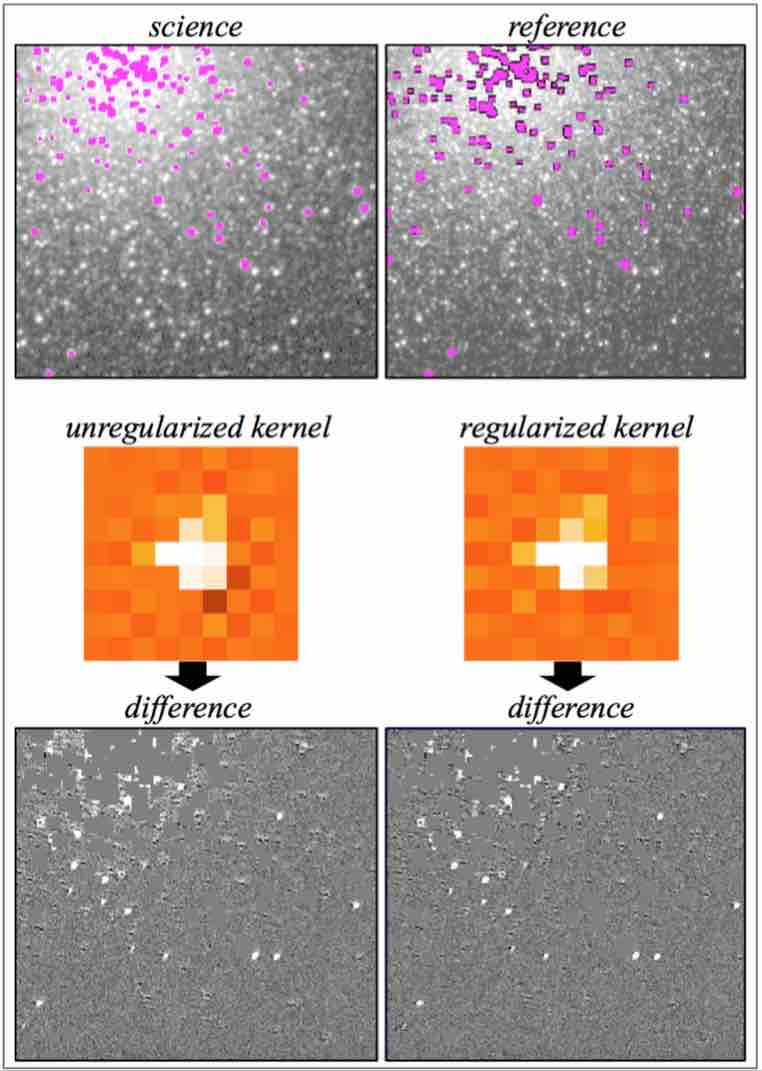}
\caption{{\it Top left and right}: zoomed-in region on a preprocessed
         science and resampled reference image respectively containing
         a portion of the M13 Globular Cluster and measuring
         $\simeq 5^{\prime}$ on a side. The magenta regions are
         saturated pixels. {\it Middle left and right}: PSF-matching
         kernels based on {\it no} regularization (from a
         na{\"i}ve inversion of equation~\ref{ax}) and with
         regularization respectively (Section~\ref{pick}). These are
         enlarged for clarity. {\it Bottom left and right}: difference images
         resulting from ``science {\it minus} kernel-convolved reference''
         for the unregularized and regularized kernels respectively.  
         Saturated pixel regions were reset to zero in the difference
         images. These were also spatially-expanded from their original
         size (top images) due to the convolution process.
         The bright source-like residuals in the difference images
         are on the locations of known RR-Lyrae variable stars.
         For an animation of products for the entire CCD
         across multiple epochs, see \texttt{\protect\url{http://web.ipac.caltech.edu/staff/fmasci/home/idemovies/d4335ccd8f2movie.html}}. }
\label{fig:diffeg}
\end{center}
\end{figure*}

When a PSF-matching kernel image with estimates $K_{lm}$ is available,
a measure of the relative residual gain between the science and
reference image pixels for a specific image partition (from which
$I_{ij}$ and $R_{ij}$ were extracted) is given by the sum:
\begin{equation}\label{ksum}
K_{sum} = \sum_l\sum_m K_{lm}.
\end{equation}
The $K_{sum}$ values (across all image partitions) can be used to assess
the accuracy of the global relative gain correction computed upstream
(Section~\ref{gm}). More importantly, they provide local estimates of
any residual photometric gain where if significant, can be used to
refine the gain factor at the image partition level prior to
differencing (see Section~\ref{appkern}). $K_{sum}$ also provides
a diagnostic to assess the quality of the kernel solution. For example,  
an image partition with a $K_{sum}$ that significantly deviates from
unity compared to that of its neighboring partitions could indicate
a problem in the estimation process, perhaps triggered by bad or low-quality
input PSFs. For details, see the discussion on quality assurance
in Section~\ref{diffqa}.

To summarize, we have extended the free form delta-function model
representation to derive PSF-matching kernels by performing a simple
regularization of the matrix system used for the least-squares solution.
This uses an eigendecomposition to retain the most significant basis vectors
within a statistically validated threshold. We have coined this
the PiCK method. Figure~\ref{fig:diffeg} shows examples of input images,
kernels, and the resulting difference images for a relatively
dense field with and without regularization included. The relative change in
$\chi^2$ (equation~\ref{chi2new}) going from the unregularized to regularized
solution for the entire image shown in Figure~\ref{fig:diffeg}
is $\approx -3.5\%$. The PiCK method leads to smoother PSF-matching kernels
and hence difference images in general. It is also robust against contamination
in the input PSFs ($I_{ij}$ and $R_{ij}$), for example when constructed
from high source-density regions.

\subsection{Kernel Application and Difference Image Products}\label{appkern}

The PSF-matching convolution kernel is first normalized to unity to yield
\begin{equation}\label{knorm}
\widetilde{K}_{lm}\,=\,\frac{K_{lm}}{K_{sum}},
\end{equation}
where $K_{sum}$ was defined by equation (\ref{ksum}).
$\widetilde{K}_{lm}$ is then convolved with the specific
image partition that was initially selected for convolution using the method
in Section~\ref{pkern}, i.e., defining the convolution direction. The reason
for decoupling $K_{sum}$ (the local relative gain factor) from the raw kernel
$K_{lm}$ is so that any residual gain correction can be refactored and applied
as a multiplicative correction on the science image pixels only, and not
the resampled reference image, regardless of the convolution direction.
This is consistent with our {\it modus operandi} in PTFIDE: all corrections
are applied to the science image pixels, in order to match the
photometrically and astrometrically calibrated reference image
as best as possible.

If the reference image was selected for convolution, the difference image 
pixel values for an image partition from which the kernel and
differential background estimates were derived can be written:
\begin{equation}\label{diffrefc} 
D_{ij} = \left[\frac{I_{ij} - dB_o}{K_{sum}}\right] - 
         \sum_l\sum_m \widetilde{K}_{lm} R_{(i+l)(j+m)}. 
\end{equation}
If the science image was selected for convolution, the difference image
pixel values for the image partition can be written:
\begin{equation}\label{diffscic}
D_{ij} = \left[dB_o + K_{sum}\sum_l\sum_m
         \widetilde{K}_{lm} I_{(i+l)(j+m)}\right] - R_{ij}.
\end{equation}

Two difference images per science, reference image pair are generated,
a positive ({\it sci -- ref}) and negative ({\it ref -- sci}) difference
image. As described in Section~\ref{prodsum}, this is to enable the
detection of transients and variables that happen to be below the
reference-image baseline level at any observation epoch. The positive and
negative difference images are written to FITS formatted files with
filename suffixes {\it \_pmtchscimref.fits} and {\it \_pmtchrefmsci.fits}
respectively. If the debug switch was set, ancillary products representing the
different components of equations (\ref{diffrefc}) or (\ref{diffscic}) prior to
differencing can also be generated (see Table~\ref{tab:sout}).
These are the final science image, convolved or not with 
$K_{sum}$ and $dB_o$ applied: {\it \_scibefdiff.fits}, and the 
convolved counterpart: either $\widetilde{K}\otimes R_{ij}$ or
$\widetilde{K}\otimes I_{ij}$: {\it \_pmtchconvref.fits} or
{\it \_pmtchconvsci.fits} respectively.

The science and reference image bad-pixel masks generated upstream
(Section~\ref{bp}) include the effects of convolution where bad pixel regions
are expanded accordingly. These are combined to produce a final
effective bad-pixel mask for both the positive and negative difference
images. If the debug switch was set, this can also be written to FITS
format with filename suffix {\it \_pmtchdiffmsk.fits}. Furthermore, all
bad pixels  in the difference images are tagged with value -999999.

An image of the 1-$\sigma$ uncertainties corresponding to the $D_{ij}$
images (equation~\ref{diffrefc} or \ref{diffscic}) is also generated.  
These uncertainties are estimated by RSS'ing the input uncertainties
for the science and reference images (equations~\ref{sigsci} and 
\ref{sigref} respectively in Section~\ref{punc}) with a correction
for correlated-noise:
\begin{equation}\label{sigdif}
\sigma_{D_{ij}} = F_c\sqrt{\sigma_{sci}^2 + \sigma_{ref}^2}.
\end{equation}
The correlated-noise correction factor $F_c$ accounts for the
{\it diminution} in the pixel RMS noise (lost to covariance) due to the
convolution process. Recall that this convolution may have been performed
on either the science or reference image (see above). $F_c$ is approximated
as the ratio of the robust pixel RMS in the actual difference image
to the RSS'd background RMSs in the science and reference images
prior to any convolution:
\begin{equation}\label{fc}
F_c\approx\frac{\sigma_{bckdiff}}{\sqrt{\sigma_{bcksci}^2 + \sigma_{bckref}^2}}.
\end{equation}
The background variances in the denominator of equation (\ref{fc}) are the
same estimates used in equations (\ref{sigsci}) and (\ref{sigref}) of
Section~\ref{punc}. The presence of correlated-noise in the difference
image will underestimate the photometric uncertainties of extracted sources
therefrom if not properly accounted for on the spatial scales of interest
(i.e., the spatial-extent on which the photometry is performed).
Note that $F_c$ does not represent any correction at the source
level. Its purpose is to capture any modification to the input 
uncertainties ($\sigma_{sci}$ and $\sigma_{ref}$) due to smoothing
from the PSF-matching process. The source level correction is
computed and applied during the source extraction step 
(see Section~\ref{psffit}).

\subsubsection{Photometric Zeropoint Quality Check}\label{czp}

Following application of the spatially-dependent convolution kernels and
associated gain-corrections ($K_{sum}$) to the science image pixels
(equations \ref{diffrefc} and \ref{diffscic}), we compute a new photometric
zeropoint for the adjusted science image. This ZP will also be
applicable to the difference images generated therefrom. It provides a
sanity check on the global-image {\it ZPSCI} value computed upstream
using the input {\it absolutely} calibrated reference image photometry
(see Section~\ref{rzp}). As mentioned, {\it ZPSCI} enables a more accurate
absolute calibration of either PSF-fit or big-aperture photometry.

The new ZP is only computed if the {\it \textendash{phtcaldif}} switch
was specified on input. If so, it is written as the {\it ZPDIF} keyword
with accompanying RMS {\it ZPDIFRMS} to the FITS headers of both the positive
and negative difference images. The computation uses exactly the same
methodology as outlined in Section~\ref{rzp}. If the debug switch was
set, a catalog of the reference-to-science image matches used to compute
{\it ZPDIF} is generated with suffix {\it \_sx\_scibefdiff.tbl}. When the
PSF-matching and additional gain-refinement steps perform as intended,
{\it ZPDIF} is generally consistent with {\it ZPSCI}, within random
measurement error. Large deviations in $\Delta ZP = ZPSCI - ZPDIF$ usually
imply a problem with either the PSF-matching kernel(s), the input astrometry,
or its later refinement since astrometric accuracy will indirectly
affect the source matching step used to estimate the ZPs. $\Delta ZP$
therefore provides a powerful quality assurance metric.

\subsubsection{Difference-Image Quality Assurance Metrics}\label{diffqa}

Metrics and diagnostics for a difference image product are shown
in Table~\ref{tab:iqa}. These encompass information on photometric
zeropoints; pixel statistics before and after PSF-matching;
properties of the PSF-matching kernels; statistics on the input science and
reference images; global image FWHM values; and the number of
candidates extracted (Section~\ref{ctp}). These metrics are used to
support later machine-learned vetting (Section~\ref{ml}).

In PTFIDE processing, two checks are performed to assess the quality
of the difference image: (i) ``atrocious'' and unusable
for extracting candidates, and (ii) simply ``bad'' and warranting visual
examination before use. PTFIDE does not extract candidates if (i) is
satisfied, but does proceed to extract candidates if (ii) is satisfied.
The indicator flag for either condition is the {\it status} flag
in Table~\ref{tab:iqa} $[= 0$ (bad) or 1 (good)$]$.
Even though candidates are extracted under (ii) during processing,
these are not loaded into the database (Section~\ref{schema}). The rationale
is that bad subtractions will lead to thousands of spurious
candidates and strain both database loading as well as machine-learned
vetting downstream. The conditions for (i) and (ii) are
determined using one-dimensional cuts on a number of metrics from
Table~\ref{tab:iqa} as follows.

First, a difference image is flagged {\it atrocious} if the following criteria
are satisfied:
\begin{equation}\label{icuts1}
\begin{aligned}
   diffpctbad                     \, & > thres\_a1        \;\;\; \text{or} \\
   (\Delta\chi^2_{med}            \, & > thres\_a2        \;\;\; \&        \\
    \Delta\chi^2_{avg}            \, & > thres\_a3)       \;\;\; \text{or} \\
   diffsigpix                     \, & > thres\_a4        \;\;\; \text{or} \\
   |1 - medksum|                  \, & > thres\_a5        \;\;\; \text{or} \\
   medkpr                         \, & > thres\_a6        \;\;\; \text{or} \\
   \left(maxksum - minksum\right) \, & > thres\_a7,
\end{aligned}
\end{equation}
where in terms of the metrics listed in Table~\ref{tab:iqa},
\begin{equation}\label{dchisq}
\begin{aligned}
\Delta\chi^2_{med} \, & = \frac{chisqmedaft - chisqmedbef}{chisqmedbef}, \\
\Delta\chi^2_{avg} \, & = \frac{chisqavgaft - chisqavgbef}{chisqavgbef}
\end{aligned}
\end{equation}
and the seven thresholds ($thres\_ai$) are specified by the
{\it \textendash{uglydiff}} input string with defaults defined in
Table~\ref{tab:inp}.  

\begin{deluxetable}{lcc}
\tabletypesize{\small}
\tablecaption{Difference Image Quality Statistics\label{atfrac}}
\tablewidth{0pt}
\tablehead{
\colhead{Quality} &
\colhead{$|b| \leq 5^\circ$;  $320^\circ\leq l\leq 40^\circ$} &
\colhead{$|b| \geq 70^\circ$; $0^\circ\leq l < 360^\circ$} \\
\colhead{} &
\colhead{(1824 diff. images)} &
\colhead{(7629 diff. images)} \\ 
}
\startdata
atrocious\tablenotemark{a}   &   25.0\%  &  0.42\%  \\
bad\tablenotemark{b}         &   34.5\%  &  11.18\% \\
good\tablenotemark{c}        &   40.5\%  &  88.40\% \\
\enddata
\tablenotetext{a}{Flagged according to the criteria in (\ref{icuts1}).}
\tablenotetext{b}{Flagged according to the criteria in (\ref{icuts2}).}
\tablenotetext{c}{Satisfy neither (\ref{icuts1}) or (\ref{icuts2}).}
\end{deluxetable}

If the difference image is not labelled atrocious according to the criteria in
(\ref{icuts1}), it is subject to the less-severe criteria below. These are
applied after transient candidates are extracted. The metrics used here
are therefore a subset of those from above (but with lower ``badness''
thresholds) and others related to the number and properties of candidates
extracted. See Table~\ref{tab:iqa} for definitions.
\begin{equation}\label{icuts2}
\begin{aligned}
   diffpctbad                     \, & > thres\_b1        \;\;\; \text{or} \\
   (\Delta\chi^2_{med}            \, & > thres\_b2        \;\;\; \&        \\ 
    \Delta\chi^2_{avg}            \, & > thres\_b3)       \;\;\; \text{or} \\
   diffsigpix                     \, & < thres\_b4        \;\;\; \text{or} \\
   diffsigpix                     \, & > thres\_b5        \;\;\; \text{or} \\
   |1 - medksum|                  \, & > thres\_b6        \;\;\; \text{or} \\
   medkpr                         \, & > thres\_b7        \;\;\; \text{or} \\
   ncandfiltrat (posdiff)         \, & > thres\_b8        \;\;\; \text{or} \\
   ncandfiltrat (negdiff)         \, & > thres\_b9        \;\;\; \text{or} \\
   \frac{refinpseeing^a}{sciinpseeing}  \, & > thres\_b10 \;\;\; \text{or} \\
   \frac{refconvseeing^b}{sciinpseeing} \, & > thres\_b11. 
\end{aligned}
\end{equation}
{\let\thefootnote\relax\footnote{\textsuperscript{a}
This ratio is used if the reference image was convolved, otherwise the inverse
of this ratio is used.}}
{\let\thefootnote\relax\footnote{\textsuperscript{b}
This ratio is used if the reference image was convolved, otherwise the ratio is
$refconvseeing/refinpseeing$.}} 
The eleven thresholds ($thres\_bi$) are specified by the
{\it \textendash{baddiff}} input string with defaults defined in
Table~\ref{tab:inp}. If the criteria in (\ref{icuts2}) are satisfied,
{\it status} $= 0$ is assigned to indicate a possibly bad difference.
This flag is propagated to the {\it subtractions} table of the
transients database (Section~\ref{schema}).

Table~\ref{atfrac} summarizes the percentages of {\it atrocious}, {\it bad},
and {\it good} difference images obtained from real-time processing for two
sky regions spanning different galactic latitudes and longitudes.
These regions sample the extremes in source-density: near the galactic
center and bulge, and high galactic latitudes. These were covered by iPTF 
from August 2015 to January 2016. The high fraction of failures near
the galactic center (or bulge) can be attributed to adverse effects from
high source confusion on the processing steps prior to differencing,
for example, astrometric calibration, and/or gain and PSF-matching. All these
steps depend on source-matching of some kind between the science and reference
images, and is severely challenged in regions of high source density.
For examples of bad or unusable difference images and their causes,
see Section~\ref{train}.

If the debug switch was set, an image of the locally-smoothed pixel
chi-square is generated for both positive and negative difference images.
This is analogous to a {\it reduced} $\chi^2$ and is defined as:
\begin{equation}\label{chimg}
\chi^2_d = \Bigg\langle \frac{D^2_{ij}}{\sigma^2_{D_{ij}}} \Bigg\rangle_{ij},
\end{equation}
where the angled brackets denote boxcar averaging over pixels $i,j$
that fall within $8\times8$ pixel bins over the difference image.
The image generated from equation (\ref{chimg}) is written to a FITS
formatted file with suffix {\it \_pmtchdiffchisq.fits}. Large values in
$\chi^2_d$ (significantly above unity) may indicate residuals from
imperfect instrumental calibrations, underestimated pixel
uncertainties, or the presence of real transient sources. Small values
($\chi^2_d\ll 1$) will imply that the pixel uncertainties are overestimated.

\begin{deluxetable}{p{4.5cm}p{12cm}}
\tabletypesize{\footnotesize}
\tablecaption{Difference image-based Quality Assurance Metrics from
              PTFIDE\label{tab:iqa}}
\tablewidth{0pt}
\tablehead{
\colhead{Metric\tablenotemark{a}} &
\colhead{Description} \\
\colhead{} &
\colhead{}
}
\startdata
zpmaginpsci &
Photometric zeropoint (ZP) estimate of science image {\it before}
rescaling to reference [mag] \\ 

zpmaginpsciunc &
1-$\sigma$ uncertainty in zpmaginpsci [mag] \\

zpmagcormed &
Median ZP correction offset to zpmaginpsci over all partitioned
convolution kernel sums [mag] \\

zpmagcormin &
Min. ZP correction offset to zpmaginpsci over all partitioned
convolution kernel sums [mag] \\

zpmagcormax &
Max. ZP correction offset to zpmaginpsci over all partitioned
convolution kernel sums [mag] \\

zpmaginpscicor &
Refined photometric zeropoint (ZP) of science image following application
of kernel sums: ``zpmaginpsci $+$ zpmagcormed'' [mag] \\

zpfacinpsci &
Photometric scale factor applied to science image to match reference
image ZP \\

zpfacinpsciunc &
1-$\sigma$ uncertainty in zpfacinpsci \\

zpref &
Photometric zeropoint (ZP) of input reference image [mag] \\ 

nmatch &
Number of sources matched within 3.0 pixels between science and
reference images to support initial gain-matching and astrometric
refinement \\

fluxrat &
Median flux ratio of matched sources prior to global gain-matching:
$fluxsci / fluxref$ \\

pctfluxrat &
$5^{th} - 95^{th}$ percentile range in 'fluxrat' values of matched sources \\

deltax &
Median positional difference along X-axis using matched sources:
$X_{ref} - X_{sci}$ [pixels] \\

sigdeltax &
1-$\sigma$ uncertainty in deltax [pixels] \\

pctdeltax &
$5^{th} - 95^{th}$ percentile range in deltax values across matched
sources [pixels] \\

deltay &
Median positional difference along Y-axis using matched sources:
$Y_{ref} - Y_{sci}$ [pixels] \\

sigdeltay &
1-$\sigma$ uncertainty in deltay [pixels] \\

pctdeltay &
$5^{th} - 95^{th}$ percentile range in deltay values across matched
sources [pixels] \\

medksum &
Median pixel-sum of all image-partitioned raw convolution kernel sums \\

minksum &
Minimum pixel-sum of all image-partitioned raw convolution kernel sums \\

maxksum &
Maximum pixel-sum of all image-partitioned raw convolution kernel sums \\

medkdb &
Median differential background over all image-partitioned raw convolution
kernels [DN] \\

minkdb &
Minimum differential background over all image-partitioned raw convolution
kernels [DN] \\

maxkdb &
Maximum differential background over all image-partitioned raw convolution
kernels [DN] \\

medkpr &
Median $5^{th} - 95^{th}$ percentile pixel range of all image-partitioned
raw convolution kernels \\

minkpr &
Minimum $5^{th} - 95^{th}$ percentile pixel range of all image-partitioned raw
convolution kernels \\

maxkpr &
Maximum $5^{th} - 95^{th}$ percentile pixel range of all image-partitioned raw
convolution kernels \\

zpdiff &
Photometric zero point of difference image [mag] \\

ngoodpixbef &
Number of good pixels in difference image before PSF-matching [pixels] \\

ngoodpixaft &
Number of good pixels in difference image after PSF-matching [pixels] \\

nbadpixbef &
Number of bad pixels in difference image before PSF-matching [pixels] \\

nbadpixaft &
Number of bad pixels in difference image after PSF-matching [pixels] \\

medlevbef & 	
Difference image median level before PSF-matching [DN] \\

medlevaft & 	
Difference image median level after PSF-matching [DN] \\

avglevbef & 	
Difference image average level before PSF-matching [DN] \\

avglevaft & 	
Difference image average level after PSF-matching [DN] \\

medsqbef & 	
Median of squared differences before PSF-matching [$\text{DN}^2$] \\

medsqaft & 	
Median of squared differences after PSF-matching [$\text{DN}^2$] \\

avgsqbef & 	
Average of squared differences before PSF-matching [$\text{DN}^2$] \\

avgsqaft & 	
Average of squared differences after PSF-matching [$\text{DN}^2$] \\

chisqmedbef &
Difference image chi-square using median before PSF-matching \\

chisqmedaft &
Difference image chi-square using median after PSF-matching \\

chisqavgbef &
Difference image chi-square using average before PSF-matching \\

chisqavgaft &
Difference image chi-square using average after PSF-matching \\

scibckgnd &
Modal background level in science image after gain and background
matching [DN] \\

refbckgnd &
Modal background level in ref-image after gain, background matching,
and resampling [DN] \\

scisigpix &
Robust sigma per pixel in science image after gain and background
matching [DN] \\

refsigpix &
Robust sigma per pixel in ref-image after gain, background matching,
and resampling [DN] \\

scigain &
Effective electronic gain in science image after gain-matching [e-/DN] \\
\tablebreak

scisat &
Saturation level in science image after gain-matching [DN] \\

refsat &
Saturation level in reference image after resampling [DN] \\

scimaglim &
Expected 5-$\sigma$ magnitude limit of science image after gain and
background matching [mag] \\

refmaglim &
Expected 5-$\sigma$ mag. limit of ref-image after gain,
background matching, and resampling [mag] \\

diffbckgnd &
Median background level in difference image [DN] \\

diffpctbad &
Percentage of difference image pixels that are bad/unusable [\%] \\

diffsigpix &
Robust sigma per pixel in difference image [DN] \\

diffmaglim &
Expected 5-$\sigma$ magnitude limit of difference image [mag] \\

sciinpseeing &
Seeing (point source FWHM) of input science image [pixels] \\

refinpseeing &
Seeing (point source FWHM) of input reference image [pixels] \\

refconvseeing &
Seeing (point source FWHM) of reference image after convolution [pixels] \\

ncandraw &
Number of candidates extracted from difference image before any internal
filtering (for positive and negative difference) \\

ncandfilt &
Number of candidates extracted from difference image after internal filtering
using {\it chi}, {\it sharp}, and {\it snr} source metrics; this is the
actual number loaded into database (for positive and negative difference) \\
                
ncandgood &
Number of candidates from difference image likely to be real using 1-D cuts
on several extracted source features (for positive and negative difference) \\

nrefsrcstodifflim &
Number of reference image extractions to difference-image mag limit
(diffmaglim) \\

ncandfiltrat &
ratio: ncandfilt / nrefsrcstodifflim (for positive and negative difference) \\

status &
Good/bad difference image status flag ($=$ 1 or 0); based on combining a number
of internal image metrics (see Section~\ref{diffqa}). Only candidates from
status $=$ 1 subtractions are loaded into database for vetting \\
\enddata

\tablenotetext{a}{A majority of these are loaded into the
                  {\it subtractions} relational database table
                  (Section~\ref{schema}) to support trending
                  and machine-learned vetting (Section~\ref{ml}).}
\end{deluxetable}

\clearpage

\subsection{Candidate Transient Detection and Photometry}\label{ctp}

The detection and photometry of transient candidates on both the positive
and negative difference images is performed using an automated implementation
of the {\it DAOPhot} tool \citep{stetson87,stetson00}. This includes
the subsidiary program {\it Allstar} which performs PSF-fit photometry
on the detections found by {\it DAOPhot}. The primary output photometry for
characterizing transient candidates is PSF-fitting. Fixed concentric
aperture photometry is also generated as a diagnostic.
The benefits of PSF-fit photometry cannot be stressed
enough, at least for detecting transient events. For example, this provides
better photometric accuracy to faint fluxes; the ability to de-blend confused
sources; and simple metrics to distinguish point (PSF-like) sources from
artifacts. These metrics can be used to maximize the reliability of candidates. 

{\it Perl} routines were written to automate all decision-related and
processing steps in {\it DAOPhot}. These steps would have been
done interactively in classic {\it DAOPhot}. This includes all file I/O,
parameter handling, checking of outputs, and quality assurance. The steps
are summarized below.

\subsubsection{Point Source Detection}\label{srcdet}

{\it DAOPhot} first detects sources for input into either the 
PSF-determination step or final PSF-fitting step using a matched filter
via its {\it find} algorithm. The filter is constructed from a 
Gaussian with a proxy for the image FWHM computed upstream using
{\it SExtractor} on the science image. This Gaussian is
convolved with the image in question to construct the point-source matched
filter, i.e., an internal product that is optimized for point-source detection.
This assumes that the PSF is approximately spatially uniform over the
image, and for the purpose of detection, the penalty in detection
S/N by not using the precise spatially-varying PSF is negligible. This image is
then background-subtracted using local estimates of the background. 
The pixel-uncertainty product from equation (\ref{sigdif}), which effectively 
includes contributions from detector read-noise, background, and Poisson
noise at the location of sources, is internally readjusted to account
for the matched filtering. The ratio of the match-filtered,
background-subtracted image and its corresponding uncertainty image is then
thresholded. Either the {\it \textendash{tdetpsf}} or
{\it \textendash{tdetdao}} S/N parameter threshold is used depending if
PSF-determination or final dectection is desired respectively. If the debug
switch was set, the detection tables from these steps are written to files with
extension {\it .coo} (see Table~\ref{tab:sout} for filenames). 

\subsubsection{PSF Determination, PSF-fitting and Aperture
               Photometry}\label{psffit}

PSF generation and PSF-fit photometry (or source extraction of any form)
are only attempted if the input difference image was determined to be of
sufficient quality according to a number of quality metrics
(see Section~\ref{diffqa}). If not, PTFIDE processing terminates
gracefully after the image differencing step. Furthermore, PSF-fit photometry
can be intentionally turned off by omitting the {\it \textendash{psffit}}
switch.

The base parameters specific to {\it DAOPhot} and {\it Allstar} reside in
configuration files specified by {\it \textendash{cfgdao}} and
{\it \textendash{cfgpht}}, with some of the more important (threshold-like)
parameters therein overridden by the command-line inputs:
{\it \textendash{tmaxpsf}}, {\it \textendash{tdetpsf}}, and
{\it \textendash{tmaxdao}}. Also, some parameters are computed dynamically
within {\it ptfide.pl} and override those from both the input files and
command-line. These parameters are those that depend on the image noise,
usable pixel range, and image FWHM: {\it RE, LO, HI, FW, PS, FI,} and
nested aperture radii $A_i$ where $i = 1\ldots 6$. The parameters that depend
on the input FWHM ({\it FW}) are the linear-half-size of the PSF image stamp,
{\it PS} (for PSF-creation only); the PSF-fitting radius, {\it FI}; and the
aperture radii $A_i$, all in units of pixels. Respectively, these parameters
are adjusted according to:
\begin{equation}\label{daodyn}
\begin{aligned}
PS =\, & \text{min}\left(19, \text{int}\{\text{max}\left[9, 6FW/2.355\right]
            + 0.5\}\right), \\
FI =\, & \text{min}\left(7, \text{max}\left[3, FW\right]\right), \\
A_i =\, & \text{min}\left(15, 1.5 \text{max}\left[3, FW\right]\right) + i - 1,
\end{aligned}
\end{equation}
where $i = 1\ldots 6$, and min, max, int denote the minimum,
maximum, and integer part of the argument respectively.

When {\it ptfide.pl} is run in PSF-fit photometry mode (with the
{\it \textendash{psffit}} switch), only one aperture measurement
corresponding to a single fixed aperture is written to the primary output
tables: file suffixes {\it \_pmtchscimrefpsffit.tbl} and
{\it \_pmtchrefmscipsffit.tbl} for the positive and negative difference
images respectively.
This is the aperture number $i$ corresponding to the user-specified
parameter {\it \textendash{apnum}} (currently $= 3$). If however, the
{\it \textendash{apphot}} switch was also specified, all nested aperture
measurements ($i = 1\ldots 6$) are written to separate tables with file
suffixes {\it \_pmtchscimrefapphot.tbl} and {\it \_pmtchrefmsciapphot.tbl}.
These products are not currently generated in production.
It's important to note that the aperture measurements are not curve-of-growth
corrected to account for the variable seeing. They only serve as diagnostics
(or source features; Table~\ref{tab:sqa}) to support machine-learned
vetting (Section~\ref{ml}). 

The PSF used for PSF-fitting is estimated automatically using utilities 
within {\it DAOPhot} and {\it Allstar}. Its shape over a CCD image is 
modeled as a linear function in $x, y$. This dependence is sufficient to
catch spatial variations in the PSF and avoids introducing too 
many free parameters. The PSF is always estimated from the 
{\it resampled} and possibly convolved reference image (following
PSF-matching). This is because the reference image (convolved or not)
is expected to have a higher S/N than the science image. In the end, we
want to estimate the PSF on an image whose point-source profiles match
those in the difference image (either positive or negative). In theory,
given that the only operations performed on the input images prior to
differencing are gain, background and PSF-matching, and given that 
differencing is a linear operation, either the science or reference image
would have sufficed for PSF estimation.

For iPTF, no more than the brightest 200 point sources with magnitudes
$\geq 15.5$ are automatically selected per CCD image to estimate an initial
spatially-varying PSF. This is refined in a second iteration by subtracting
neighboring sources (in the wings) from the initially chosen PSF stars
and then re-estimating the PSF. This second iteration can be rather slow if
there are many neighbors since it involves PSF-fitting to obtain the fluxes
and positions of the neighbors to subtract. To speed up the process, we
regulate the number of neighbors to consider by only subtracting the 
brightest 1000 neighbors to all the initially PSF-picked stars. Therefore,
given a random distribution of sources and $\leq 200$ sources picked for 
PSF generation, $\gtrsim 5$ (brightest) sources on average will be subtracted
prior to refinement of the PSF in the second iteration. This makes the
PSF estimation relatively fast and robust, and there are always enough stars
in an image to yield a reasonably accurate PSF model.

The PSF model is stored in the default {\it DAOPhot} format, with output
file suffix {\it \_pmtchconvrefdao.psf} if the reference image was convolved,
otherwise {\it \_resamprefdao.psf} if the science image was convolved. 
This file consists of a look-up table of corrections to a best-fitting
Gaussian basis model over the image. Other basis functions are available,
but we found a Gaussian works reasonably well for PTF data (in terms of
photometric accuracy). This basis representation also has the least
number of free parameters.  
If the debug switch was set, the {\it DAOPhot}-formatted PSF file is
converted to a FITS image of $16\times32$ PSF-stamps for visualization
(output file suffix {\it \_pmtchconvrefdaopsf.fits}). Other ancillary
files, e.g., the list of PSF-picked stars, their neighbors, and DS9
region files are also generated (see Table~\ref{tab:sout}).

Following PSF estimation, sources are detected on the positive and negative
difference images above a specific threshold as described in
Section~\ref{srcdet}. PSF-fit photometry is then performed using
{\it Allstar}. This program estimates seven quantities per input detection:
flux; flux uncertainty (to be rescaled later, see below); refined image
{\it x, y} position; uncertainties in {\it x, y}; and two metrics from
the fit: {\it chi}, and {\it sharp} (see below). We do not iteratively
subtract the PSF to uncover new sources that were missed in the first
detection pass, e.g., because they were hidden in the wings of brighter
sources. This is commonly done for crowded fields. Given we are extracting
sources from difference images, source-crowding (or rather, blending of 
transient candidates) is largely absent, even in areas of high source
density. One can argue however for cases where instrumental residuals
could blend with real transient sources. This is also rare, but nonetheless
the components of a blend only need to be detected in the first place
so that simultaneous PSF-fitting can yield fluxes and other metrics for
further examination downstream.

The PSF-fit fluxes are converted to absolute calibrated magnitudes using the
{\it ZPSCI} image zeropoint computed upstream (Section~\ref{rzp}). 
This places the photometry on the PTF filter system \citep{ofek12}.
Note that the supposedly refined {\it ZPDIF} value
(Section~\ref{czp}) is not used since analyses have shown that it
exhibits a greater variance than {\it ZPSCI}. 
This arises from systematics in the PSF-matching and relative
gain-matching process. The source {\it x, y} positions are converted
to R.A., Dec. using the difference image WCS (inherited and refined from the
reference image WCS). This information is written to the transient-candidate
extraction tables (one for each difference image; see above). DS9 region
files also accompany these tables (see Table~\ref{tab:pout} for filenames).

\subsubsection{Correcting Uncertainties for Correlated Noise}

Another detail is correcting the flux uncertiainties from PSF-fitting for
correlated pixel-noise in the difference image. As discussed in
Sections~\ref{punc} and \ref{appkern}, correlated noise arises from the
reference-image construction process (interpolation and resampling), but the
more dominant effect is from convolution by the PSF-matching kernel, where
either the science or reference image may have been convolved. 
Given that the reference image has a higher S/N in general, and that
this is the image that is usually convolved (by design), the fraction 
of the pixel noise that is correlated in a difference image is likely
to be small. For example, if the reference image is made from a stack
of $N_f$ science images all with similar pixel noise-variances, the fractional
contribution to the difference image pixel-noise from the (convolved)
reference would be $\simeq (1 + N_f)^{-0.5}$. If the science image were
convolved however, the fractional contribution would be greater, i.e.,
$\simeq (1 + [1/N_f])^{-0.5}$, approaching 100\% if $N_f$ is large.

We use a simple method to correct the source-flux uncertainties from PSF-fit
photometry before writing them to the photometry tables. The flux 
uncertainties from {\it Allstar} are estimated using some combination
of the difference image pixel uncertainties (equation \ref{sigdif})
within the effective fitting area of the PSF. Following convolution
(smoothing) of the one of the images, these will underestimate the true  
source-flux uncertainty. The true uncertainty contributed by the convolved
image can be recovered by scaling its pixel uncertainties by the
effective number of {\it noise pixels}\footnote{\label{notenp}see \brokenurl{http://web.ipac.caltech.edu/staff/fmasci/home/}{mystats/ApPhotUncert\_corr.pdf}}
defining the convolution kernel: 
\begin{equation}\label{npk}
N_k\,=\,\left[\sum_l\sum_m \widetilde{K}_{lm}^2\right]^{-1}, 
\end{equation}
where $\widetilde{K}_{lm}$ are the unit-normalized pixel values of the
kernel image (equation~\ref{knorm}).
$N_k$ effectively represents the correlation length (or in this case the
correlation area) in pixels. Therefore, we seek a correction factor $C$ for
scaling the difference image source flux variance which can also be
written in terms of its science and reference image contributions, after 
convolution by $\widetilde{K}_{lm}$:
\begin{equation}\label{srcvar}
C\sigma_{srcdiff}^2 = \sigma_{srcsci}^2 + N_k\sigma_{srcref}^2, 
\end{equation}
where we assume (for illustration) that the reference image was
convolved. If the science image was convolved, $N_k$ would be multiplying
the science image variance term. Since all the flux-variance terms in
equation (\ref{srcvar}) are some weighted combination (or effectively an RSS)
of the pixel noise variances, in the limit of background-dominated
noise, the correction factor can be estimated from: 
\begin{equation}\label{corrfac}
C\,\approx\,\frac{\sigma_{bcksci}^2 + N_k\sigma_{bckref[conv]}^2}
                 {\sigma_{bckdiff}^2},
\end{equation}
where the variances now denote robust estimates of the background pixel RMS in
the science, $[convolved]$ reference, and difference images. These quantities
are estimated within respective image partitions, i.e., the same
partitions used to compute the kernel images (and hence $N_k$).
This construction was verified using Monte Carlo
simulations\textsuperscript{\ref{notenp}}.

\subsubsection{Coarse Filtering of Raw Candidates}\label{filt}

The raw candidates initially extracted from the difference images
as described in Section~\ref{psffit} undergo loose filtering prior to
writing them to the transient-candidate tables (file suffixes
{\it \_pmtchscimrefpsffit.tbl} and {\it \_pmtchrefmscipsffit.tbl} for
positive and negative difference images respectively). The intent is to
catch the most deviant non-PSF-like residuals from the difference images
and remove them. This somewhat relieves traffic on database loads 
(Section~\ref{schema}) and the machined-learned vetting step
(Section~\ref{ml}) which also heavily relies on interactions with the database.

This simple filtering can result in reductions of up to a factor of five
in the number of raw transient candidates initially {\it detected} from the
point-source match-filtered image down to S/N $\simeq 3.5$
({\it \textendash{tdetdao}} input threshold; Table~\ref{tab:inp}).
It's worth mentioning that this relatively low initial detection
threshold is to ensure completeness prior to further filtering and 
analysis downstream, including photometric S/N. The filtering
applied to the raw detected candidates is
$chi \leq -tchi$; $|sharp| \leq -tshp$; $snrpsf > -tsnr$, where the
thresholds on the right are currently 8, 4, and 4 respectively
(from Table~\ref{tab:inp}).

All these parameters are source-based metrics from
PSF-fitting and are defined as follows: $chi$ -- the ratio of the RMS in
PSF-fit residuals to that expected using prior pixel uncertainties; $sharp$ --
effectively the difference between the squared FWHM of the source light profile
(from a 2D Gaussian fit) and that expected from the PSF template model
derived for the image, i.e., $FWHM_{obs}^2 - FWHM_{psf}^2$; and $snrpsf$ -- the
photometric signal-to-noise ratio using the PSF-fitted flux and uncertainty.
Relatively large values of $chi$ and/or $|sharp|$ indicate deviations from
the nominal PSF template estimated upstream for the difference image
(Section~\ref{psffit}). The optimum values of $chi$ and $sharp$ are
1 and 0 respectively. For example, a cosmic-ray spike would yield
$sharp\ll 0$ and an extended source $sharp\gg 0$, as well as $chi\gg 1$
for both these cases. The number of candidates that satisfy
initial filtering using $chi$, $sharp$, and $snrpsf$ are recorded as
{\it ncandfilt} in the QA output table (Table~\ref{tab:iqa}).

\begin{figure}[ht]
\includegraphics[scale=0.3]{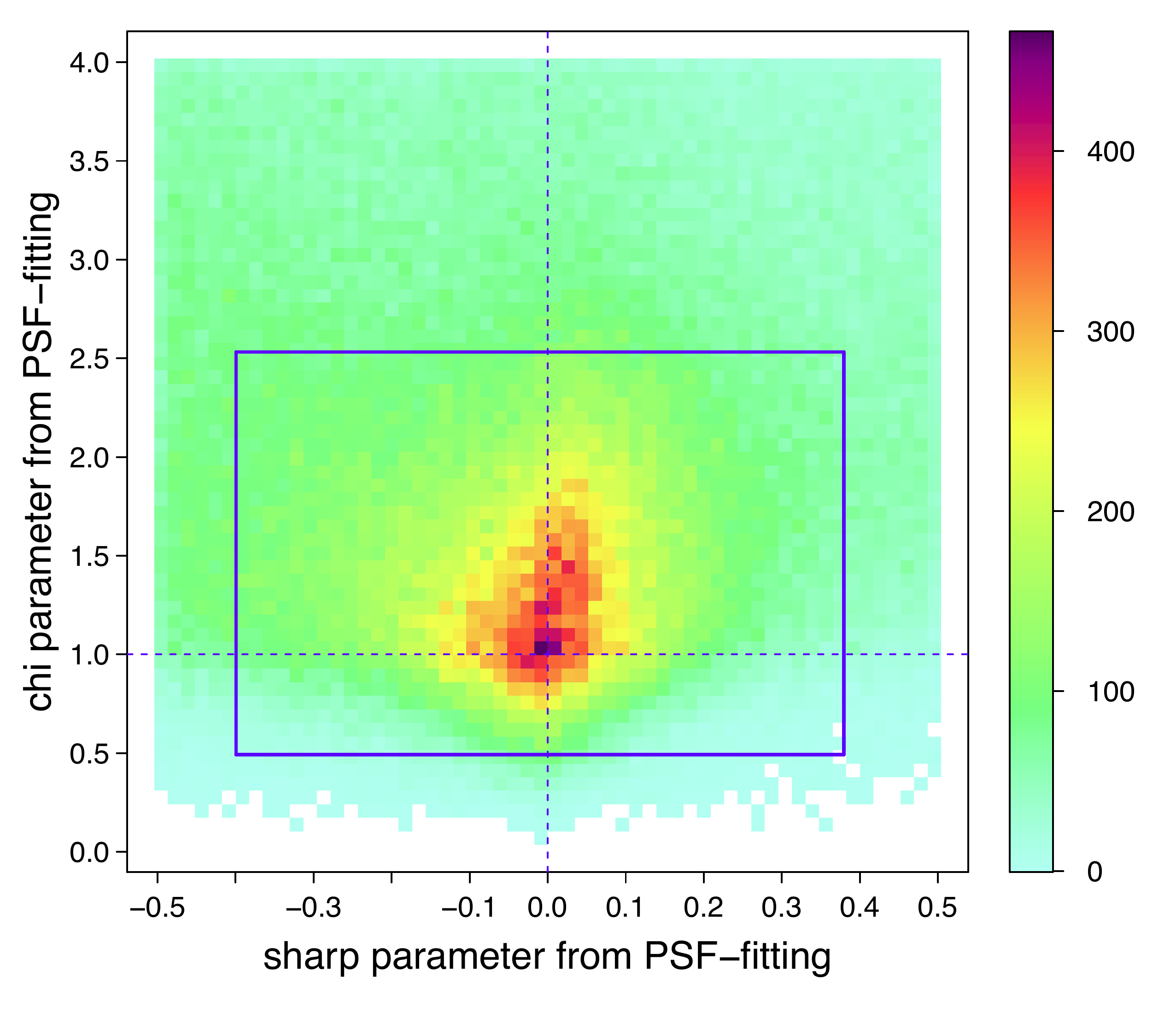}
\caption{{\it chi} versus {\it sharp} parameters from PSF-fitting
         (using {\it DAOPhot}) for transient candidates extracted from
         $\simeq 30,000$ difference images. Only candidates with
         $14.0\leq R_{PTF}\leq 18.5$ and S/N $> 5$ were used. The solid
         rectangle spans the 1 - 99 percentile range along
         each independent axis for candidates with a {\it real-bogus}
         machine-learned reliability score of $\geq 0.8$
         (see Section~\ref{filt} for details).}
\label{fig:chishp}
\end{figure}

Figure~\ref{fig:chishp} shows an example of the distribution in {\it chi}
versus {\it sharp} for raw transient candidates extracted from a
collection of iPTF difference images acquired during late 2015 to
early 2016, mostly at high galactic latitudes. The solid rectangle
shows the region covered by reliable ({\it likely-real}) candidates
according to a machine-learned classification ({\it realbogus}) score of
$\geq 0.8$ (see Section~\ref{ml} for details). The rectangle boundaries
span 1 - 99 percentile ranges along each axis for these likely-real
candidates only. Note that the machined-learned classifier uses over
40 metrics (or features from Tables~\ref{tab:iqa} and~\ref{tab:sqa})
for scoring, of which $chi$ and $sharp$ are the most important.

\subsubsection{Candidate Source Metadata and Features}\label{srcqa}

In addition to the measurements and metrics from PSF-fitting (see above),
the transient-candidate extraction tables are augmented with more
metrics and features to support machine-learned vetting (Section~\ref{ml}).
Most of these metrics are listed in Table~\ref{tab:sqa}. The shape metrics,
for example {\it aimage}, {\it bimage}, {\it elong}, are computed  
by first executing {\it SExtractor}, then associating the extractions
(with metadata) with those from {\it DAOPhot} above. Furthermore, the 
source metrics with names that end in {\it nr} in Table~\ref{tab:sqa}
refer to those of the {\it nearest} reference image source. These are
assigned by associating the {\it DAOPhot} extractions with the input
reference-image catalog, also originally from {\it SExtractor}. 

Prior to implementation of the machine-learned vetting (Section~\ref{ml}),
we resorted to simple one-dimensional cuts on a number of metrics in
Table~\ref{tab:sqa} in order to isolate those candidates as probably
{\it real} transients. Unlike machine-learning, this involved no probabilistic
classification of candidates into likely {\it reals} and {\it bogus}
transients. It only provided a means to isolate candidates for further
follow-up. The most powerful metrics and filtering logic we have found 
to label a candidate as ``interesting'' are as follows:

\begin{equation}\label{scuts}
\begin{aligned}
        chi           \, &  <     thres\_s1  \;\;\;    \& \\
        |sharp|       \, &  <     thres\_s2  \;\;\;    \& \\
        snrpsf        \, &  >     thres\_s3  \;\;\;    \& \\
        magfromlim    \, &  \geq  thres\_s4  \;\;\;    \& \\
        nneg          \, &  \leq  thres\_s5  \;\;\;    \& \\
        nneg          \, &  \neq  -999       \;\;\;    \& \\
        nbad          \, &  \leq  thres\_s6  \;\;\;    \& \\
        nbad          \, &  \neq  -999       \;\;\;    \& \\
        |magdiff|     \, &  \leq  thres\_s7  \;\;\;    \& \\
        mindistoedge  \, &  >     thres\_s8  \;\;\;    \& \\
        magnr         \, &  >     thres\_s9  \;\;\;    \& \\
        distnr        \, &  >     thres\_s10 \;\;\;    \& \\
        elong         \, &  <     thres\_s11 \;\;\;    \& \\
        |1-ksum|      \, &  \leq  thres\_s12 \;\;\;    \& \\
        kpr           \, &  <     thres\_s13 
\end{aligned}
\end{equation}
where the metric names on the left are defined in Table~\ref{tab:sqa} and the
corresponding thresholds ($thres\_si$) are specified by the
{\it \textendash{goodcuts}} input string with defaults defined in
Table~\ref{tab:inp}. The number of candidates that satisfy the above criteria
per difference image are recorded as {\it ncandgood} in the QA output table
(Table~\ref{tab:iqa}).

\begin{figure*}[ht]
\begin{center}
\includegraphics[scale=0.3]{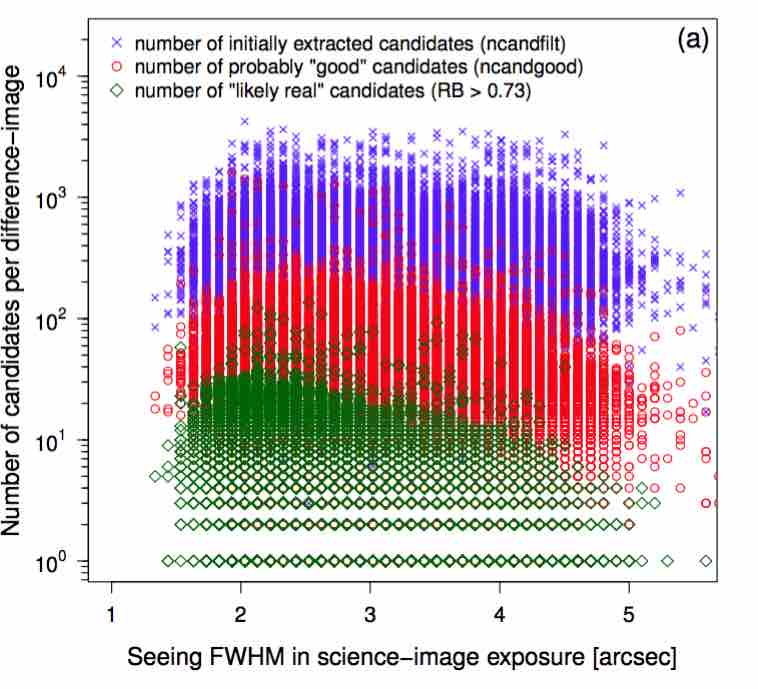}
\includegraphics[scale=0.3]{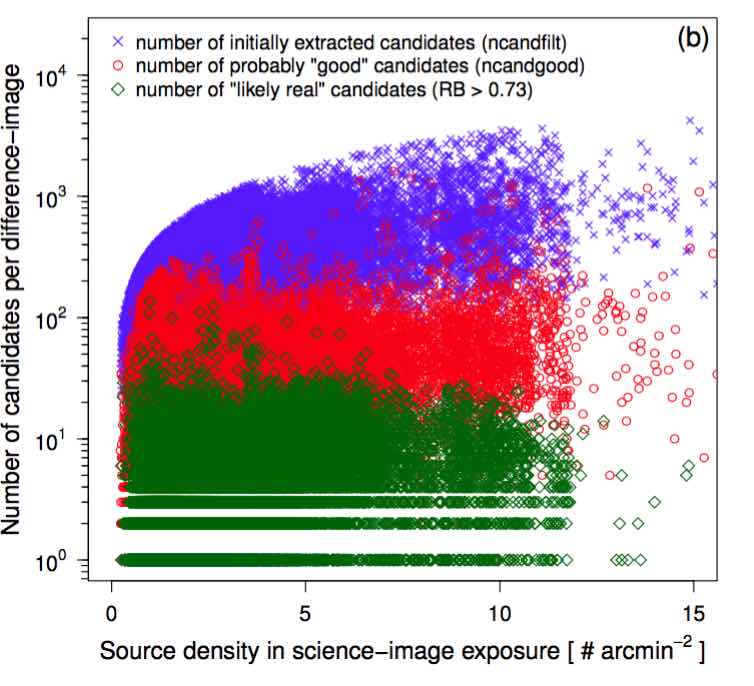}
\caption{Number of transient candidates extracted per {\it positive}
         difference image as a function of (a) seeing FWHM and (b)
         integrated source density to $R_{PTF} \simeq 20.5$ mag in the
         corresponding science image exposure. Extractions are from
         $\simeq 30,000$ difference images. To estimate the 
         number of candidates per $\text{deg}^2$, multiply the
         vertical axes by $\simeq 1.56$ $\text{deg}^{-2}$.
         See Section~\ref{srcqa} for details.}
\label{fig:ncand}
\end{center}
\end{figure*}

Figure~\ref{fig:ncand} shows the number of transient candidates extracted
from a collection of iPTF difference images acquired during late 2015 to
early 2016, mostly at high galactic latitudes. These are shown as
a function of the (point-source) FWHM and density of sources extracted
from the science image using {\it SExtractor} to a fixed magnitude limit
of $R_{PTF} \simeq 20.5$ mag. Three different levels of candidate
filtering are shown: first, the number initially extracted using
the loose filtering described in Section~\ref{filt} ({\it ncandfilt});
second, the number resulting from the simple one-dimensional cuts
in Section~\ref{srcqa} (equation~\ref{scuts}; {\it ncandgood});
and third, the number satisfying a machine-learned
classification ({\it realbogus}) score of $> 0.73$. This score
corresponds to a false positive rate of $\lesssim 1$\% (see Section~\ref{ml}). 
A noteworthy feature in Figure~\ref{fig:ncand}b is the significant reduction
in the number of initial candidates at relatively high source densities
following simple 1-D filtering (yielding {\it ncandgood}) or
machine-learned vetting. In the end, the number that are visually-scanned
and subject to scrutiny before further follow-up are those that were 
machine-learned vetted (diamonds in Figure~\ref{fig:ncand}).

We have constructed animations of difference-images for two PTF CCD footprints
made from image data acquired at $> 200$ observation epochs. These can
be accessed from the following URLs: 
\begin{list}{$\bullet$}{\leftmargin=1em \itemindent=0em}
\item{Field containing the M13 Globular Cluster: \brokenurl{http://web.ipac.caltech.edu/staff/fmasci}{/home/idemovies/d4335ccd8f2movie.html}}
\item{Field heavily used for supernova searches: \brokenurl{http://web.ipac.caltech.edu/staff/fmasci}{/home/idemovies/d4450ccd2f2movie.html}}
\end{list}
Each epochal difference image is annotated with the two candidate numbers
mentioned above (and shown in Figure~\ref{fig:ncand}): {\it ncandfilt}
(or $Nraw$ in the animation frames) and the number following machine-learned
vetting above a {\it realbogus} ($rb$) cut: $N(rb>0.73)$.

\subsubsection{Photometric Performance}

One way to assess the photometric accuracy of the difference-image 
extractions is to examine the repeatability in their photometry under
different observing conditions and/or input noise assumptions. Instead,
we explore the photometric repeatability empirically at {\it prior} source
positions across a stack of difference images generated from 
spatially-overlapping exposures acquired over a range of observation epochs. 
Unfortunately we cannot perform this test on real
flux-transients and variables because they intrinsically vary and
will confuse repeatability statistics. We have resorted to exploring the
scatter in photometric measurements from {\it forced} photometry on a list of
prior source positions detected in a reference image co-add. The majority of
the sources here will be non-variable and {\it non-detected} in the difference
images. Even though undetected, their photometric residuals will persist,
therefore providing sensitive probes of all random (and systematic)
errors affecting the end-to-end difference image construction process in a
relative sense across epochs. These residuals would arise from image
misalignments, erroneous photometric-gain matching, flat-fielding errors,
PSF-matching errors, Poisson noise from the science and reference images,
and other instrumental/detector noise. At some level, there are also
airmass-dependent color-refraction effects and astrometric scintillations.

It's important to note that these residuals will also include systematics
from the forced-photometry process itself. For example, for forced 
{\it PSF-fit} photometry, these would include errors in the PSF estimates
for each epochal image and their placement on the purported source
positions in the difference image (the prior positions selected from the
reference image). Therefore, the photometric variance inferred
using {\it forced} PSF-fit photometry is likely to overestimate the
true variance (or relative photometric accuracy) of {\it detected}
transients in a difference image. As a reminder, the primary photometry
measured for detected transients in PTFIDE is PSF-fit photometry
(Section~\ref{psffit}) where both flux and position are estimated
per source. In forced photometry on {\it prior} positions, only fluxes
are estimated.

\begin{figure*}
\begin{center}
\includegraphics[scale=0.44]{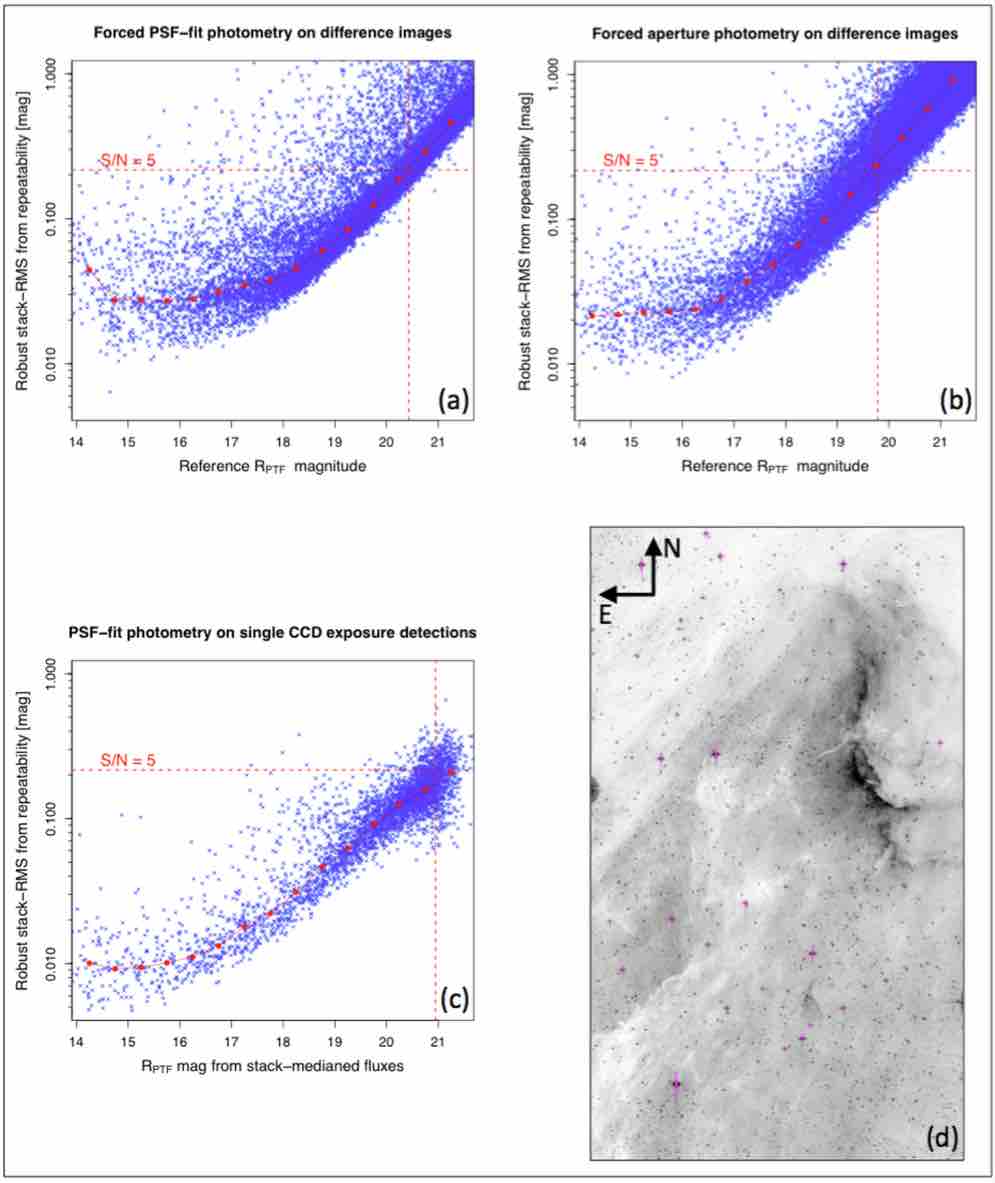}
\caption{(a) RMS in photometric repeatability from {\it forced PSF-fit}
         photometry on difference images overlapping a
         $\simeq 35^\prime\times 70^\prime$ CCD footprint falling on the
         North America Nebula. Prior positions used to seed the forced
         photometry are from the reference-image (co-add); (b) Same as (a)
         but using {\it forced aperture} photometry on the same positions;
         (c) RMS in photometric repeatability from single-exposure
         PSF-fit extractions (not forced on prior positions) from 
         a nearby CCD footprint;
         (d) A $\simeq 30^\prime\times 55^\prime$ portion of
         the reference image containing part of the North America Nebula
         and from which source positions where used for (a) and (b).
         The vertical dashed lines in (a), (b), and (c) indicate the
         approximate 5-$\sigma$ magnitude limits. The linked solid circles
         are binned medians.}
\label{fig:photperf}
\end{center}
\end{figure*}

Figure~\ref{fig:photperf}a shows an example of the {\it robust} RMS in repeated
forced {\it PSF-fit} photometry using 26,385 targets selected to $R\simeq22$
mag from the reference image co-add. A portion of this reference image is
shown in Figure~\ref{fig:photperf}d where it contains part of the
North America Nebula.
This was created from 20 good-seeing CCD images. Note that the complexity
of this field is atypical of iPTF in general, but it provides a good test
of image differencing in complex environments. Each reference-target position
probes $> 150$ difference images, generated from CCD images acquired over
several months. The robust stack RMS is based on half the $84.13 - 15.86$
percentile difference in all photometric measurements per target position.
This measure is relatively immune to outliers. This RMS is shown as a function
of reference image magnitude and can be interpreted in the context of {\it real}
transients extracted from difference images as follows. A real transient
with PSF-fit magnitude $R_{PTF}$ is expected to have a 1-$\sigma$ uncertainty
in the {\it frequentist} sense no larger than the RMS shown in 
Figure~\ref{fig:photperf}a (i.e., relative to repeated measurement if
the same event with the same intrinsic flux were re-observed).

Figure~\ref{fig:photperf}b shows stack-RMS estimates for the same prior target
positions, but using forced {\it aperture} photometry instead. A fixed
aperture of radius 6 arcsec was used throughout. Comparing with the
forced {\it PSF-fit} photometry in Figure~\ref{fig:photperf}a, there are
two noteworthy differences: first, aperture photometry results in a higher
photometric precision at bright fluxes; and second, aperture photometry has a
shallower limiting magnitude (5-$\sigma$ limits are depicted by the vertical
dashed lines). The converse of these applies to
PSF-fit photometry. PSF-fit photometry lacks precision (relative to
aperture photometry) at bright fluxes because knowledge of the underlying
PSF is more critical. Systematic
errors in the shape of the PSF-template and/or its centroiding will inflate
errors in the photometry by a greater amount. Aperture photometry is more
immune to these effects. The encouraging observation is that PSF-fit
photometry leads to a fainter sensitivity limit (or more accurate photometry
at fainter fluxes), in this case by $\simeq0.7$ magnitudes.

For comparison, Figure~\ref{fig:photperf}c shows the performance of PSF-fit
photometry extracted directly from a set of science image CCD exposures falling
in a field flanking the North America Nebula. This has a background that
is not as complex. As expected, the precision at bright fluxes is considerably
higher that that inferred from {\it forced} PSF-fit photometry on difference
images (Figure~\ref{fig:photperf}a). This is also higher than that achieved by
{\it forced} aperture photometry (Figure~\ref{fig:photperf}b).
Furthermore, the limiting magnitude from PSF-fitting on single CCD
exposures is in general deeper (by at least 0.8 mag) than all other
types of photometry performed for iPTF, for example  
{\it SExtractor}'s $mag\_auto$ measure \citep[Section~\ref{refcon};][]{ofek12}.

\clearpage

\begin{deluxetable}{p{4.5cm}p{11cm}}
\tabletypesize{\footnotesize}
\tablecaption{Transient Candidate Source Metrics and Features from
              PTFIDE\label{tab:sqa}}
\tablewidth{0pt}
\tablehead{
\colhead{Metric\tablenotemark{a}} &
\colhead{Description} \\
\colhead{} &
\colhead{}
}
\startdata
xpos &
X-image coordinate [one-based pixels] \\

ypos &
Y-image coordinate [one-based pixels] \\

ra &
J2000 Right ascension [degrees] \\

dec &
J2000 Declination [degrees] \\

magpsf &
Magnitude from PSF fit [mag] \\

sigmagpsf &
1-$\sigma$ uncertainty in PSF-fit magnitude [mag] \\

flxpsf &
Flux from PSF fit [DN] \\

sigflxpsf &
1-$\sigma$ uncertainty in PSF-fit flux [DN] \\

magap &
Magnitude from aperture photometry [mag] \\

sigmagap &
1-$\sigma$ uncertainty in magap [mag] \\

flxap &
Flux from aperture photometry [DN] \\

sigflxap &
1-$\sigma$ uncertainty in flxap [DN] \\

snrpsf &
Ratio: flxpsf / sigflxpsf \\

sky &
Local sky background level [DN] \\

nneg &
Number of negative pixels in a 7 x 7 box \\

nbad &
Number of bad pixels in a 7 x 7 box \\

distnr &
Distance to nearest reference image extraction [arcsec] \\

magnr &
Magnitude of nearest reference image extraction [mag] \\

sigmagnr &
1-$\sigma$ uncertainty in magnr [mag] \\

arefnr &
aimage (major axis RMS) of nearest reference image extraction [pixels] \\

brefnr &
bimage (minor axis RMS) of nearest reference image extraction [pixels] \\

normfwhmrefnr &
Ratio: (fwhm of nearest ref-image extraction) / (average fwhm of ref-image) \\

elongnr &
Elongation of nearest reference image extraction ($=$ arefnr / brefnr) \\

chi &
Chi value from PSF fit \\

sharp &
Sharpness value from PSF fit \\

nneg2 &
Number of negative pixels in a 5 x 5 box \\

nbad2 &
Number of bad pixels in a 5 x 5 box \\

magdiff &
Magnitude difference: magap - magpsf [mag] \\

fwhm &
FWHM from Gaussian profile fit [pixels] \\

aimage &
Windowed RMS along major axis of source profile [pixels] \\

aimagerat &
Ratio: aimage / fwhm \\

bimage &
Windowed RMS along minor axis of source profile [pixels] \\

bimagerat &
Ratio: bimage / fwhm \\

elong &
Elongation $=$ aimage / bimage \\

seeratio &
Ratio: fwhm / (average fwhm of science image) \\

mindistoedge &
Distance to nearest edge in frame [pixels] \\

magfromlim &
Magnitude difference: diffmaglim - magpsf [mag] \\
\tablebreak

ksum &
Pixel sum of psf-matching kernel for image partition containing source \\

kdb &
Differential background associated with psf-matching kernel estimate for
image partition containing source [DN] \\

kpr &
$5^{th} - 95^{th}$ percentile range of pixel values in psf-matching kernel
for image partition containing source \\

rb\tablenotemark{b} &
Real-bogus quality score from machine-learned vetting \\

strid\tablenotemark{c} &
Primary key from Stars table, if match is available \\

luid\tablenotemark{c} & 
Primary key from LU (Local Universe) table, if match is available \\

cvsid\tablenotemark{c} &
Primary key from CVs (Cataclysmic Variable Stars) table, if match is
available \\

qsoid\tablenotemark{c} &
Primary key from QSOs (Quasi-Stellar Objects) table, if match is available \\

lcid\tablenotemark{c} &
Primary key from LCs (Light Curves) table, if match is available \\

rockid\tablenotemark{c} &
Primary key from Rocks (Asteroids) table, if match is available \\
\enddata

\tablenotetext{a}{A majority of these are loaded into the {\it candidates}
                  and {\it features} relational database tables
                  (Section~\ref{schema}) to support trending
                  and machine-learned vetting (Section~\ref{ml}).}
\tablenotetext{b}{The real-bogus score is assigned following the loading of
                  of all source metrics, features, and difference image-based
                  metrics (Table~\ref{tab:iqa}).}
\tablenotetext{c}{This is assigned following an association with a pre-loaded
                  static database table (See Figure~\ref{fig:schema}).}
\end{deluxetable}

\clearpage

\subsection{Forced, sub-image (archival) Mode}\label{force}

PTFIDE can be executed in a mode where it operates exclusively on
square sub-image cutouts. This is enabled if the {\it \textendash{forced}}
switch is specified. The science and reference image cutouts
have a center (equatorial) position and side-length (pixels)
specified by {\it \textendash{forceparams}} (Table~\ref{tab:inp}).
In this mode, PTFIDE also expects as input a pre-computed (archived)
PSF-matching kernel image FITS-cube, for example initially generated
with suffix {\it \_pmtchkerncube.fits} from a prior run of PTFIDE.
The planes of this cube store the convolution kernels corresponding to
partitions of the parent image (Section~\ref{pkern}) and is supplied
via the {\it \textendash{kerlst}} input. This cube also stores metadata
on each kernel, for example, the parent-image partition pixel ranges to which 
these apply, including all gain-correction factors. 
The appropriate kernel image for the parent-image partition containing
the image cutouts (science and reference) is then applied (as in
Section~\ref{appkern}) to match their PSFs. Image-differencing is then
performed on the cutouts. 

Only a positive ({\it science -- reference}) difference image stamp with
accompanying uncertainty and mask image are generated in
this mode. There is no source detection. Further outputs are
generated if the debug switch was set (see end of Table~\ref{tab:sout}).
The purpose of this mode is to support later {\it forced photometry}
on a target position of interest. This position would be 
the same used to generate the initial image cutouts. Operating on
stamp cutouts using pre-existing kernel images is very fast, particularly
when difference-images containing specific source positions over a
historical observation range are needed.
This avoids regenerating entire difference images,
including all associated PSF-matching kernels and corrections.
It also avoids archiving entire difference images in the first place.

\section{Transients Database Schema}\label{schema}

\begin{figure*}
\begin{center}
\includegraphics[scale=0.58]{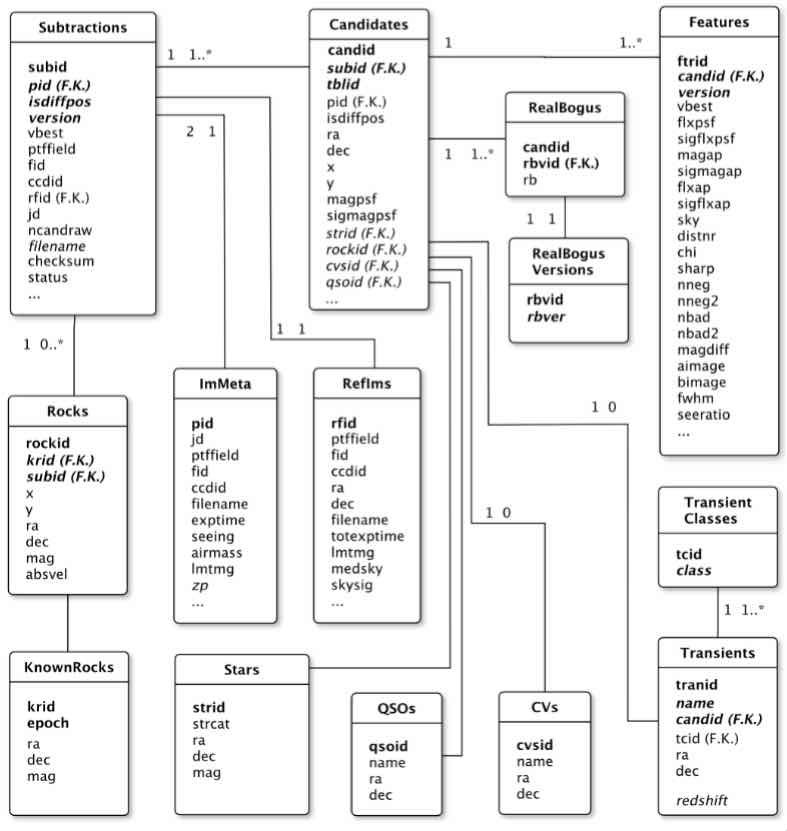}
\caption{Schema design for the iPTF Transients Database in current operations.
         See Section~\ref{schema} for details.}
\label{fig:schema}
\end{center}
\end{figure*}

This section describes the schema and related workings of the iPTF Transients
Database in current operations, along with how it will be improved for ZTF
in future. Figure~\ref{fig:schema} depicts a simplified snapshot of the
schema, in which each box represents a separate database table with a
given name and some number of columns. The major table columns are listed.
The columns in bold font are the table's primary keys.  
The columns in bold-italicized font are the alternate primary keys.  
The columns in regular font are not-{\it null} columns, 
and those in regular-italicized font are {\it null} columns 
(in which {\it null} values may also be stored). {\it F.K.} represents a
foreign key and ``1~{\jot 24pt}~1*..'' indicates a relationship of one
record to many records.

The {\it Subtractions}\/ table stores a variety of metadata for both the
positive and negative difference images generated for each processed CCD
image. The boolean {\it isdiffpos}\/ discriminates between the two
subtraction types. Many of the QA metrics from PTFIDE defined in
Table~\ref{tab:iqa} are included in this database table.
Associated with each record are the observation time ({\it jd} is the Julian
date) and foreign-key database IDs of the corresponding field, filter,
chip, reference image, and pre-processed science image.
A stored difference-image filename contains the full pathname,
and in its record, is accompanied by an MD5 checksum,
a timestamp for when the record was created, and a {\it status} flag
for whether the subtraction satisfied a number of QA criteria:
$status=1$ if so, $status=0$ if not (see Section~\ref{diffqa}).
Multiple versions of difference images are possible if the
pipelines are rerun, and therefore the $version$\/ number and $vbest$\/ flag
for which version is best (generally the latest) are also stored.
The primary key for this table is {\it subid}, while the alternate
primary keys are {\it pid}, {\it isdiffpos}, and {\it version}.

The {\it Candidates}\/ and {\it Features}\/ tables hold metadata for all
transient candidates extracted from difference images with $status=1$ only.
The records in these tables store many of PTFIDE's source-based metrics and
features, which are listed in Table~\ref{tab:sqa}.
Included are {\it null} columns for storing the foreign-key database IDs of 
the closest positional matches to astronomical objects of interest 
(as discussed below). These are updated after the corresponding
record is loaded, where a {\it null} value indicates either there is no
match or no match was attempted. While {\it candid}\/ is a unique
candidate index assigned in sequence by the database,
{\it tblid}\/ is a relative number assigned to each candidate 
for a given difference image. The versioning mechanism in the
{\it Features}\/ table is not used at this time.

The {\it RealBogus}\/ table stores the scores computed by the 
machine-learned classifier for all transient candidates (see Section~\ref{ml}).
The RB score is a value in the range [0, 1] where a higher value
indicates the candidate is more likely to be a {\it real} transient.
Improved versions of the classifier are forthcoming, and a
mechanism for tracking them in this database has been implemented.

The {\it KnownRocks}\/ table contains the predicted positions (ephemerides) and
magnitudes of all known numbered asteroids through the end of year 2019.
The predictions are spaced 1~day apart. This table contains 
$\approx 400,000$~asteroids and $\approx 1500$~epochs for each, giving a
total of $\approx 600$~million rows. For each positive subtraction,
the known rocks within its sky-footprint are found and loaded into
the {\it Rocks}\/ table. A 30\arcsec\ cone search (accounting for uncertainty)
is used to find the nearest rock to each candidate, and the corresponding
{\it rockid} is updated in the {\it Candidates}\/ record.

The {\it Stars}\/ table holds the positions and magnitudes derived from PTF
$R$-band reference-image catalogs for sources that have been classified to be
stars by a star/galaxy classifier \citep{miller16}.
Candidates are matched to stars in this table to within 1\arcsec, and the
resulting {\it Candidates}\/ record is updated with primary key {\it strid}
of the match.

A typical database query would be for transient candidates within an
observation-time and RB-score range, with an additional constraint
that each be associated with at least another candidate (potentially from 
the same source) at approximately the same sky position, but with
observation-times and magnitudes that differ by specified tolerances.
Transients that have been spectroscopically confirmed and classified are
inserted into the {\it Transients}\/ table. The iPTF Transients Marshal,
which is a web-based tool for analyzing lightcurves, is updated
continuously with records from this table. Approximately 64 thousand
classified transients are currently stored.

The database schema will be streamlined in future for ZTF so that it is
scalable and more efficient.  Sets of complete {\it Candidates}\/ records
will be fully constructed ahead of time for bulk database-loading.
This will require the unique candidate IDs to be formed predictively
rather than from a database sequence. Any redundant and unnecessary  
columns in the {\it Features}\/ table will be eliminated. Also, the
{\it Candidates}\/ and {\it Features}\/ tables may be combined into a single
table. Partitioning the {\it Candidates}, {\it Features}, and perhaps other
tables into child tables that isolate different observation-time ranges
will also be considered, as this will allow for faster queries by 
taking advantage of constraint exclusion in the PostgreSQL database.

\section{Machine-Learned Vetting}\label{ml}

The machine-learned vetting of sources is necessitated by the overwhelming
number of artifacts produced by image subtraction and subsequently extracted
during source finding.  The true ratio of real astronomical sources (referred
to as {\it reals}) versus artifacts generated by image subtraction
(referred to as {\it boguses}) is unknown since the majority of sources
extracted are unexamined.  We estimate the bogus to real ratio for PTFIDE
is typically greater than 10 to 1. This necessitates the use of
automated systems to discriminate between {\it boguses} and {\it reals}
in order to filter out unreliable candidates and prioritize detections for
further study.

We refer to systems that perform this task as ``RealBogus'' systems, a term
coined by \citet{bloom12}. The use of machine learning for the vetting of
astronomical transients extends back to PTF \citep{bloom12}.
Machine learning systems are typically statistical classifiers that are
able to score candidates on a spectrum from zero ({\it bogus}) to one
({\it real}). Classifiers are trained with annotated data exemplars as opposed
to expert-specified rules. Three machine learning systems are currently in
use for iPTF \citep{brink13,wozniak13,rebba15} for vetting outputs from the
image differencing pipeline at NERSC. The results therefrom are
combined to minimize missed detections. Machine-learned
vetting is also being used for other surveys that use image differencing for
transient discovery, for example, the Dark Energy Survey \citep{goldstein15}
and Pan-STARRS \citep{wright15}.

Here we briefly describe the construction and evaluation of the
{\it RealBogus} system for PTFIDE and leave the details to a future paper.

\subsection{Classifier Description}\label{mldes}

The {\it RealBogus} system for PTFIDE, like all its predecessors, is based
on a random forest classifier. For an overview of random forests, see
\citet{breiman01}, \citet{hastie09}, and \citet{masci14}. We use an ensemble
of 300 trees trained on 10,000 {\it real} and 10,000 {\it bogus} candidates.
The proportion of trees reporting a classification
of {\it real} is reported as the candidate's score.  Each candidate is
described via a set of features that forms an input vector into the
classifier.  The current set of 89 features are outputs from PTFIDE and
consist of both image-based and source-based features (Tables \ref{tab:iqa}
and \ref{tab:sqa} respectively). This excludes irrelevant and trivial 
information such as source IDs, database counters, positions (i.e., ra, dec),
and the constant photometric zeropoints.

\begin{figure}[ht]
\includegraphics[scale=0.32]{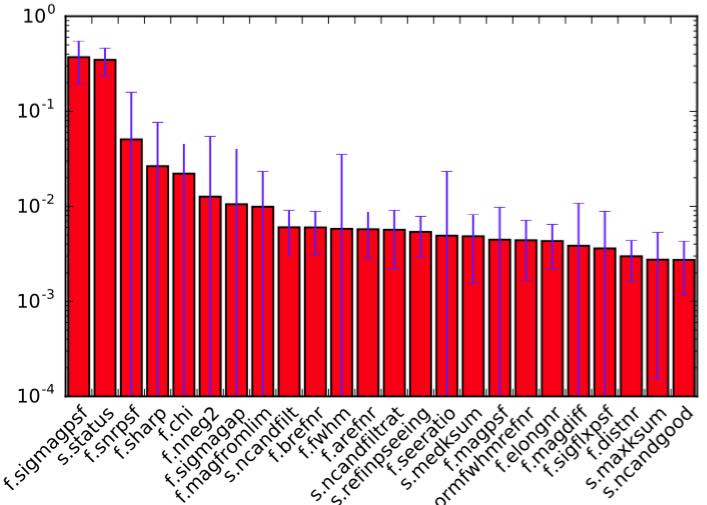}
\caption{Relative importance of the first twenty most-important features. 
         A higher ``importance value'' implies the feature is better at
         discriminating between {\it real} and {\it bogus} transients.}
\label{fig:feature_importance}
\end{figure}

Figure~\ref{fig:feature_importance} shows
a feature importance diagram that provides an estimate of the relative
importance of each feature to the classification process.  Only the twenty most
important features are displayed.  Despite the high number of features,
the top two features for discriminating between {\it real} and {\it bogus}
candidates are {\it sigmagpsf} (the 1-$\sigma$ uncertainty in the
PSF-fit magnitude) and {\it status} (a flag for whether the difference image
satisfied a number of QA criteria; see Section~\ref{diffqa}). 
It's important to note that the relative feature importance ranking
in Figure~\ref{fig:feature_importance}
does not account for any correlation between features. For example,
the {\it chi} and {\it sharp} features are expected to be highly correlated. 
Removal of one of these features will still result in a good classifier,
while removing both will not. I.e., the presence of either one but
not necessarily both is important for overall classifier performance.

\subsection{Training Data}\label{train}

The training data must be well sampled with respect to the true distributions
of {\it real} and {\it bogus} candidates on any given night.
The {\it RealBogus} system for
PTFIDE benefited from predecessor systems at NERSC in that we could
reprocess images containing real candidates discovered at NERSC and
recover them with PTFIDE.  We queried the NERSC database for all objects
that were spectroscopically-confirmed to be a supernova, variable star,
gap transient, cataclysmic variable or nova.  This resulted in 372 candidates.
We augmented this set with data belonging to the lightcurves of
these candidates at all observation epochs. This resulted in 15,168 real
candidates in total. All images containing this candidate set were
reprocessed with PTFIDE. Of these 15,168 candidates, 2,153 were
lost due to PTFIDE failures, of which 11 were
spectroscopically-confirmed. Of the remaining recoverable
13,015 candidates, we recovered matching transients for 11,075
($\simeq$85.1\%) with PTFIDE. 361 of these were spectroscopically-confirmed
of which we recovered 310 ($\simeq$85.8\%).
We reserved 10,000 of the 11,075 for training and reserved the remaining
1,075 as an independent test set for final validation.

\begin{figure*}[ht]
\begin{center}
\includegraphics[scale=0.33]{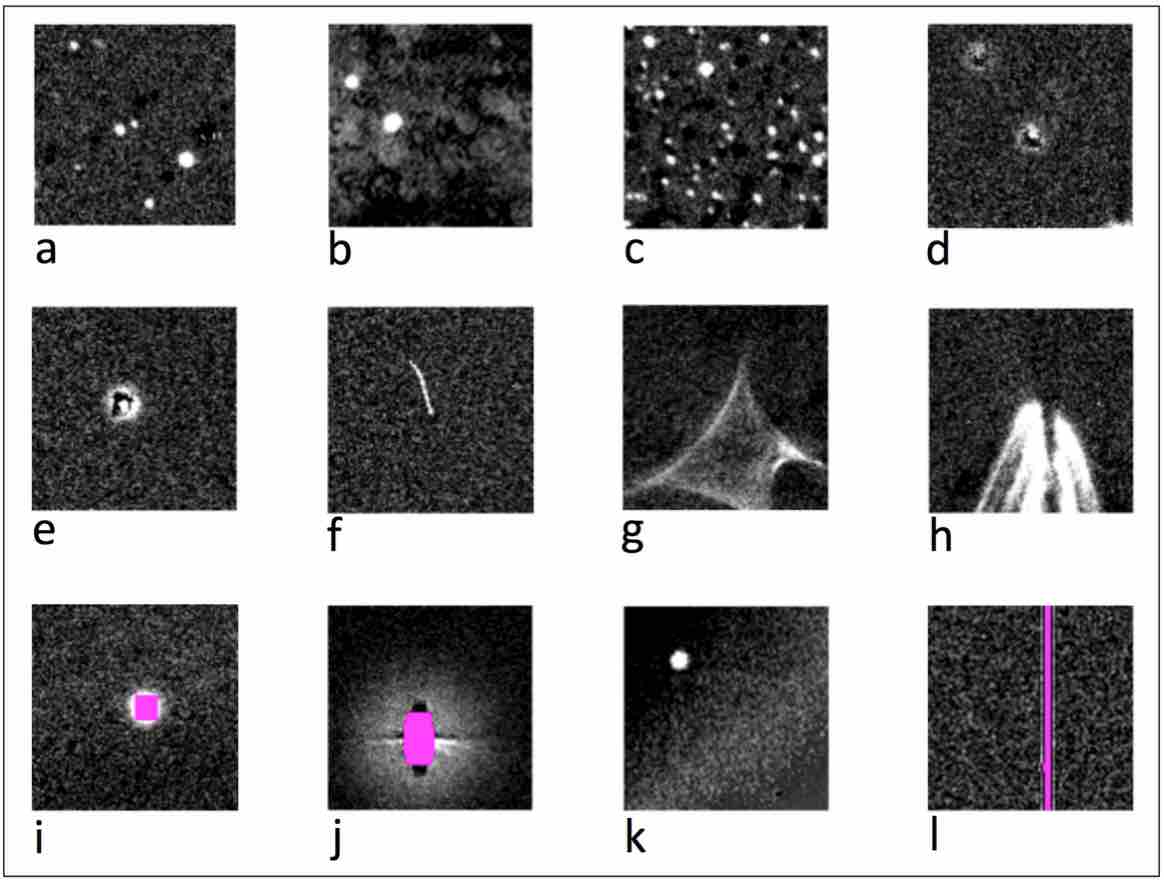}
\caption{A sample of ``bad'' difference images with ``bogus'' transients.
         Their causes are as follows:
         (a) bad astrometric calibration; 
         (b) bad photometric-throughput (gain) matching brought about by
             spatial variations in either atmospheric transparency or
             bad flat-fielding;
         (c) bad astrometric calibration in a high source-density field; 
         (d) bad PSF-matching brought about by bad seeing and
             inability of the PSF-matching kernel to accommodate the disparity
             between science and reference image FWHM values;
         (e) same as (d); 
         (f) moving-object streak;
         (g) bright-source halo artifact;
         (h) bright-source glint;
         (i) incomplete masking of a saturated source;
         (j) incomplete masking of a bright source and its halo and charge
             bleed artifacts;  
         (k) bad background matching and photometric-throughput 
             (gain) matching;
         (l) incomplete masking of the edges of a bad-pixel column.}
\label{fig:bogus}
\end{center}
\end{figure*}

{\it Bogus} candidates are pipeline artifacts that must be sampled directly
from the PTFIDE database. We randomly sampled 20,000 candidates exclusive
of known reals and declare them as {\it bogus}. We reserve 10,000 for training
and another 10,000 as an independent test set. Figure~\ref{fig:bogus} shows an
example of the various kinds of {\it bogus} transients extracted. Some are
induced from bad or inaccurate upstream instrumental calibrations, while
others are due to inadvertently unmasked detector glitches or artifacts
from the optical system.

It is impossible to ensure
the purity of our samples without examining each candidate individually.
The {\it bogus} set may include missed detections while our labeled sets of
real candidates may contain artifacts. Future work includes plans to
assess training set contamination via a machine learning technique
called active learning \citep[e.g.,][]{richards12}.

\begin{figure}[ht]
\includegraphics[scale=0.34]{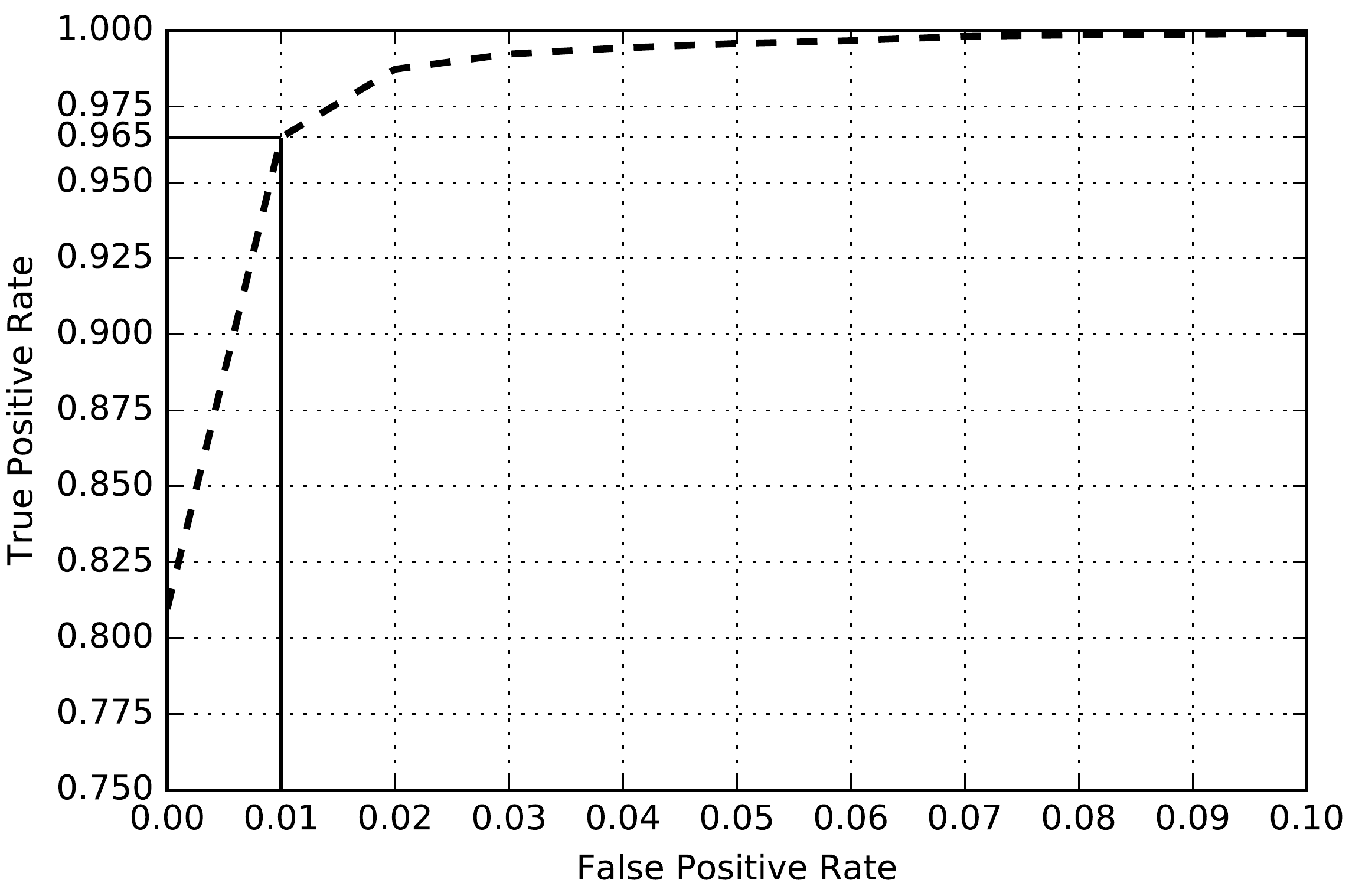}
\caption{Receiver operating characteristic (or ROC) curve. See
         Section~\ref{perfml} for details.}
\label{fig:roc}
\end{figure}

\subsection{Evaluation and Setting Decision Thresholds}\label{perfml}

We have two methods of evaluation.  The first is to use ten-fold
cross-validation in order to form a receiver operating characteristic
(ROC) curve that measures the false positive rate (FPR) and false
negative rate (FNR) at a continuum of decision thresholds from 0 to 1.
In ten-fold cross-validation, all 20,000 labeled examples (exclusive of the
independent test sets) are randomly split into 10 groups, where one fold is
held out as the test fold and the classifier is trained on the remaining nine.
The test fold is rotated and the predicted outcomes from the ten classifiers are
averaged across the folds. Methodologically, cross-validation usually assumes
examples are independent and identically-distributed.  That is not the case
here, since light curve observations from the same source (especially variable
stars) may span both the training and test folds. We use cross-validation as
a guide when comparing competing versions of the classifier, rather than for
assessing the system's overall performance.  In fact, cross-validation is
prone to overfitting if the labeled population does not well represent the
general population of candidates. Figure~\ref{fig:roc} shows the ROC
curve for the classifier using cross-validation, where the {\it y}-axis
shows the true positive rate (TPR $= 1 -$ FNR). The FNR at 1\% FPR, the
maximum FPR tolerated by the science teams, was 3.51\% from cross-validation.  

\begin{figure*}[ht]
\begin{center}
\includegraphics[scale=0.45]{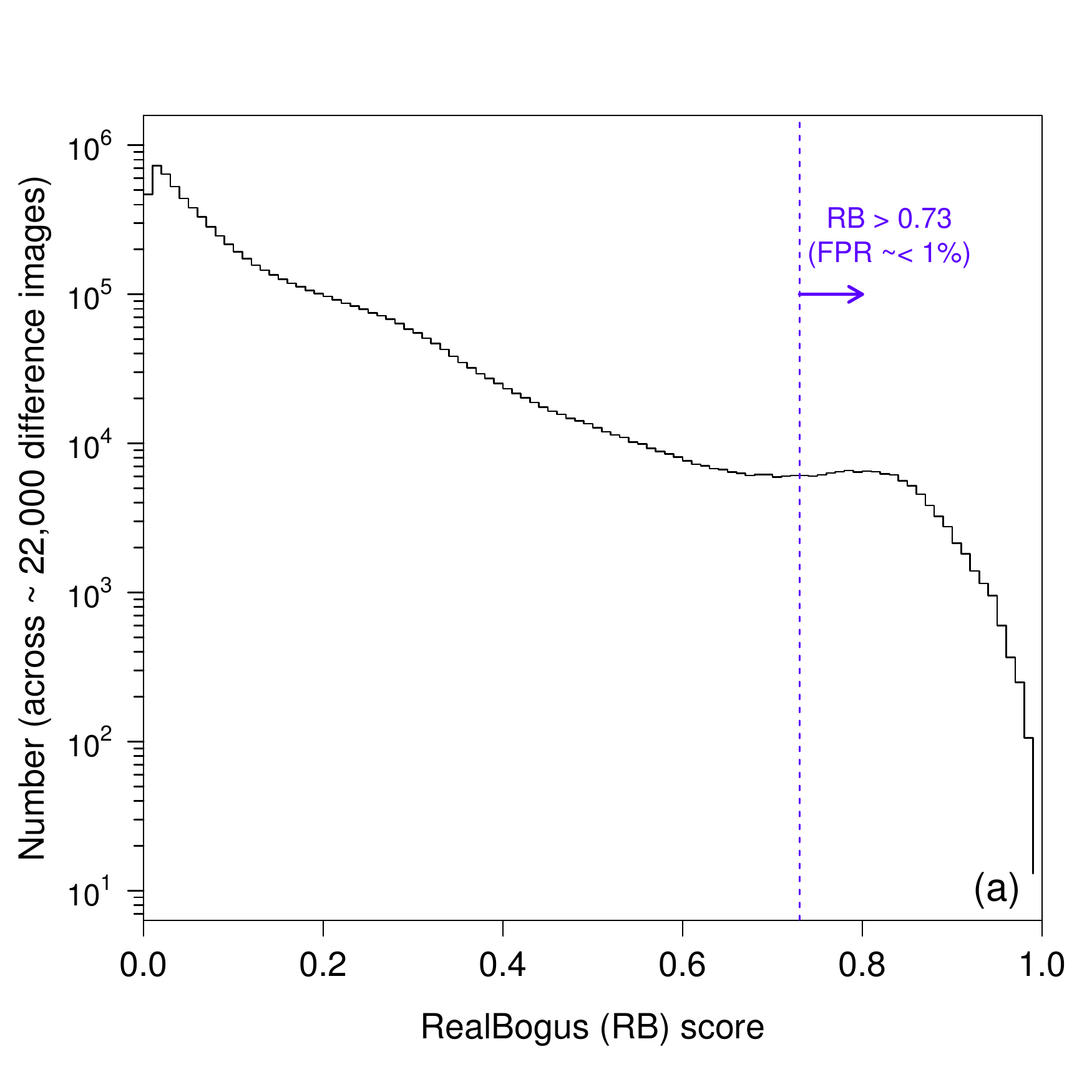}
\includegraphics[scale=0.45]{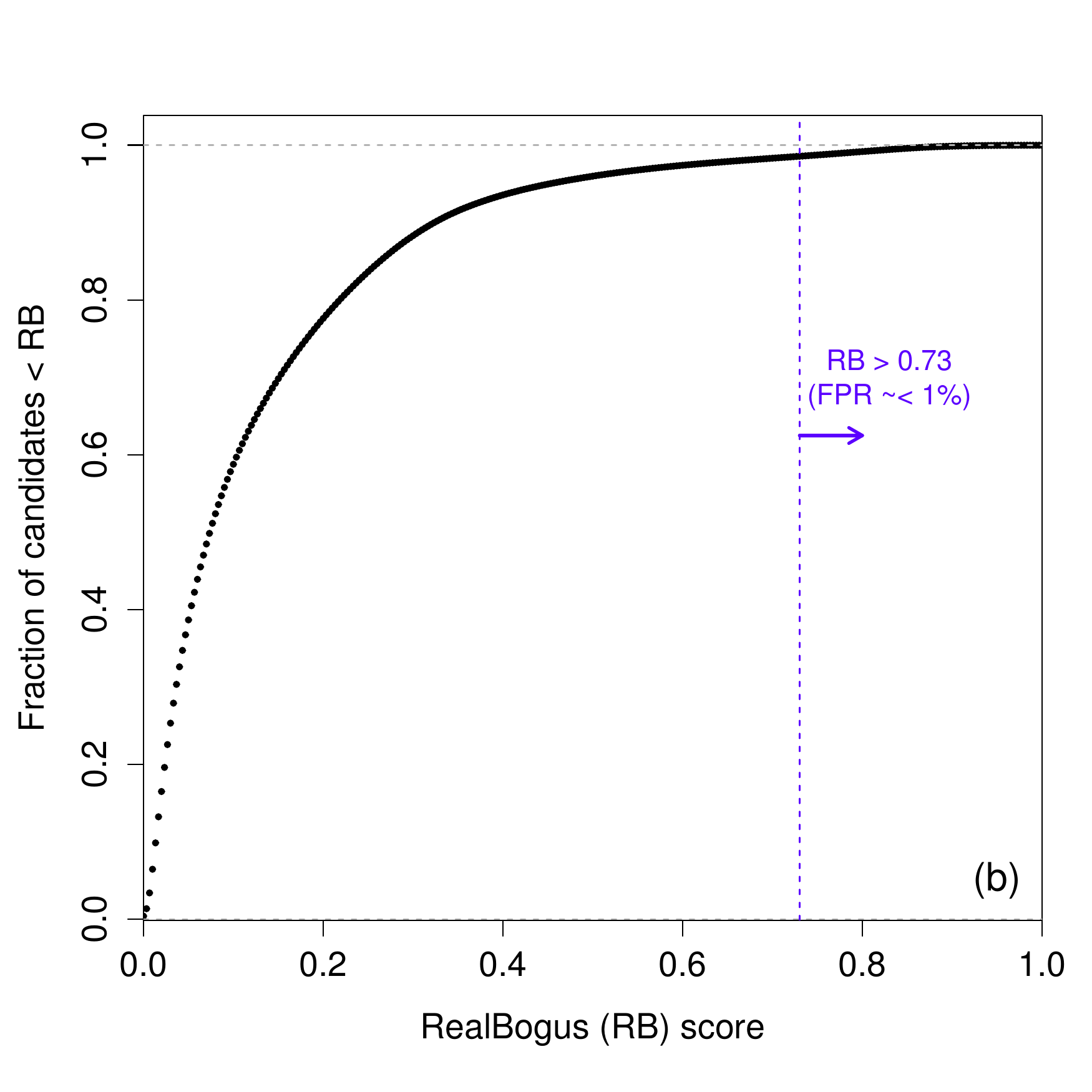}
\caption{(a) Histogram of the {\it RealBogus} (RB) scores for transient
         candidates extracted from $\simeq22,000$ $R$-band ({\it positive})
         difference images down to S/N $\simeq$ 4. These are from exposures
         acquired from Dec 2015 to Feb 2016, mostly at moderate to high
         galactic latitude; (b) Cumulative fraction as a function of RB.
         See Section~\ref{perfml} for details.}
\label{fig:ncandhist}
\end{center}
\end{figure*}

The second evaluation of the system looks at score distributions on the two
independent test sets of {\it real} and {\it bogus} examples. This gauges
performance and determines the system's decision threshold. Candidates that
score above the decision threshold are presented to the science teams for
inspection, while those below will likely remain unexamined. We classify the
candidates in the test set of randomly selected candidates and note the
threshold that admits only 1\% of the set as false positives.  Similarly, using
the independent test set of known reals, we classify this set and note
the threshold that admits only 5\% as false negatives (or missed detections).
The thresholds resulting in a 1\% FPR and 5\% FNR were 0.735 and 0.724
respectively.
As a result, the decision threshold for the {\it RealBogus} classifier was
set to 0.73. Figure~\ref{fig:ncandhist} shows the distribution of 
{\it RealBogus} scores obtained from PTFIDE run on iPTF data.   
As seen in Figure~\ref{fig:ncandhist}a, the decision threshold of
0.73 corresponds to a plateau followed by a knee in the distribution
above which the FPR falls below 1\%.

\subsection{Further Work on Machine-Learned Vetting}\label{mlfw}

Areas of future work include plans for identifying and eliminating training
set contamination.  We have built an active learning prototype that
identifies candidates that are likely mislabeled, and present those to the
science teams for cleaning.  We have also developed a new way to randomly
sample against the PTFIDE candidates database to ensure our {\it bogus} sample
is not overly biased towards certain types of artifacts and is representative
of the full distribution of observing conditions.  Finally, we have used machine
learning to analyze the frequency of certain types of {\it bogus} artifacts
produced by PTFIDE in order to understand the software and
faciliate improvements.  Details on these methods will be published
in a forthcoming paper.

\section{Lessons Learned}\label{lessons}

The iPTF realtime processing system has evolved considerably over the last
few years through feedback received from the various science programs.
The system was developed by a small team with limited resources.
Communication between the development team and users of the products
was paramount. The success of a large astronomical survey requires
(i) a clear definition of the science goals and deliverables needed to
achieve these; (ii) a tractable system engineering plan to enable (i),
given the available resources. Below we list some of the challenges
and pitfalls encountered over the course of development and how these
were addressed. We also present thoughts on how specific aspects 
could have been improved if resources allowed.

\begin{enumerate}
\item{Tuning and optimization of all processing components is an iterative
      process that requires real on-sky data acquired with your survey
      instrument. Do not rely on the commissioning period to optimize
      everything to perfection. Resources (and schedule) are limited.
      More eyes on the data, the better, and this can only occur by
      harnessing the expertise of the scientific community. Immediate
      visibility to data products is therefore crucial in the early
      phases of the survey.}
\item{Implement offline versions of your primary production
      pipelines to allow for experimentation and ongoing tuning. This
      ensures minimal disruption to the production system that is serving
      users. Communicate all planned updates in advance and only deploy
      when the stakeholders have confirmed and understood the updates.}
\item{Allow versioning control of all pipeline software parameters and
      instrumental calibrations {\it together} with the software versions
      they were optimized for. This will allow for easier traceability
      and reproducibility of specific science products in future.}      
\item{Have a well defined quality-assurance plan for the entire observing
      and data-processing system. This entails implementing QA metrics
      for each subsystem and a means to communicate these across 
      all subsystems. For example, implementing automated checks at the
      raw-data level (close to realtime) alerts the team of bad data
      and that no products are expected downstream.
      A modular and {\it visible} end-to-end QA/alerting system allows for
      easier accountability of missing products, traceability of errors,
      and recovery.}
\item{Related to the previous point, routine monitoring should include
      overall performance of the network, data-transfer rates, and all
      aspects of the compute cluster.}      
\item{Assign ownership and responsibility of key components of
      the system to individuals of the team. For example, this may consist of
      collecting performance metrics, analysis reports, and/or providing
      feedback for improving aspects of the data-processing system. Having all
      communication channels defined at the outset will allow priorities to
      be better managed.}
\item{Time-domain surveys are dynamic, literally. For example, science goals
      or their priorities may change over time in response to
      scientific outcomes or analyses early in the survey. Goals may also
      change in response to unforeseen problems in instrumentation, hardware,
      and/or algorithmic or processing details. Be prepared to adapt.
      {\it Modularity} in all pipeline software and hardware components to
      accomodate change and growth is therefore crucial.}            
\item{In the context of realtime discovery using image-differencing,
      the accuracy of upstream calibrations is crucial. This primarily refers
      to astrometric and photometric calibration. Flat-fielding in particular
      is important for the latter. Proper trending and QA of all calibration
      products prior to use is therefore necessary to avoid a flood of
      false-positives downstream.
      Early notification of bad calibrations and a means to either
      recreate them on-the-fly or fallback to archived versions should
      be planned.}      
\item{Have a plan to assess the overall performance of image-differencing and
      transient-discovery over the course of the survey. This may
      involve for example tracking the relative efficiency and reliability
      of candidates extracted from pre-defined survey fields known to contain
      a good statistical sample of variable stars. Alternatively, one could
      inject synthetic episodic transients for offline analysis. The important
      thing is that the appropriate metrics and methodologies to enable
      this monitoring be identified and implemented prior to
      commencement of the survey.}      
\item{Image-differencing is a game of (i) minimizing false-positives at the
      expense of also maintaining a low false-negative rate, and (ii)
      maximizing the photometric accuracy.
      These sensitively depend on the initial extraction S/N.
      The power of using machine learning to probabilistically classify
      candidates into either {\it real} or {\it bogus} cannot be overstated.
      False-positives are inevitable. We advise setting a
      maximum tolerable threshold for their occurence to
      enable tuning of the relevant extraction and scoring thresholds.
      These will evolve as algorithms and software improve.}   
\item{The efficacy of a machine-learned vetting infrastructure crucially depends
      on the data it was trained on (in the context of
      {\it supervised} learning). This should be kept in check over the
      course of a survey according to the different science applications and
      possible changes in survey design. For example, if a decision is made
      to survey more of the galactic plane, the machined-learned classifier
      should be retrained accordingly using data from the same region.
      This minimize biases when predicting the reliability of transient
      candidates.}
\item{A {\it supervised} machine-learned classifier will have been
      trained on products from a specific version of pipeline software.
      Any algorithmic or parameter changes in the pipeline usually requires
      a retraining of the classifier. This dependency will incur
      a delay in the software delivery cycle and must be accommodated.
      We have not yet streamlined this delivery and integration process since
      classifier (re)training can be time-consuming. We advise
      that any classifier-retraining be performed on a stable version of
      the pipeline software. Both can then be updated as shortcomings are
      identified during the survey.}
\item{Plan on reprocessing any or all of your data, for example, to recover
      from failures in processing and/or hardware outages. This also enables
      one to regenerate products for a future archive using a consistent set
      of pipeline parameters and software. These may have evolved over the
      course of the survey.}
\end{enumerate}

\section{Enhancements and Future Work}\label{enhance}

A number of shortcomings were identified over the course of development of
the iPTF Discovery Engine. Some of these are at the algorithmic level and some
relate to data management practices. In 2017, iPTF will be replaced by
the Zwicky Transient Facility (ZTF; see Section~\ref{intro}). The higher
data rates and volumes will require a redesign of some of the subsystems
to minimize processing latencies and the delivery of transient candidates
for scanning. The planned upgrades are as follows:

\begin{enumerate}
\item{Improve the efficiency of loading and retrieval of candidates into/from
      the Transients Database described in Section~\ref{schema}. The plan is to
      periodically construct lists of pre-machine-vetted candidates
      ahead of time at intervals throughout a night and batch-load
      them. Source features and metrics will be consolidated into single
      (flatter) database tables. We will also consider retaining only
      candidates and associated metadata for the last 30 nights or more to
      enable more efficient near real-time discovery and lightcurve generation.
      Older transient candidates will roll-off to a growing archival database.
      The reason for this is to keep the number of candidate records for
      active-querying (close to their discovery epoch) relatively small.}
\item{Transient candidates and image-cutouts for human scanning will need to
      be delivered for external viewing using pre-defined
      (or cached) database queries submitted by an automated process at
      regular intervals throughout a night. Currently for iPTF, there is no
      limit on how many queries can be submitted. Having many
      scanners submit similar queries in an uncoordinated manner has
      led to severe bottlenecks.}
\item{Astrometric calibration needs to be made more robust against changes
      in source density, seeing, depth, atmospheric refraction, and telescope
      tracking. This includes the ability to properly model and capture
      time-dependent distortion effects from the optical system and atmosphere.}
\item{Absolute photometric calibration will need to be performed on a
      {\it per-image} basis in the {\it realtime} pipeline. The use of
      PSF-fit photometry in particular is paramount. PSF-fitting will
      automatically account for seeing variations and regions with high
      source-confusion through its de-blending ability. This will ensure
      that photometric zeropoints can be accurately determined for a
      larger fraction of the images. As discussed in Section~\ref{gm}, 
      the zeropoints are refined using big-aperture photometry to enable
      more accurate gain-matching prior to image-differencing. This method
      is fragile and has not been reliable.}
\item{Extend PTFIDE to include some of the optimal methodologies for
      co-addition, image-differencing, source detection, and photometry
      presented in the detailed study by \cite{zackay16}.}
\item{Consider using pre-classified star catalogs constructed initially
      from reference image catalogs (e.g., via machine-learning) to properly
      seed inputs for deriving PSF-matching kernels.}
\item{Consider dynamic updates to reference image products as the survey
      proceeds in order to use the best-quality epochal data acquired to date. 
      In other words, the reference-image library can be progressively
      refined to ensure optimal image-differencing.}
\end{enumerate}

\section{Conclusions}\label{conc}

We have described a transient-discovery engine (IDE),
currently in use to support near real-time discovery for iPTF at IPAC/Caltech.
A refined version will be used for ZTF in future. Guided by previous
implementations of the image-subtraction problem, this paper reviews our
algorithms, optimization strategies and machine-learned vetting scheme.
Once tuned, the pipeline requires little intervention and is resilient
to bad input data and/or inaccurate instrumental calibrations.
Our development approach was to make all processing steps as modular
as possible to allow for easier debugging and tractability.

Our goal has been reliable transient discovery and robustness. The methods
were refined using the knowledge gained from 6$+$ years of archived
science-quality PTF data. The elements we find that are most crucial
to image-differencing performance, and hence the efficiency and 
reliability of transient discoveries are:
(i) astrometric calibration; (ii) flat-fielding; and
(iii) related to this, photometric calibration (either relative or absolute).
Having these calibrations optimized (in the maximal S/N sense) paves the way
to more accurate PSF-matching and image-differencing.
This also relieves the amount of work needed downstream to weed out
false positives, by both human and machine. Despite differences in the details
of instrumentation, image quality and/or survey design, IDE provides a
testbed for future large time-domain surveys.

\begin{acknowledgments}

This work was funded in part by the iPTF and ZTF projects at the California
Institute of Technology.
iPTF is a partnership led by the California Institute of Technology and 
includes the Infrared Processing \& Astronomical Center; Los Alamos 
National Laboratory; University of Wisconsin at Milwaukee; 
Oskar-Klein Center of the University of Stockholm, Sweden; Weizmann Institute 
of Sciences, Israel; University System of Taiwan, Taiwan; the Institute for 
Physics \& Mathematics of the Universe, Japan; Lawrence Berkeley National
Laboratory and the University of California, Berkeley.
ZTF is funded by the National Science Foundation under grant no. AST-144034.
MMK acknowledges support from the National Science Foundation PIRE GROWTH
award. AAM acknowledges support for this work by NASA from a Hubble Fellowship
grant: HST-HF-51325.01, awarded by STScI, operated by AURA, Inc., for NASA,
under contract NAS 5-26555. Part of the research was carried out at the Jet
Propulsion Laboratory, California Institute of Technology, under a contract
with NASA.

The pipelines use a number of software packages from other institutions
and past projects (see Table~\ref{tab:sw}), for which we are indebted.
Portions of the analysis presented here made use of the Perl Data Language
(PDL) developed by K. Glazebrook, J. Brinchmann, J. Cerney, C. DeForest,
D. Hunt, T. Jenness, T. Lukka, R. Schwebel, and C. Soeller and can be
obtained from \brokenurl{http://pdl.perl.org}{} PDL provides a
high-level numerical functionality for the Perl scripting
language \citep{glaze97}.

We thank the anonymous referee for valuable comments that helped improve
the quality of this manuscript.

\end{acknowledgments}

{\it Facilities:} \facility{PO:1.2m}



\end{document}